\documentclass[letterpaper,11pt]{article}
\usepackage{pb-diagram}
\usepackage[psamsfonts]{amssymb} %see CTAN
\usepackage{graphicx}
\input{epsf} % this works with UNIX at LANL

\newcommand{\defeq}{{\buildrel \rm def\over=}}
 \newcommand{\n}[1]{\index{#1}}
 
 \def\wt{\widetilde}
 \def\hra{\hookrightarrow}
 \def\fra{\leftrightsquigarrow}
\def\bfq{{\bf q}}
\def\bft{{\bf t}}
\def\wrt{with respect to\ }
\def\ifff{if, and only if,\ }
\def\po{{\cal P}}
\def\lam{{q}}%was\lambda
\def\pd{polar decomposition\ }
\def\ot{\!\otimes\!}
\def\tf{.\,\dot{ }\, . }
\def\Ra{\Rightarrow}

\def\ra{\rightarrow}
\def\lra{\longrightarrow} 
\def\1f{ { }_1\hskip-1pt F_2 }

\def\1{ \mathbb 1 }

\def\gg{ \Gamma}
\def\H{ \mathbb H }

\def\Q{ \mathbb   Q}
\def\S{ \mathbb S }

\def\U{ \mathbb U }
\def\CircleAccent{\mathaccent"7017 }
\def\ca{\CircleAccent}

\def\nin {\noindent }
\def\nl {\newline }
\def\w{\widetilde}
\def\d {,\dots,}

\newcommand{\R}{{\mathbb R}}
\newcommand{\Z}{{\mathbb Z}}
\newcommand{\C}{{\mathbb C}}

\newcommand{\ie}{{\it i.e.\ }}        
\newcommand{\eg}{{\it e.g.\ }}

\newcommand{\cf}{{\it cf.\ }}

\def\ss{\textsf{S}}
\def\v {\overline}
\def\w {\widetilde}

\def\sk{{\sum_{k}}}
\def\SC{{Schr\"odinger}}
\def\lk{\lam_k}
\def\sgk{{\sum_{k} \lam_{k}\Gamma_{k}}}
\def\bo{{\bf 0}}
\def\b1{{\bf 1}}
 \def\bfA{{\bf A}}
 
\def\bfB{{\bf B}}

\def\bfD{{\bf D}}
\def\bfe{{\bf e}}
\def\bff{{\bf f}}
\def\bfg{{\bf g}}

\def\bfi{{\bf i}}
\def\bfk{{\bf k}}

\def\bfj{{\bf j}}
\def\bfI{{\bf I}}

\def\bfn{{\bf n}}
\def\bfP{{\bf P}}
\def\bfp{{\bf p}}
\def\bfr{{\bf r}}
\def\bfS{{\bf S}}
\def\bfs{{\bf s}}
\def\bfT{{\bf T}}

\def\bfu{{\bf u}}
\def\bfv{{\bf v}}
\def\bfw{{\bf w}}

\def\bfx{ {\bf x}}

\def\bfz{ {\bf z}}
\def\({ \left(}
\def\){ \right)}
\def\<{ \langle}
\def\>{ \rangle}
\def\tf{ \therefore\ }
\def\mf{ \mathfrak  }
\newcommand{\deq}{{\,{\buildrel {\rm def}\over =}\,}}

\def\ON{{\rm orthonormal}}
\def\HS{{\rm Hilbert-Schmidt}}
\def\Hsp{{\rm Hilbert space}}
\def\ham{{\rm Hamiltonian}}
\def\hor{{\rm horizontal}}

\def\all4{{\rm\ for\ all }}
\makeatletter
\usepackage{theorem}
\def\ifundefined#1{\expandafter\ifx\csname#1\endcsname\relax}
     \theorembodyfont{\slshape}
        \newtheorem{thm}{Theorem}[section]
     \newtheorem{prop}[thm]{Proposition}
     \newtheorem{lem}[thm]{Lemma}
     \newtheorem{cor}[thm]{Corollary}

     \theorembodyfont{\upshape}

\ifundefined{proof}

\newenvironment{proof}[1][\proofname]{\par
  \normalfont
  \topsep6\p@\@plus6\p@ \trivlist
  \item[\hskip\labelsep\scshape
    #1{.}]\ignorespaces
}{%
  $\qed$\endtrivlist
}
\newcommand{\proofname}{Proof}
\fi

\newsavebox{\savepar}

\newcommand{\AMSMSC}[2]{\begingroup \def \protect 
{\noexpand \protect \noexpand }\xdef \@thefnmark { }\endgroup 
\@footnotetext{{1991 \it Mathematical Subject Classification.\/} Primary: 
#1; Secondary: #2.}}

\def\p@enumi{}
\makeatother

\newcommand{\comment}[1]{}

\def\qed{\blacksquare}
\ifundefined{qed}
    \DeclareMathSymbol{\qed}{0}{AMSa}{"03}
\fi

\newcommand{\eqref}[1]{\textup{Eq.~(\ref{#1})}}

 \title{Extension of   Quantum Mechanics to Individual Systems}
  \author{James Ax\thanks{Visiting Research Scholar,\ E-mail:
\texttt{jimax@princeton.edu}}\\
   \and
    Simon Kochen\thanks{E-mail: \texttt{kochen@math.princeton.edu}}   }

\begin{document}

\maketitle
\vskip-.25in
\centerline{Mathematics Department, Princeton University}
\setlength{\baselineskip}{5mm}
%\setlength{\baselinestretch}{1}

%ееееееееее

\begin{abstract}
\setlength{\baselineskip}{4mm}

 The Copenhagen interpretation describes individual
 systems
 using the same Hilbert space formalism  as does the statistical 
ensemble interpretation (SQM) . This leads to the
well-known paradoxes surrounding the Measurement Problem.
% This paper extends
We extend this common
 mathematical structure to encompass certain natural
bundles with Hamiltonian-dependent connections over the Hilbert sphere $\bfS$. This
permits a consistent extension   of the statistical interpretation to
interacting individual systems.

Suppose  \textsf{V} is a physical system in interaction with 
 another system \textsf{W}.
 The standard state vector
 $\gg(t)$ of the  two interacting systems 
 has a set of polar decompositions
$\gg=\sum_k \lam_k\,\phi_k\ot \psi_k, $ with the  $\lam_k$ {\it complex}.
 These are
parameterized by the right toroid $T$ of amplitudes $\bfq=({\lam_k})_k$
 and  comprise a {\it singular} bundle over  $\bfS$,  
 the enlarged state space of $ \textsf{U}= \textsf{V}+ \textsf{W}$.
The evolution  of $\bfq$ is determined via the connection on this bundle. 
  We prove that each fiber $T$ has a unique natural
 convex 
partition $\{{\mf p}_1,{\mf p}_2,\cdots\}$
yielding the correct SQM probabilities, since the  
 circle of unit vectors which generate the ray 
 corresponding to the  SQM state of $\gg$ 
 intersects ${\mf p}_j$ in an arc of length $|\lam_j|^2.$
 In the extended theory, 
$ \textsf{V}$ is in the  state $\phi_j$ (and synchronously 
$ \textsf{W}$ is in the  state $\psi_j$) precisely when $\bfq\in
{\mf p}_j$. This refines the assertion of SQM which assigns to $ \textsf{V}$ only
the mixed state $\sum_k |\lam_k|^2\, |\phi_k\>\<\phi_k|.$

In the new interpretation, rays in Hilbert space correspond to ensembles,
while unit vectors in a ray correspond  to individual members of such an ensemble. 
 The apparent indeterminism  of SQM is thus 
 attributable to 
the effectively random distribution 
of initial phases. 
\end{abstract}

%%%%%%%%%%%%%%%%%%%%%%%%%%%%%
\bigskip
\bigskip
\bigskip
\bigskip

\noindent American Mathematical Society Subject Classification:
                   81Qxx    % QM 

\noindent Physics and Astronomy Classification Scheme:
                   03.65.Bz, % Foundations, theory of measurement, 
                             %  miscellaneous theories (including
                             %  Aharonov-Bohm effect, Bell 
                             %  inequalities, Berry's phase)
            
\smallskip
\smallskip
\noindent {Key Words:}   Berry's phase, polar decomposition, bundles, connections,
convex partitions, Kochen-Specker theorem
%\keywords{ Kochen-Specker theorem, hidden  variable theories, gauge theory}
%%%%%%%%%%%%%%%%%%%%%%%%%%%%%

\newpage
%еееееееееееееTCTCTCTCTTCTCTTCTCTCTCTTCTCTTCTCTCTCTTCTCT
\tableofcontents
%еееееееееееееTCTCTCTCTTCTCTTCTCTCTCTTCTCTTCTCTCTCTTCTCT
\listoffigures
\newpage
\section{Introduction}\label{intro}
\setcounter{equation}{0}

\begin{quotation}%p111 of Dictionary of Mathematical Quotations
There is geometry in the humming of the strings.
\emph{Pythagoras} 
\end{quotation}

The purpose of this paper is to extend the formalism of quantum mechanics to model
the dynamics of individuals interacting with one another. 
The extension will first require  a modest re-interpretation and clarification of the existing
formalism. In the new interpretation, rays in Hilbert space ${\cal H}$, (\ie the elements of
the projective space $\bfP:=\bfP({\cal H})$) will correspond to ensembles,
while unit vectors in a ray will correspond  to individual members of such an ensemble. 
Second, we need to 
augment the unit sphere $\bfS:=\bfS({\cal H})$ in   ${\cal H}$ 
by equipping it with
 a toroidal bundle ${\po}$ over
$\bfS$ together with a Hamiltonian-dependent
 connection on ${\po}\times \R$. We call ${\po}$ \textbf{the polar bundle.}
A  key  aspect  of this extension of the mathematical
structure   consists of a novel but natural way of
partitioning each of the toroidal  fibers of ${\po}$.

\subsection{ IQM State =  SQM State  + Phases}\label{sqm+}
{\bf We denote by IQM  the proposed new theory 
 of individual physical systems, 
while we use SQM to denote the standard statistical assertions of quantum mechanics,
which IQM extends.}

For much of this paper we will adopt a ``top-down'' perspective: we start from a
total system \ss\ and consider a subsystem $\ss_1$ together with its
complement $\ss_2$,  symmetrically.
The IQM state of a composite system $ \ss=\ss_1+\ss_2$ consists of an element of the fiber
${\po}_\gg$
above a  vector $\gg$ in the unit sphere $\bfS$ of ${\cal H}.$  
This fiber can be
thought of as the set  of  all the possible complex {\bf polar decompositions} of $\gg$:
\begin{equation}\label{PD0}%еееееееееееееееееееееееееееееееее
\Gamma=\sum_{k} \lam_{k}\,\Gamma_{k}\,, 
\end{equation}%ееееееееееееееееее      еееееееееееееее
where the  $\lam_{k}\in\C,$ and the $\gg_k$ are bi-orthonormal. Thus each 
$\gg_k$ is of the form $\phi_k\ot\psi_k$ where  the  $\phi_{k}$ and the $\psi_{k}$ are
orthonormal. We are using  here the conventional practice of taking  our (pure) 
vector
 states for
 $\ss_1+\ss_2$ from the Hilbert  space ${\cal H}_1\ot{\cal H}_2.$

The requisite properties of the polar decompositions   of $\Gamma$ are
elaborated   in Appendix~\ref{AppPolar}. We note that their totality forms a
{\bf right toroid}
$\bfT(\bfr,\gg)$ isometric to
$\prod_k \bfS^1(r_k),$ the product of circles with radii $r_k:=|\lam_k|.$ 
(We reserve the term {\bf torus} for the equal radii case.) Thus the
additional
 information  they
carry about the state of $\ss$ beyond $\gg$ is a compounding
of the extra phase data utilized in the
first step of refining the ray to a  unit vector $\gg$ in the ray.
 In other words, the polar bundle
${\po}$ can be regarded as the natural amalgamation of the 
 polar decomposition  with  the Hopf bundle  $\bfS\ra\bfP.$

We stress that the state represented by an element of $\po$ is a state  of the 
{\it composition} $\ss=\ss_1+\ss_2$; it depends on all three of $\ss,\ss_1,\ss_2$, 
although any two of them determine the other. To emphasize this contextuality, 
 we refer to these states  as {\bf polar states} of  $\ss=\ss_1+\ss_2$. Another name 
that might be appropriate is {\it joint } state, although this does not signal the special 
way it is formed.

Throughout this paper, whenever $\alpha$ is a non-zero vector in a Hilbert space, 
 $[\alpha]$ will  denote the ray $\C\alpha$ it   generates. 
 In  the polar  decomposition, only $[\phi_{k}]$ and
$[\psi_{k}]$   are  determined, \ie 
$
\phi_{k}$ and
$\psi_{k}$  are only determined up
  to phase factors but once they are fixed then the
  $\lam_k\in\C$ are uniquely determined.
  Let $\gg=\gg(0)$ be specified at $t=0.$ Then the $r_k=|\lam_k|$ are determined.
  There is precisely a
  toroid $\bfT(\bfr)$ of possibilities for the ${\bfq}:=(\lam_1,\cdots).$
  Suppose $\bfq=\bfq(0)$ is fixed at $t=0.$ Then the
  $\gg_k(0):=\gg_k:=\phi_k\ot\psi_k$ are also determined, although $\phi_k$ can still
  be multiplied by an arbitrary phase factor as long as $\psi_k$ is multiplied by its reciprocal.
 
 In Section~\ref{section3EnlargingStates}, we define the
{\bf dynamical connection} on the bundle $\bfS\times\R\ra\bfP\times\R $ as 
the unique connection  with the property that
for any unitary evolution and any $\gg\in\bfS$, the paths $t\ra(\gg(t),t)$ are
horizontal.
This connection defines a unique evolution of the bi-orthonormal frame
$\gg_k(0),\, t\mapsto\gg_k(t)$, compatible with  the \SC\ evolution $t\ra\gg(t)$  of
$\gg(0).$ 
For the $[\gg_k(t)]$ are uniquely determined by $\gg(t)$, and so once the $\gg_k(0)$ 
are specified, the connection determines the  $\gg_k(t)$. 
This, in turn, defines an evolution of the
amplitudes
$\lam_k=\lam_k(0)$, viz.  $t\ra\bfq(t):=\(\lam_1(t),\cdots\).$ Now $\bfq(t)$
lies in the toroid $\bfT(\bfr,\gg(t))$ above $\gg(t)$.
  
  We prove in Section~\ref{Pyth}  a precise
theorem whose rough content is as follows:

{\it   There is a unique natural way of partitioning any  n-dimensional right toroid 
  with basepoint ${\bf1}$ into $n$ convex subsets so that
  the $k$-th member ${\mf p}_k$ of the partition  
 ${\mf P}$ meets  every
  translate of the diagonal circle in an arc of length  $2\pi r_k^2.$
}

 Each $\mf p_k$ is,
 essentially, a convex neighborhood of the circle $ C_k\deq\bfS^1(r_k)\gg_k+\sum_{j\ne k} r_j
\gg_j$ in $\bfT({\gg(t)}).$  
  It is a property of $\mf P$ that, regarding the initial phase of $\gg(0)$
as  randomly uniformly distributed on the circle $\bfS^1(1)\gg,$ the probability that
$\ss_1$ is in the state $\phi_k$ is  $|\lam_k|^2.$ This implies that IQM is consistent
with SQM. Thus, the probabilistic nature of SQM derives from the indeterminacy of the initial 
data. Therefore, we may, {\it  without contradicting SQM}, go beyond it, and hypothesize that
{\it
  ${\textsf S}_1$ is in the  projective  state $[\phi_k]$ 
 and simultaneously  
  ${\textsf S}_2$ is in the   projective state $[\psi_k]$  whenever
$\bfq(t)$ is in the interior of $ {\mf p}_k$  
}
  
 We discuss the 
relation of this hypothesis to the current interpretation of SQM and its new
implications in greater detail in Section~\ref{section5Discussion}

Thus a joint   or polar state of $\ss=\ss_1+\ss_2$ 
assigns a {\bf  conditional spectral} state $[\phi_k]$ to $\ss_1$. 
We call it {\it spectral}  since $[\phi_k]$ corresponds bi-uniquely to the 
one dimensional spectral projection $P_{\C\phi_k}$ of the mixed state
 $\sum_k |\lam_k|^2 |\phi_k\>\<\phi_k|$,    which is all that SQM assigns  
to $\ss_1$.

Even if it is granted that $\ss_1$ has a pure state, why should it be one of
the eigenprojections of the density operator? After all, as first stressed by Fano in \cite{Fano},
the mixed state can be realized by many convex combinations of various pure states,
some of which can be made to naturally arise in experiments.
The reason for choosing a {\it  spectral} projection is that it is the only {natural} choice. 
This somewhat vague statement is intuitively obvious, {and} its precise 
formulation and proof are given in Appendix~\ref{AppCats}. We shall see in 
Section~\ref{epr} that the experiment described by Fano can be handled using 
spectral projection{\bf s}.

The extension IQM  of SQM to
individual systems  is complete, natural, thus symmetry-preserving,   and
essentially unique.  Our main purpose is to rationalize the foundations of QM (hopefully, only
in the good sense of that word), but it is this uniqueness property which may be of 
interest in ``real'' physics.

We believe that what is canonically constructed in the mathematical formalism of a
 physical  model has a physical meaning in that model. The history of physics, 
from Maxwell's equations to Einstein's gravitational theory and the  
Dirac equation, bears out this principle. It is this idea that was formulated 
by   Dirac in \cite[p.60 ] {DiracQ}.  
 
  \begin{quotation}The most powerful method of advance that can be suggested at present is
  to employ all the resources of pure mathematics in attempts to perfect and generalize
  the mathematical formalism that forms the existing basis of theoretical physics, and
  after each success in this direction, to try to interpret the new mathematical
  formalism in terms of physical entities.
  \end{quotation}

Our main approach has been to  follow the path indicated by Dirac; we 
 will refer back to it after 
taking  
 some steps along these lines.  
 
  \subsection{ SQM=Standard or Statistical Quantum Mechanics.}%SUBSUBеееееееееееSUBSUBеее
  
  Our point of departure is the standard Hilbert space formulation of quantum mechanics, 
  which we conservatively take to apply to statistical ensembles.
  Ensembles are  idealized  objects which can  be  realized, to a good approximation,
  by sufficiently weak beams.
  
  The irrefinable ensembles or pure states ${\cal S}$ of $\ss$ correspond to the set  
  $\bfP({\cal H})$ of
  rays  in
  ${\cal H}$, with the transition probability given by $|\<\alpha,\beta\>|^2$ for unit vectors 
 $\alpha,\beta$
  representing
  pure states. The  model is extended to encompass mixed states, observables,
  dynamics, symmetry, etc. 
  
  The  mixed states  
    are modeled by the positive trace 1 operators on ${\cal H},$ 
%${\rm Herm}_{+1}({\cal H}) ,$ 
together with
  the transition probability $ {\rm Trace}(XY).$ 
  This extends  the previous pure version where the   rays in ${\cal H}$ correspond to the
  1-dimensional orthogonal projections on them.
  
  To a composite system $\textsf{S}=\textsf{S}_1+\textsf{S}_2$, SQM 
associates a tensor product
  of their respective Hilbert spaces,
  ${\cal H}={\cal H}_1\ot{\cal H}_2,$ provided the states of $\ss_1$ and of $\ss_2$ are
distinguishable. The case where some states of each may be indistinguishable is also very
important, especially for  the theory of identical particles.  These require special treatment
which  we do not carry out in this paper except, briefly in Appendix~\ref{appident}, 
for the {key} case of two identical systems.
  
  As long as the theory is applied to the computation of   probabilities  which are
  physically realized as  statistical relative frequencies, no interpretational problems arise.
  Of course, the question of the nature and behavior of the individuals comprising the ensembles
  is not touched in this formulation. 

  Even  measurement in this setting is unproblematical. The usual
  formulation goes as follows. SQM models the measurement of some quantity such as the spin of
  the atoms in some pure beam  with state $[\Phi]\in\bfP$, by considering an ensemble of systems
  $\textsf{S}$ composed of the atom $\textsf{S}_1$ and, say,  a Stern-Gerlach
  apparatus
  $\textsf{S}_2.$ The ensemble of atoms is  sufficiently well represented by the beam (although
  finer experiments would undoubtedly reveal unwanted correlations unless the beam was very
  sparse.)  A single macroscopic apparatus at a sequence of  slightly different times 
can be interpreted as
  the second components $\textsf{S}_2$ of the ensemble.
  
  It is assumed initially that $\phi = \sum_k\lk \phi_k$ is a superposition of the spin states
$\phi_k$.
  The mathematical formulation of the measurement process then considers a 
 unitary evolution
  of $\textsf{S}$ with  initial state $[\phi\ot\psi]$
   at time $t=0$ so that
  
  $$
  \gg(0):=\phi\ot\psi\ {\rm evolves\  to\ }
  \gg(1) =\sum_k \lk \phi_k\ot\psi_k.
   $$
  In  the last equation, the $\phi_k$ and the $\psi_k$ are orthonormal.
   There is a well-defined mixed state attributed to $\textsf{S}_1$ after the measurement,
  namely the reduced trace, ${\rm Tred}_1(P_\gg),$ where {$P_\gg=P_{\C\gg}$} 
 denotes
projection on
  $\C\gg.$ In this situation the density operator is
   $\sum_k |\lk|^2 P_{\phi_k}.$
  The mixed state ${\rm Tred}_1(P_\gg)$ is interpreted to mean  that
  with probability $|\lk|^2$, $\textsf{S}_1$ is in state $[\phi_k].$ 

The paper of Kochen \cite{Kochen} was the first to define a pure state of an 
interacting system as a spectral ray $[\phi_k]$ of the polar decomposition, thereby 
extending the interpretation of SQM for the explicit purpose of resolving the measurement
problem.  There was however no proposal for the dynamics leading to
 the particular $[\phi_k]$.  
The present  extension of the mathematical formalism  of SQM to IQM supplies the 
dynamics and toroidal partitions, leading to a natural choice of $k$ for each time.

  \subsection{ Individual systems.}%SUBSUBеееееееееееSUBSUBеее
  Our main goal is to find a consistent extension IQM of SQM which models
  individual systems such as single atoms. 
  Of course, it is generally taken that the mathematical formalism outlined
  in the previous section does double-duty in modeling  individuals as well as
statistical ensembles.
  
  The interpretational difficulties of this ambiguous usage begin with 
 the notion that $|\lk|^2$ is the
  {\it probability} that an {\it individual} is in the state $[\phi_k].$ In applications, it
  is interpreted, as usual, in terms of the relative frequency  of an ensemble
  of systems all in state $[\phi].$ This subjective interpretation of probability  is
 awkward but, by itself,
  consistent. However,  this is only the beginning of the  difficulties; they
  end with the need to somehow identify the 
  pure state $\sum_k \lam_k\, \phi_k\ot\psi_k$
  with the mixed state  $\sum_k |\lk|^2 P_{\phi_k\ot\psi_k}.$
 This identification is usually called ``collapse of the wave function''.

  From this brief description, it can be seen that the heart of the problem is to consistently 
  attribute a pure state to an individual system, including one such as $\textsf{S}_1$ which
  is a subsystem of $\textsf{S}$ interacting with $\textsf{S}_2,$ and then to mathematically
  model these pure states. The possibility of such a model is formally equivalent to the
  consistency of IQM, but the acceptability of the model will depend upon its naturalness,
  uniqueness, and intrinsic interest.
  
  The first step in constructing the new model is to  represent states of individuals, in part, 
  by unit vectors instead of rays. In  all but the most foundational studies of QM,  
  states are already so represented. Usually, there is an early, one-time warning that vectors
  differing by a phase factor give the same pure state.
  Basically,  authors pay lip service to $\bfP$ but, in practice, almost  always  use $\bfS$ or 
  ${\cal H}.$

  Relative phases occur in the literature, along with some controversy as to what
observable or self-adjoint operator should be used to represent them. The absolute
phases we are invoking could be avoided
  but they clearly foreshadow the additional ``moduli'' necessary for a complete 
  individual state description: a  torus of phases.
 
  \section{Enlarging the  State Space}\label{section3EnlargingStates}
\setcounter{equation}{0}
 
In the introduction, we already indicated that IQM extends the SQM state space of rays to unit
vectors representing individuals. 
   The mathematical  relation 
  of the unit vector $\phi$ being in   the ray $[\phi]$
   mirrors the physical relation of the individual system in  state $\phi$ 
  being in the ensemble with state $ [\phi].$
  We can think of an ensemble with pure state $[\phi]$ as being composed of individuals
  having state vectors of the form $\zeta \phi,$ where the  $\zeta $ are 
uniformly distributed random phase  factors. 
 Perhaps a reason this minor extension of the model for SQM has not 
explicitly appeared in the literature is that it requires  acceptance of the 
non-applicability of SQM to  individuals.

In this section, we  must compound this extension and simultaneously include all 
  the phases arising in polar decompositions of vectors in a tensor product:
$$
\gg=\sum \lam_k\,\gg_k=\sum \lam_k\,\phi_k\ot\psi_k
$$ 
 \subsection{An  {\it ad hoc} construction of the polar bundle ${\po}$. }
  
We consider the set of all possible  pairs
   $(\gg,\bfq)=(\gg(t),\bfq(t))$ which can arise; we form 
  the (right) toroids $\bfT(\bfr,\gg)$  consisting of any $\bfq$ arising from %c%
  a possible polar decomposition $\sgk.$ We thus get a family of toroids parameterized by
   the $\gg\in\bfS$ for which they occur
  as the polar decomposition coefficients.
  
  Let 
  $$\bfT\deq\left\{(\gg,\bfq) \;|\; \(\exists\, \gg_k\)\; \gg=\sgk {\rm\ is\ a\
  polar\ decomposition\ }\right\}=\prod_{\gg\in \bfS}\bfT(\bfr,\gg).$$
  Then $\bfT$ is the total space of a generalized bundle 
  ${\po}\deq [\bfT{{\buildrel\Pi\over\ra}}\bfS],$ where $\Pi(\gg,\bfq)=\gg.$ 
  We are talking then of a {\bf bundle} in its most
  general form, a map between spaces. 
    We call ${\po}$  a {\bf toroidal 
  bundle} since each fiber is a toroid. We may  consider ${\po}$ as a 
  {\bf singular  torus bundle}, with torus group 
  $\bfT^n({\bf1})\deq\prod_k\bfS^1(\bf1).$ Indeed,  just as the unit circle $\bfS^1(1)$
  acts on any circle $\bfS^1(r)\subset \C$ by
   $\bfS^1(r)\ni r e^{i\tau}\;{{\buildrel{e^{i\theta}}\over\lra}}\; r
  e^{i(\tau+\theta)}\in\bfS^1(r),$ (even if $r=0$) so does $\bfT^n(\bf1)$ act on
  $\bfT^n(\bfr)$, provided a {\bf bouquet}  of generating circles $\bfS^1(r_k)$ is
  distinguished. This distinguishing is automatic when the $r_k$ are all distinct. 
  Thus, if we restrict the bundle to the set
  $\bfS^{\rm reg}$ of {\bf regular} 
   $\gg$, i.e. those with positive distinct  $r_k$, then 
  ${\po}^{\rm reg}\deq\left[\Pi^{-1}(\bfS^{\rm reg}){{\buildrel\Pi\over\ra}}\bfS^{\rm
  reg}\right]$ is a (standard) smooth principal $\bfT^n-$bundle on which we have put a 
metric on each fiber. 
  
  The coefficient $\lam_k(t)$ might be thought of as ``the amplitude that
  $\ss$ is in  the state $ \gg_{k}(t)$'', using
  typical QM textbook language. 
  
  We will take as our IQM state space of $\ss=\ss_1+\ss_2$ the total space 
  $\bfT=\bfT({\cal H},{\cal H}_1,{\cal H}_2)$  of the bundle ${\po}.$  
This bundle is not a locally
trivial bundle,
  since the base space is connected, while the fibers do not have constant dimension.
 Less trivially, if we restrict to the stable points   $\bfS^{\rm stab}$ of $\bfS,$ 
\ie those for which no $r_k=0,$ then the fibers
{\it do} have constant dimension, but the bundle is not locally a product at those
points where the $r_k$ are not all distinct.
If we restrict  the bundle to the regular points  $\bfS^{\rm reg}$ of $\bfS,$ then
it is indeed a trivial or product bundle. Nevertheless, even  
restricted to $\bfS^{\rm reg}$, 
the  connection $A^H$ we will employ  is not trivial; it has curvature and, 
 when the \ham\ $H\ne 0,$ even torsion.
 
  \subsection{A natural construction of  ${\po}$ using moment maps.}\label{natmomentum}

In the introduction, we stressed the importance of canonical constructions in extending 
physical models. In this section we show that the polar bundle can be realized canonically.
  The idea behind this  construction  makes use of the moment (or momentum) map
from   symplectic geometry and general bundle operations.
  
 ${\cal H}$ has a natural symplectic structure $ \omega_0$,
which, in terms of coordinates  $z_k$ with respect to an orthonormal basis,
can be written (as in \cite[p.130]{McDuff})
  $$
  \omega_0 = {i\over2}\sum_k{dz_k}\wedge \v{dz_k}.
  $$
  The action of the unitary group $\U({\cal H}) $ on ${\cal H}$ is then Hamiltonian so that,
denoting the Lie algebra ${\mf L}\(\U({\cal H})\)$ by ${\mf u}$, there
exists a moment map \cite[p.162]{McDuff}
  $$
  \mu:{\cal H}\ra {\mf u}^*\cong  {\mf u},\;
  \mu(\bfz) = {i\over2}\bfz\bfz^*={i\over2}\|\bfz\|^2 P_{\C\bfz}.
  $$
  We used here  the natural identification of the Lie algebra ${\mf u}$
with its dual
${\mf u}^*$,
  deriving from the Hilbert-Schmidt inner product.
  
  If $G$ is a subgroup of $\U({\cal H})$ then its moment map $\mu_G$ is given by composing
  $\mu$ with  projection on the subalgebra ${\mf L}(G)$, i.e.
  $\mu_G= P_{{\mf L}(G)}\circ \mu.$ We are going to apply this together with
  the fact (see Appendix~\ref{tred} ) that the reduced trace is itself, essentially, a projection 
in the space of operators with
  respect to the Hilbert-Schmidt Hilbert space inner product. 
  
  Now suppose we have a Hilbert space factorization:
  $$
  {\cal H}={\cal H}_1\ot{\cal H}_2.
  $$
  Let $G$ be the  compact group generated
  by $\U({\cal H}_1)\ot\bfI_2,\bfI_1\ot\U({\cal H}_2).$
  $G$ is almost isomorphic to the  direct product  $\U({\cal H}_1)\times \U({\cal H}_2),$ 
\ie\ 
  the bottom row of the following diagram   of Lie algebras is exact:
  
  \[
  \divide\dgARROWLENGTH by4
  \begin{diagram}
  \node[6]{{\cal H}} \arrow[2]{s,r}{\mu_G} \arrow[2]{sw,t}{\tau} 
  \\
  \\
  \node[0]{\bo}
        \arrow[1]{e,t}{} 
     \node[1]{i\R \bfI}  \arrow[2]{e,t}{\delta} 
    \node[2]{\mf u({\cal H}_1)\times\mf  u({\cal H}_2)}  \arrow[2]{e,t}{\sigma} 
   \node[2]{\mf g}  \arrow[1]{e,t}{} \node{\bo}
    \end{diagram}
  \]
  where $\mf  u({\cal H}_i)= {\cal L}(\U({\cal H}_i )), \mf g={\cal L}(G),$ 
$\delta(a\bfI)= (a\bfI_1,a\bfI_2)$ and $\sigma(u_1,u_2)= u_1\ot\bfI_2-\bfI_1\ot u_2.$
  Moreover, defining 
$$\tau(\bfz)= {i\over2}\({\rm Tred}_1(P_{\C\bfz}),{\rm
Tred}_2(P_{\C\bfz})\),
$$
 the triangle
  commutes. If we pass to the corresponding projective spaces we get a moment map
  $\widetilde{\tau}:\bfP({\cal H})\ra \widetilde{\mf g}\cong {\mf su({\cal H}_1)
\times\mf  su({\cal H}_2)}
  =: {\mf L_{12}}$
  given  in finite dimensions by the formula   
$$
\widetilde{\tau}([\bfz])={i\over 2}\({\rm Tred}_1(P_{\C\bfz})-{1\over
n_1}\bfI_1,{\rm  Tred}_2(P_{\C\bfz})-{1\over n_2}\bfI_2\)$$
 with $\bfz\in\bfS.$
    
  The pair of reduced traces  gives a map to the equi-spectral density operators.
  We have shown that this map is essentially the moment map for the group $G$ of 
 automorphisms of $\bfP({\cal H})$
  preserving the tensor product decomposition: $G$ is $\S\U({\cal H}_1)\times\S\U({\cal H}_2)$ 
modulo its  center.

  If $ \Gamma=\sum_k \lam_k\,\phi_k\ot\psi_k$ is  a polar decomposition of $\gg\in\bfS$, then 
   $$
  \w{\tau}([\gg])={i\over2}\(\sum_k |\lam_k|^2P_{\C\phi_k}-{1\over n_1}\bfI_1,
  \sum_k |\lam_k|^2P_{\C\psi_k}-{1\over n_2}\bfI_2\).
  $$
  In any case, the fibers of $\w{\tau}$ are the toroids  
   $\bfT(\gg) \deq \{ \sum_k \lam'_k\phi_k\ot\psi_k\;|\; |\lam'_k|=|\lam_k|\}.$
  The  (generalized) bundle
   $ \widetilde{\tau}: \bfP\ra {\mf L_{12}}$
induces a bundle,
  with the same fibers,  over any
  space  $M$ mapping to ${\mf L_{12}}$. So $M{{\buildrel\eta\over \ra}}{\mf L_{12}}$ yields
 a commutative diagram
  \[
  \begin{diagram}
  \node{ \eta^*\bfP}
        \arrow[2]{e,t}{\eta^*} \arrow[2]{s,l}{ \eta^*(\w{\tau})} 
     \node[2]{\bfP} \arrow[2]{s,r}{\w{\tau}}
  \\
  \\
  \node{M}\arrow[2]{e,t}{\eta}
        \node[2]{{\mf L_{12}}}
  \end{diagram}
  \]
  We can, in particular, apply this construction to the case where $M=\bfP, \eta=\w{\tau}.$
  In general, this  bundle induced by the projection map (${\w{\tau}}$ in this case)
 is called the {\bf square} of the original bundle \cite[p.49]{Steenrod}.  It alway leads to a
bundle with a cross-section which is as smooth as is locally possible. In the present
  case, it yields a toroidal bundle $B$ over $\bfP.$   
  We have the Hopf map $\bfS\ra \bfP.$  We can use the Hopf map to induce from $[B\ra \bfP]$, a
  bundle $[\bfT\ra\bfS ]$, which is our polar bundle ${\po}.$
Using this squaring  operation, it
is not hard to justify the statements made at the end of the previous section about 
product bundles.
\nl{\bf Review  of the natural construction of $\po.$}  We start with the essentially classical notion of
the moment map from $\bfP\ra  {\mf L_{12}}$, the Lie algebra of the natural  group acting on the tensor 
product.  We then take the square of this bundle, using traditional terminology. Thinking of the bundle 
as the {\it map}  ${\w{\tau}}$ this makes the space of the new  bundle, our 
polar state space, a kind of square-root of
$\bfP$.
  The final step is
 the familiar Hopf bundle construction,  $\bfS\ra\bfP$. The last two steps
could be carried out  in the opposite order. We have not explicitly carried 
 out the natural construction in the
infinite  dimensional case, although the naturality implies 
 this is possible by a limiting procedure. At any
rate,  the final bundle is well-defined in infinite, as well as finite, dimensions.

The idea that quantum mechanics arises as a kind of  square-root of  standard structures is
an intuition which is here made precise. A precise representation-theoretic version of this 
square-root operation was 
 established in \cite[Ax]{Ax}.

  We note that if ${\cal H}_1={\cal H}, {\cal H}_2=\C,$ then the bundle $B\ra\bfP$ is equivalent to 
  $\bfS\ra\bfP.$  When squared, this becomes 
diffeomorphic to the product bundle $\bfS\times \bfS^1\ra \bfS.$

 \subsection{ The Geometric and Dynamical Connections}
  
  In the original papers of Berry and Simon,  their
successors (and predecessor
  Pancharatnam), collected in  \cite{Shapere}, the physical
  and mathematical significance of the canonical unique lifting of smooth curves in  
  $\bfP$ to $\bfS$ was explicated. 
  This lifting is by means of the {\bf canonical connection}. 
A  good reference for connections with a physics orientation is Frankel's book
\cite{Frankel}.
  We recall that one way of specifying a connection on a principal bundle is by giving a 1-form $A$
  on the total space which is compatible with its principal bundle structure. In the present case, this
  1-form $A$ is on the total space $\bfS$ of the principal $\bfS^1$-bundle $\bfS{\buildrel \pi\over\ra}\bfP.$
  It is given by the following formula.
  \begin{equation}\label{CanonicalConn0}%еееееееееееееееееееееееееееееееее
  A :=A^0\deq \<\bfz, d\bfz\>= \sum \bar z_{k} d z_{k}, {\rm\ for\ } \|\bfz\|=1.
  \end{equation}%ееееееееееееееееее      еееееееееееееее
  A curve $t\ra \bfz(t)\in \bfS$ is {\bf horizontal} (over $\bfP$
   \wrt the canonical connection) \ifff the induced 1-form on $\R$ vanishes, \ie
  \begin{equation}\label{7CanonicalConn0}%еееееееееееееееееееееееееееееееее
  \forall t\;\<\bfz(t), \dot\bfz(t)\>=0.
  \end{equation}%ееееееееееееееееее      еееееееееееееее
  The curve $t\ra \bfz(t)\in \bfS$ is the unique horizontal lift of the curve $t\ra [\bfz(t)]\in \bfP$,
  which begins at $\bfz(t)$ for $t=0.$
  
  This 1-form $A$ takes values in $i\R$ which is naturally identified with the Lie algebra
   ${\cal L}\bfS^1$ of $\bfS^1.$
  It is easy to see that
  \nl (i) $A$ is invariant under the natural action of $\bfS^1$ on $\bfS.$ 
  \nl Moreover, for
  any fixed $\bfw\in\bfS$ and variable $\zeta=e^{i\theta}\in\bfS^1$, we have
  \nl (ii) $ \<\zeta\bfw, d(\zeta\bfw)\>= \bar\zeta\,d\zeta=id\theta.$
  
  Since, we are going to be dealing with some slight generalizations of $A^0,$
  we indicate the two compatibility conditions required for a 1-form $A$ on a general principal
  $G-$bundle  $P{\buildrel \pi\over\ra}M$ (where $G$ is a Lie group and $M$ is a manifold.)
   $A$ is now required to be $\mf G:={\cal L}G-$valued. Thus $G$ acts naturally on $A$ by combining
  its action on $P$ with its Ad-action on $\mf G.$
   These two conditions  are:
  \nl 1) $A$ is invariant under the action of $G$;
  \nl 2) For all points $p\in P, $ the pull-back $\gamma^*(A)$ to a 1-form on $4G$ is the
right-invariant
  Maurer-Cartan form on $G$ ( which is $id\theta$ when $G=\bfS^1.)$
  \nl For the canonical connection, these conditions are satisfied, in view of (i) and (ii).
  
  The horizontal lift of a closed loop in $\bfP(\cal H)$ need not be closed in $\bfS(\cal H),$
  but will differ at its endpoints by a phase factor. The argument of this phase factor is 
  {\bf Berry's  phase.}
  The original papers, and so far as we know, all subsequent papers on the subject
  separated the total phase change of various kinds of liftings of curves in 
  $\bfP$ to $\bfS$ as being the sum of a geometrical phase and a dynamical 
   phase.
  The first is
   Berry's phase, determined by the canonical connection, the second is the
 remaining phase angle needed to
  comply with the Schr\"odinger  evolution of  a vector in ${\cal H}.$ 
  
  We are going to reverse this procedure because
  we prefer an equivalent but  more unified treatment of these phase factors by getting the total
  phase from a  new connection, which we  call the {\bf dynamical  connection }
  This will be useful in the sequel, where we study certain horizontal
   liftings of {\it non-evolutionary curves} in
  $\bfP.$ 
   But it  is clear from the   consideration of an energy eigenvector that no such connection can 
  exist on $\bfP$! Namely,  if $H \Gamma(0)= E \Gamma(0),$ then    $\Gamma(t)=
  e^{-i E t}\Gamma(0).$  Of course, no phase determined by a connection  can be involved here, since
  $[\Gamma(t)] $ is the constant curve. 
  
   For this reason we will work with the corresponding contact manifold
  $\bfP\times \R$. This is also a convenient space for the consideration of time-dependent
  Hamiltonians $H(t).$ Its analog appears in classical mechanics, where it is sometimes called
  the ``extended phase space'', as in   \cite[p.236]{Arno}.
  
  \subsubsection{The dynamical  connection on time extended phase space.}\label{schrcon}
  
   \begin{lem}\label{lem:contact} We suppose given a (possibly time-dependent)
  bounded Hamiltonian $H(t)$ on ${\cal H}.$
  The  evolutions $\ t\ra \Gamma(t)$ in $\bfP(\cal H)$ correspond
  bijectively to the maps
  $t\ra (\Gamma(t),t)$ in $\bfP\times \R.$
  
  \nin There exists a  connection on $\bfS\times \R{\buildrel \pi\times \b1\over\ra}\bfP\times
  \R$ so that every curve $t\ra (\Gamma(t),t)$ corresponding to a \SC\  evolution
  $t\ra 
  \Gamma(t)$ is horizontal. 
   It is given  by 
    the 1-form:
   \begin{equation}\label{8CanonicalConnS}%еееееееееееееееееееееееееееееееее
   A^H\deq\<\bfz,d\bfz\>+ i\<\bfz,H(t)\bfz\> dt=: A^0+i{\cal E}^H(\bfz)dt%=: A^0+i{\cal E}(\bfz)dt
   \end{equation}%ееееееееееееееееее      еееееееееееееее
  We can characterize $A^H$ as the  unique connection compatible with the Schr\"odinger evolution
   which agrees with $A^0$ on constant time slices.
  \nl The condition for a curve  $t\ra(\gg(t),t)\in \bfS\times \R$ to be horizontal with respect to
$A^H$ 
  is the
  following variation of \eqref{7CanonicalConn0}, with which it coincides when $H(t)\equiv0$.
   \begin{equation}\label{9CanonicalConnS}%еееееееееееееееееееееееееееееееее
  \forall t\;\<\gg(t),\dot\gg(t)\>=- i\<\gg(t),H(t)\gg(t)\> .
   \end{equation}%ееееееееееееееееее      еееееееееееееее
    \end{lem}
  {\bf Proof.} 
  \eqref{8CanonicalConnS} defines a 
  1-form $A^H$ on $\bfS\times\R$ which clearly satisfies compatibility condition 1).
  To be more precise, in this equation we should  actually replace the 1-form $A^0$ by its
pullback to
  $\bfS\times \R$. This form is independent of $t$ and it has  the same expression
  $ \<\bfz, d\bfz\>$ as does $A^0$, so we neglect this nicety.
  To see that 2) is also satisfied, we  note that $A^H$ has the same pullback to $\bfS^1$ as does 
  $A^0$. Thus $A^H$ {\it does} define a connection.
  
  To obtain \eqref{9CanonicalConnS} we take the tangent vector 
  $(\dot\gg(t),\partial_t)$ to the time-extended phase space
  and apply $A^H$, as defined by
  \eqref{8CanonicalConnS}. 
  
  Now suppose $t\ra(\gg'(t),t)\in \bfS\times \R$ is a horizontal lift of $t\ra[\gg(t)]\times\R,$ where
  $\gg(t)$ satisfies the Schr\"odinger equation. Then from
  \eqref{9CanonicalConnS}, we have 
  $\<\gg',\partial_t \gg'\>= -i\<\gg',H(t)\gg'\>= -i\<\gg,H(t)\gg\>.$
  Since $\gg'(t)$ differs from $\gg(t)$ only by a phase factor  $\zeta(t)$, it follows that
  $\<\zeta\gg,\partial_t (\zeta \gg)\>= -i\<\gg,H(t)\gg\>\Ra
  \<\gg,\partial_t \gg\>+\bar\zeta\dot\zeta=-i\<\gg,H(t)\gg\>$. It follows that $ \dot\zeta=0 $
  And hence  $\gg'(t)$ and $\gg(t)$ differ only by a {\it constant} phase factor. Thus $t\ra  \Gamma(t)$ is
  horizontal. 
  
  The uniqueness of $A^H$ follows from the  fact that the tangent vectors at a point
  $(\bfz,t) $ are generated by $\partial_t$ and  the tangent space of the constant time slice. 
  \nl$\qed$
  
  A  technique for finding $A^H-$horizontal versions $(\Omega(t),t)\in \bfS\times\R$ of a
  $A^0$-horizontal curve
  $\Omega^0(t)\in\bfS $ follows from the  argumentation of the proof. Namely, 
  write $\Omega(t)= \zeta(t)\Omega^0(t)$ and take inner products:
  $$
  \<\Omega,\partial_t \Omega\>=\<\zeta\Omega^0,\partial_t (\zeta \Omega^0)\>= 
  \<\Omega^0,\partial_t \Omega^0\>+\bar\zeta\dot\zeta=\bar\zeta\dot\zeta.
  $$ 
  Now  we want \eqref{9CanonicalConnS} to hold, so we require
    \begin{equation}\label{Shorizontalize}%еееееееееееееееееееееееееееееееее
  \bar\zeta\dot\zeta=-i\<\Omega,H \Omega\> =-i\<\Omega^0,H \Omega^0\>
  \Ra \partial_t \ln\zeta(t) =-i\<\Omega^0(t),H(t) \Omega^0(t)\>.
   \end{equation}%еееееееееееееееееееееееееееееееее
  Thus we can express the $A^H$-horizontalizing phase factor $\zeta$ in terms of the given $\Omega^0:$
    \begin{equation}\label{ShorizontalizingZeta}%еееееееееееееееееееееееееееееееее
  \zeta(t)
  = e^{-i\int^t_0\<\Omega^0(s),H(s) \Omega^0(s)\>ds}.
   \end{equation}%еееееееееееееееееееееееееееееееее
  We  express this result in a lemma for  reference.
   \begin{lem}\label{horlem} If $t\ra\Omega^0(t) \in \bfS $ is a 
  curve horizontal with respect to the canonical connection 
    and if
    \begin{equation}\label{ShorizontalizingZeta8}%еееееееееееееее
    \Omega^H(t)= \zeta(t)\Omega^0(t)
    \end{equation}%еееееееееееееееееееееееееее
  then $\Omega^H(0)=\Omega^0(0)$ and $( \Omega^H(t),t)$  is horizontal with respect to $A^H$ if and
only if 
   \begin{equation}\label{ShorizontalizingZeta9}%еееееееееееееееееееееееееееееееее
  \zeta(t) 
  = e^{-i\int^t_0\<\Omega^0(s),H(s) \Omega^0(s)\>ds}.
   \end{equation}%еееееееееееееееееееееееееееееееее
      \end{lem}
$\qed$
\pagebreak
  
  If we make the substitution $\zeta(t):=e^{i\alpha(t)}, $
    \begin{equation}\label{ShorizontalizingZeta3}%еееееееееееееееееееееееееееееееее
   \alpha(t) = -\int^t_0 \<\Omega^0(s),H(s) \Omega^0(s)\>ds.
   \end{equation}%еееееееееееееееееееееееееееееееее
    
  We can compare this equation with  Equation (3) in the paper of 
Aharanov-Anandan, contained in  the  previously referenced collection   
\cite{Shapere}. There we see  the total phase
  $\phi$ represented as the geometric phase $\beta$ plus $\alpha.$ While that paper and
  others are at pains to separate out $\beta$ and obtain it  from the canonical
  connection, we reverse this procedure so as to get the total phase from a connection.
  
The canonical connection is the  dynamical  connection formed with
a trivial \ham. The following lemma is therefore a generalization of the previous
one and is proved similarly.
   \begin{lem}\label{horlemgen} If $t\ra\Omega'(t) \in \bfS $ is a 
  curve horizontal with respect to the dynamical  connection formed \wrt\ the \ham\ 
$H'$, and if $\Omega''(0)=\Omega'(0)$, then (dropping the second component) $t\ra\Omega''(t)
=\zeta(t)\Omega'(t)
$ is  horizontal with respect to the dynamical  connection formed \wrt\ the \ham\ 
$H=H'+H''$
    \ifff\ 
  \begin{equation}\label{ShorizontalizingZeta12}%еееееееееееееееееееееееееееееееее
  \zeta(t) 
  = e^{-i\int^t_0\<\Omega'(s),H''(s) \Omega'(s)\>ds}.
   \end{equation}%еееееееееееееееееееееееееееееееее
      \end{lem}
   \nin$\qed$
  
  \subsubsection{The dynamical  connection as a Lagrangian.}
  
  Suppose $C:=C_\gg :=[0,1]\ni t\ra(\gg(t),t)\in \bfS\times\R$ is a critical
   curve for the curve-functional
  $$
  {\cal S}(C):= \int_0^1 C^*( A^H) =\int_0^1 ( \<\gg(t),\dot\gg(t)\>+ i\<\gg(t),H(t)\gg(t)\>)dt=: 
  \int_0^1\textsf{L}(\gg,\dot \gg)dt.
  $$
Here $ C^*$ is the pullback map.
  $$
  \hskip-1in{\rm\ By\ the\ Euler\!-\!Lagrange\ equation: }
   {\partial  \textsf{L}\over \partial
  \gg}={d\over dt}\({\partial\textsf{L}\over \partial \dot\gg}\), 
  {\rm\ we\ get\ the\ }
  {\rm  {Schr\ddot{o}dinger }\ equation:}
  $$
   \begin{equation}\label{EulerLS}%еееееееееееееееееееееееееееееееее
  \v{\dot\gg(t)}=i \v{H(t)\gg(t)}\Ra {\dot\gg(t)}=-i {H(t)\gg(t)}.
   \end{equation}%ееееееееееееееееее      еееееееееееееее
  We thus see that the connection form $A^H$ is a Lagrangian form.
   The defining property of the 1-form $A^H$ is that ${\rm 
   {Schr\ddot{o}dinger }}$ evolutions are horizontal. It then turns out that
  ${\rm  {Schr\ddot{o}dinger }}$ evolutions are critical values of the $A^H$ action integral.
  
  This  is part of a very general situation, in which horizontality
  with respect to a connection on a bundle over a manifold yields critical values of a
  related Lagrangian, e.g.
  the projections of  horizontal curves in the tangent bundle of a Riemannian manifold 
   $M$ are the geodesics of $M$  as in \cite[Vol.I,Prop 6.3]{Koba}.
  
  \subsubsection{The relativistic dynamical connection.}
  
  Extending the SQM state space $\bfP$ to $\bfP\times\R$ is a convenient way to exhibit
  the phase of the constant SQM states corresponding to  energy eigenstates. But it has
  the immediate effect of ruining the symmetry of the extensions of SQM to bundles over $\bfP$.
  For suppose a group $G$ acts on $\bfP$. Then we need a natural extension of this action to $\bfP\times \R$
  which, in general, will be affine on the second ``time'' factor $\R.$ This is fine for Galilean
  relativity, but it won't do for Poincar\'e covariance. 
  
  This suggests that to make the theory relativistic we begin by extending
   the SQM phase space $\bfP$ to $\bfP\times M$, where $M:=M^4$
  \n{$M:=M^4$}
   is Minkowski space with its usual action by 
  the simply connected cover $G$ of the Poincar\'e group.  We also assume that $G$ acts
 unitarily on ${\cal H}$ and
  thence on $\bfP=\bfP({\cal H}).$ 

There is as yet no  {\it generally accepted}
 rigorous version of  the quantum field theory or even  QED which is  required to model the non-trivial 
dynamics of interacting relativistic particles, despite the best efforts of   constructive 
quantum field theorists.  
The analytic considerations of QFT are  beyond the scope of this paper, so we  
proceed merely formally. We assume that $G$ acts compatibly on the
attendant additional structures, such as  the  distribution-valued field operators, which we 
 formally treat as ordinary unbounded operators.
In particular the (unbounded) energy operator $H$ may be combined with
 the canonical connection,
as before, to produce the dynamical  connection with respect to a given inertial frame with
coordinates $ {\vec {\bf x}}= (x_0=t,x_1,x_2,x_3).$ For a curve of  vectors $\gg(t)$
 analytic  for $H$,
we can consider the connection 1-form in \eqref{8CanonicalConnS}.
 \begin{equation}\label{98CanonicalConnS}%еееееееееееееееееееееееееееееееее
   A^H\deq\<\bfz,d\bfz\>+ i\<\bfz,H\bfz\> dt=A^0+ i\<\bfz,H\bfz\> dt.
   \end{equation}%ееееееееееееееееее      еееееееееееееее
We want to formally exhibit  a manifestly covariant version of this form on $\bfP\times M$
which along a time-like line in an inertia frame agrees with this expression, interpreting
$t$ as the proper time. Let 
\[\Lambda( {\vec {\bf x}}, {\vec {\bf y}})\deq x_0y_0-x_1y_1-x_2y_2-x_3y_3
\]
be the usual Lorentzian inner product. 
Let $ {\vec {\bf p}}= (p_0=H,p_1,p_2,p_3)$ denote the 4-vector of energy momentum
operators,
\ie the generators of the translation group action on $\bfP.$ The connection form $A^0$
is already invariant under $G$, so we need only modify the imaginary
part of $A^H.$  The new, manifestly covariant form is:
 \begin{equation}\label{14CanonicalConnS}%еееееееееееееееееееееееееееееееее
 \<\bfz,d\bfz\>+ i\Lambda(\<\bfz, {\vec {\bf p}}\bfz\> , {\vec {\bf x}}).
   \end{equation}%ееееееееееееееееее      еееееееееееееее

It now follows, at least formally, 
 that $G$ acts naturally on the $A^H-$horizontal  lifts in ${\po}$ of curves in $\bfP.$ It follows
that our theory  faces no {\it new} insuperable obstacles  from special
relativity. Those already present in SQM are, of course, quite sufficient.

   \subsection{Evolution in the   polar bundle.}\label{evolggk}
  
  Let us examine the  possible polar decompositions appearing in 
    \begin{equation}\label{goodPolar}%еееееееееееееееееееееееееееееееее
  \Gamma=\sk{ {\lam_{k}  \phi_{k} \!\otimes\!\!{\psi_{k}} }}\,\;=\sgk 
    \end{equation}%\label{ShorizontalizingZeta}%еееееееееееееееееееееееееееееееее
  especially in the non-degenerate case.
 
   The expression on the right hand side of 
  \eqref{goodPolar} is not unique for it is possible to multiply the $\Gamma_{k}$ by
   any phase factor,
  and simultaneously multiply the $\lam_{k}$ by the inverse or conjugate phase factor. 
  It is tempting to try and pick some unique representation, for example by requiring the
  $\lam_{k}>0.$ This choice is affirmed by recognizing that the $|\lam_{k}|$ comprise
  the eigenvalues of the positive part $P=+\sqrt{\wt{\gg}^*\wt{\gg}}$ of
  the  polar decomposition of
   $\wt{\gg}=+\sqrt{\wt{\gg}^*\wt{\gg}}\,\,U$, where $U$ is an 
  isometric linear map $\v{{\cal H}_2}\ra {\cal H}_1$ satisfying 
$U(\v{\psi_{k}})\ra\phi_{k}.$ 
 See Appendix~\ref{AppPolar}. In the
  completely non-degenerate case, where the $\phi_{k}$ comprise an
  orthonormal basis for ${\cal H}_1$ and
  the $\v{\psi_{k}}$ comprise an orthonormal basis for $\v{{\cal H}_2},$ $U$ is a well-defined
  unitary operator:
  $ U=P^{-1}\wt{\gg}, $ since the $|\lam_{k}|$ are assumed positive and distinct.

  The trouble with this apparently natural choice, is that it is
 not sustainable under the natural
  evolution of the $\Gamma_k.$ 
This evolution is the lift by the   connection $A^H$ of the curve $t\ra [\gg_k(t)]\in \bfP,$ 
where the $\gg_k(t)$ are bi-orthonormal polar components of $\gg(t).$
   To see what is involved here, let us suppose we have some Hamiltonian ${ H}$ on 
  ${\cal H}$ generating a unitary evolution $t\mapsto U(t)=e^{-i t {{ H}}}$, taking
  units for which $\hbar=1.$
  
  Let
  \begin{equation}\label{12Polar(0)}%еееееееееееееееееееееее
  \Gamma(0)=\sum_{k} \lam_{k}(0)  \Gamma_{k}(0). 
  \end{equation}%ееееееееееееееееее      еееееееееееееее
  be an initial fixed  polar decomposition. Set $\Gamma(t)=U(t)\Gamma(0).$
  Let 
  \begin{equation}\label{12Polar(t)}%еееееееееееееееееееееееееееееееее
  \Gamma(t)=\sum_{k} \lam_{k}(t)  \Gamma_{k}(t). 
  \end{equation}%ееееееееееееееееее      еееееееееееееее
  be any  polar decomposition of $\Gamma(t),$ with some choice of smooth 
   $\lam_{k}(t) $ and $ \Gamma_{k}(t).$
    
   The curve $t\mapsto
   \gamma_{k}(t):=[\Gamma_{k}(t)]\in\bfP$ is independent of any such choice. 
  Let  $\Gamma_{k}^H(t)$ be the unique curve  horizontally lifting $\gamma_{k}(t)$ with
  respect to the connection form $A^H.$ We loosely use this expression when we really
  mean: 
  $(\Gamma_{k}^H (t),t)$ is the horizontal lift of
  $(\gamma_{k}(t),t)$ to
  $\bfS\times\R$ which begins at $(\Gamma^H_{k}(0),0).$  
  By \eqref{9CanonicalConnS}, this means
  $\<\Gamma^H_{k}(t),\dot\Gamma^H_{k}(t)\>=-i\<\gg_k^H(t),H(t)\gg_k^H(t)\> .$ Of course,
this 
  condition alone is far from characterizing the
  $\Gamma^H_{k}(t).$ This requires the additional  condition that 
  $[\Gamma^H_{k}(t)]=\gamma_{k}(t),$
  i.e. we also require $[\Gamma^H_{k}(t)]=[\phi_{k}(t)\ot{\psi_{k}(t)}].$
  
  Now the  $\lam^H_{k}(t)$ are uniquely determined by 
  finally requiring
  \begin{equation}\label{SPolar}%еееееееееееееееееееееееееееееееее
  \Gamma(t)=\sum_{k} \lam^H_{k}(t)  \gg^H_{k}(t). 
  \end{equation}%ееееееееееееееееее      еееееееееееееее
   	 In the next section, we exhibit the equations determining 
  $\Gamma^H_{k}(t)$ and $\lam^H_{k}(t)$; but we can easily see that positivity
  of the $\lam^H_{k}(t)$ is ruled out in general by
  \eqref{ShorizontalizingZeta9} because the horizontalizing factor
needed to go from
 $ A^0-$horizontal to  $ A^H-$horizontal is given by
   \begin{equation}\label{SZeta}%еееееееееееееееееееееееееееееееее
  \zeta(t) 
  = e^{-i\int_0^t\<\gg^0(s),H(s) \gg^0(s)\>}ds,
   \end{equation}%еееееееееееееееееееееееееееееееее
  which can be any phase factor for some $t$ and some $H.$

  \subsubsection{Evolutionary equations for   the amplitudes $\lam^H_{k}$.}\label{tailor}
  \begin{quotation}%p81 of dictionary
  If you are out to describe the truth, leave elegance to the tailor.
  \emph{Einstein}%\footnote{Perhaps this would look better if  \emph{Einstein}
  %were centered.}
  \end{quotation}

  Let $\gg(t):=\gg^H(t)$ evolve according to \SC 's equation with \ham\ $H.$
  We  want to determine the behavior of the $\lam^H_{k}(t)$
   when $\gg(t)=\sum_k\lam^H_{k}(t)  \gg^H_{k}(t)$ is a polar decomposition
  in which the $ (\gg^H_{k}(t),t)$ are horizontal with
  respect to $A^H.$ Using Lemma~\ref{horlem} we can get this from  the 
  easier situation
  where $\gg(t)=\sum_k\lam^0_{k}(t)  \gg^0_{k}(t)$ is a polar decomposition
  in which the $ \gg^0_{k}(t)$ evolve horizontally with
  respect to $A^0.$
  
  So we now examine the polar decompositions with reference to 
  $A^0$. A horizontal lift of $[\gg_{k}(t)]$ can be taken  of the form
  $\phi_{k}(t)\ot{\psi_{k}(t)}$ where $\phi_{k}(t){\rm\ and\ }\psi_{k}(t)$  
   are horizontal with respect to their own (lower-dimensional) canonical connections on
  $ \bfS({\cal H}_1){\rm\ and\ } \bfS({\cal H}_2).$  Indeed,
  \begin{equation}\label{20deigenvec}%еееееееееееееее
  \<\phi_{k},\partial_t\phi_{k}\>=0\;\&\<\psi_{k},\partial_t\psi_{k}\>=0\Ra
  \<\phi_{k}\ot\psi_{k},\partial_t(\phi_{k}\ot\psi_{k})\>=0.
  \end{equation}%ееееееееееееееееее      еееееееееееееее

  Let $\bfA(t):= \wt{\gg^0(t)}\wt{\gg^0(t)}^*\,,
  \bfB(t):= \wt{\gg^0(t)}^*\wt{\gg^0(t)},$ 
as in Appendix~\ref{AppPolar}
 and let the $r_j^2=|\lam_j|^2$ be
  the common eigenvalues of $\bfA(t)$ and  $\bfB(t).$ 
  Because these  $\phi_{k}(t){\rm\ and\ }\psi_{k}(t)$
  are assumed non-degenerate eigenvectors of  $\bfA(t){\rm\ and\ }\bfB(t)$, we can apply first-order
  perturbation theory (see Appendix~\ref{AppPolar}.)
   Then
  \begin{equation}\label{19deigenvec}%еееееееееееееее
  \dot\phi_j= \sum_{k\ne j } {\<\phi_{k},\dot \bfA\phi_{j}\>\over r_j^2 -r^2_{k}} \phi_{k},\;
  \dot\psi_j= \sum_{k \ne j} {\<\psi_{k},\dot \bfB\psi_{j}\>\over r^2_j -r^2_{k}} \psi_{k}.
  \end{equation}%еееееееееееееее
  Moreover we can also make $\dot\bfA(t)$ and  $\dot\bfB(t)$ more explicit
  by taking reduced traces, as in \cite[Blum, p.72]{Blum}.
  Let $P_{\gg}$
be the 1-dimensional orthogonal projection
  on $\gg\in {\cal H}_1\ot{{\cal H}_2}.$ Then
  \begin{equation}\label{23deigenvec}%еееееееееееееее
  \dot\bfA(t)={{1\over i}}{\rm Tred}_{1}
  [H,P_\Gamma(t)],\;\dot\bfB(t)={{1\over i}}{\rm Tred}_{2}
  [H,P_\Gamma(t)].
  \end{equation}%ееееееееееееееееее      еееееееееееееее

  We  make use of the abbreviations: %еееееееееееегггггггггггггггггггг
  \begin{eqnarray}\label{defH_jkmn}%еееееееееееееее
  H_{jk,mn}&:=& \< H(\phi_j\otimes\psi_k) ,\phi_m\otimes\psi_n\>.\\
 \label{defbeta{ab}}%еееееееееееееее
  \beta_{ab}&:=&-i(\overline{\lam_b}\sum_k\lam_k {H_{ab,kk}}-
  \lam_a\sum_k\overline{\lam_k} H_{kk,ba})\\
 \label{defbeta'{ab}}%еееееееееееееее
  \beta'_{ab}&:=&
  -i(\overline{\lam_b}\sum_k\lam_k {H_{ba,kk}}-
  \lam_a\sum_k\overline{\lam_k} H_{kk,ab})
  \end{eqnarray}%еееееееееееееее
  Combining these abbreviations with \eqref{19deigenvec}, \eqref{20deigenvec},
 and Lemma~\ref{lambdaEq}, 
  we  are going to prove  (using Appendix~\ref{dEigenVec} )  for  
   the canonically horizontal evolutions that  the following system of highly
  coupled,  highly non-linear (usually infinite)  autonomous  system of first order ODE's holds:
    \begin{eqnarray}%еееееееееееееее
  \label{eqxakeq1}\dot \lam_a &=& -i\sum_k H_{aa,kk}\,\lam_k \\
    \label{eqxakeq2}{\dot\phi_a} &=& \sum_{k\ne a }
   {\beta_{ka}\over |\lam_a|^2-|\lam_k|^2}{\phi_k}
   \\ 
    \label{eqxakeq3}{\dot\psi_a} &=& \sum_{k\ne a }
   {\beta'_{ka}\over  |\lam_a|^2-|\lam_k|^2}{\psi_k }.
    \end{eqnarray}%еееееееееееееееееееееееееее
  
  \begin{thm}\label{OmOm} %%еееееееTHEOREMееееееееееTTTTTTTTTTTTT
  Let $\phi_{k}(t),\psi_{k}(t)$ be horizontal
   with respect to their own (lower-dimensional) canonical connections on
  $ \bfS({\cal H}_1){\rm\ and\ } \bfS({\cal H}_2).$
  Let $\gg^0_{k}(t)=\phi_{k}(t)\ot\psi_{k}(t).$
 The $ \gg^0_{k}(t)$ are horizontal with
  respect to $A^0.$
  Let $\gg(t)=\sum_k\lam_{k}(t)  \gg^0_{k}(t)$ be a polar
  decomposition.  Then the $\lam_{a},\phi_{a},\psi_{a}$ satisfy the above system of autonomous
  ODE's.
  \end{thm}
{\bf Proof.}
Taking reduced traces, as in \cite[p.72]{Blum}, 
\begin{equation}\label{eq:rhodot}%еееееееееееееее
\dot\rho_{S_1}(t)={{1\over i }}{\rm Tred}_1
[H,P_\Gamma(t)]
\end{equation}%еееееееееееееее
Now we can express this in terms of polar bases:
\begin{equation}%еееееееееееееее
\hskip-1in  \<\phi_a,{\rm Tred}_1HP_\Gamma(\phi_b)\>
=\sum_k\<\phi_a\otimes\psi_k,HP_\Gamma
(\phi_b\otimes\psi_k)\>
\end{equation}%еееееееееееееее
\begin{equation}%еееееееееееееее
=\sum_k\<\phi_a\otimes\psi_k,\<\Gamma,
\phi_b\otimes\psi_k\>H\Gamma\>
=\sum_k\<\Gamma,
\phi_b\otimes\psi_k\>\<\phi_a\otimes\psi_k,H\Gamma\>
\end{equation}%еееееееееееееее
\begin{equation}%еееееееееееееее
=\sum_k\delta_{kb}\,\v{\lam_b}\,\<\phi_a\otimes\psi_k,H\Gamma\>
=\v{\lam_b}\,\<\phi_a\otimes\psi_b,H\Gamma\>
\end{equation}%еееееееееееееее

Similarly, for the oppositely ordered product in the commutator,
\begin{equation}%еееееееееееееее
\hskip-1in  \<\phi_a,{\rm Tred}_1P_\Gamma H(\phi_b)\>
=\sum_k\<\phi_a\otimes\psi_k,P_\Gamma
H(\phi_b\otimes\psi_k)\>
\end{equation}%еееееееееееееее
\begin{equation}%еееееееееееееее
=\sum_k\<P_\Gamma
(\phi_a\otimes\psi_k),H(\phi_b\otimes\psi_k)\>
\end{equation}%еееееееееееееее
\begin{equation}%еееееееееееееее
=\sum_k\<\<\Gamma,
\phi_a\otimes\psi_k\>\Gamma,H(\phi_b\otimes\psi_k)\>
=\sum_k\overline{\<\Gamma,
\phi_a\otimes\psi_k\>}\<H\Gamma,\phi_b\otimes\psi_k\>
\end{equation}%еееееееееееееее
\begin{equation}%еееееееееееееее
=\lam_a\<H\Gamma,\phi_b\otimes\psi_a\>
\end{equation}%еееееееееееееее
Thus we can write \eqref{eq:rhodot} as
\begin{equation}\label{eq:rhodot2}%еееееееееееееее
\<\phi_a,\dot\rho_{S_1}(\phi_b)\>={{1\over i }}
(\v{\lam_b}\,\<\phi_b\otimes\psi_a,H\Gamma\>-
\lam_a\<H\Gamma,\phi_a\otimes\psi_b\>)
\end{equation}%еееееееееееееее
We also have,
\begin{equation}%еееееееееееееее
H\Gamma=\sum_k\lam_k H \Gamma_k=\sum_k \lam_k H
(\phi_k\otimes\psi_k)
\end{equation}%еееееееееееееее
Thus
$$%еееееееееееееее
\<H\Gamma,\phi_b\otimes\psi_a\>=
\sum_k\,\v{\lam_k}\, \<H
(\phi_k\otimes\psi_k),\phi_b\otimes\psi_a\>=
\sum_k\,\v{\lam_k}\, H_{kk,ba}
$$%еееееееееееееее
$$%еееееееееееееее
\<\phi_a\otimes\psi_b,H\Gamma\>=\sum_k \lam_k\<\phi_a\ot\psi_b,H \phi_k\ot\psi_k\>
=\sum_k\lam_k {H_{ab,kk}}
$$%еееееееееееееее
\begin{equation}\label{eq:rhodotab7} %еееееееееееееее
\<\phi_a,\dot\rho_{S_1}(\phi_b)\>={{1\over i }}
(\v{\lam_b}\,\sum_k\lam_k \overline{H_{kk,ab}}-
\lam_a\sum_k\v{\lam_b}\, H_{kk,ba})
\end{equation}%еееееееееееееее
This, combined with \eqref{eq:deigen21} yields \eqref{eqxakeq2}.

For $S_2,$ we get  
\begin{equation}\label{eq:rhodotab7s2}%еееееееееееееее
\<\psi_a,\dot\rho_{S_2}(\psi_b)\>={{1\over i }}
(\v{\lam_b}\,\sum_k\lam_k \overline{H_{kk,ba}}-
\lam_a\sum_k\v{\lam_k}\, H_{kk,ab})
\end{equation}%еееееееееееееее
This yields \eqref{eqxakeq3}. The proof of \eqref{eqxakeq1} is similar, but since we want to establish  
a more general version  that will be needed later, the proof is given in Lemma~\ref{lambdaEq}.

  \nin$\qed$
  
  \begin{thm}\label{1m} %%еееееееTHEOREMееееееееееTTTTTTTTTTTTT
  Let $\gg(t)=\sum_k\lam^H_{k}(t)  \gg^H_{k}(t)$ be a polar decomposition
  in which the  $ \gg^H_{k}(t)$ evolve horizontally with
  respect to $A^H.$
   Then, up to a constant phase factor, the $\lam^H_{k}(t)$ can be expressed in terms of the 
  $\lam_{k}(t)$ in the last theorem as
  \begin{equation}\label{C2S}%еееееееееееееее
  \lam^H_{k}(t)=\lam_{k}(t)  e^{\int_0^ti H_{kk,kk}ds }
  \end{equation}%еееееееееееееее
  \end{thm}
  {\bf Proof.}
  By Lemma~\ref{horlem}, $\gg^H_{k}(t)=\gg^0_{k}(t) e^{-i\int_0^t\<\gg^0_{k},H
  \gg^0_{k}\>ds}=
  \gg^0_{k}(t) e^{-i\int_0^t H_{kk,kk}(s)ds}.$
  \nl $\qed$
  
\section{Natural Partitions of Toroids }\label{Pyth}

\begin{quotation}
I don't believe it; you've actually found a practical use for geometry!  \par
     \rightline{\emph{B. Simpson} } 
\end{quotation}

The polar state space of $\ss=\ss_1+\ss_2$ replaces { each point  of the SQM state space $\bfP$}  
  by a  toroid of phase factors. It turns out this toroid has a canonical partition into 
convex subsets, one for each  circular factor. This is exactly what is needed to associate to 
the pair $\ss_1$ and 
$\gg=\sum_k q_k\,\phi_k\ot\psi_k$ precisely one of $[\phi_k]$. The reason  for 
considering  
the  {\it Pythagorean} type of partition is discussed in Section~\ref{section5Discussion}.

In this section, we carry out this purely mathematical analysis of the partitioning of the
right toroids which have arisen  as the fibers of  the polar bundle. 
We state and discuss the precise theorems  in the next section, and the proofs  
are in the following sections. 

\subsection{Right toroids and their mappings. }
\setcounter{equation}{0}

By a {\it right} toroid $ \bfT(\bfr)$ is meant the direct product of circles. Let $I$ be an index set
which is either all positive integers  $\Z_{>0}$ or just those in $[1,n]$ where $n\in \Z_{>0}.$
 Occasionally we abuse notation to write $\Z_{>0}=[1,n]$ with $n=\infty.$  Recall,
that if  $\bfr=(r_k),
 r_1\ge r_2\ge\cdots\ge0,$ then
$$ \bfT(\bfr)=\prod_{k\in I}\bfS^1(r_k) =\prod_{k\in I}\(\R/2\pi r_k\Z\).$$
We have a canonical surjection $\varpi:\R^n\ra\bfT(\bfr),$ which is a local isometry.
 We denote the set of all these right toroids  by ${\mathfrak T}.$ 
An allowable map (or morphism) 
$$
\iota:\bfT(\bfr)\ra \bfT({\bfr}') {\rm\ is\ of\ the\ form\ }
(\zeta_j)_{j<n}{\buildrel \iota\over\ra}({\zeta}'_k)_{k<n'}
$$ 
where for every $k<n'$ there exists
$j<n$ so that
 $\zeta'_k=\zeta_j$ or $\zeta'_k=1.$
 We also require that $\iota$ be injective, so that $\iota(\bfT(\bfr))$  is just a sub-product
of $\bfT(\bfr')$. In other words,
 the morphisms {\it split}.  $\bfT(\bfr){{\buildrel \iota\over\hra}} \bfT({\bfr}') $
 is just an inclusion map, which  can usually be omitted. 
It is easy to check that if     $\bfT(\bfr){{\buildrel \iota\over\hra}}  \bfT(\bfr')$ 
is a morphism, then so is the induced map $\bfT(\bfr)\cap\bfT(\bfr')\hra \bfT(\bfr) .$

For each $k\in I$, 
$
C_k:=C_k(\bfr)\deq \{(\zeta_j)_{j\in I}\}, {\rm\ where\ } \zeta_j= 1 {\rm\ if\ } j\ne k,
$
 are the circles bijectively
corresponding to the imbedded image of
$\bfS^1(r_k)$. For example, $C_1=(S^1(r_1),1\cdots ),$ is a member of the
 {bouquet} 
 $\{C_j\:|\; j\in I\}.$ 
These right toroids arise as the amplitudes in polar decompositions, i.e. the fibers of the 
polar bundle discussed in Section~\ref{section3EnlargingStates}. 
The {\bf main diagonal} circle (or subgroup)
${\tilde D}:={\tilde D}(\bfr)\deq \{ ( r_j e^{i\theta}\;|\; \theta\in \R\}$  plays a central role.
 For any $\bft\in \bfT(\bfr)$, we call a translate
$\bft\cdot{\tilde D}\subset\bfT(\bfr)$ of ${\tilde D}$   {\bf  a} diagonal. 

\centerline {\bf Pythagorean Partitions.} 

We are interested in the natural way(s) of partitioning right toroids.
So we form  the set  ${\mf P\mf T} $ of {\bf partitioned} right toroids $\(\bfT(\bfr), {\mf P}(\bfr)\) $. 
Here the partitions ${\mf P}$ are of  the {\bf Pythagorean}
 type ${\mf P}=\{{\mf  p}_j\}_{j\in I}$; this means
 $\(\bfT(\bfr), {\mf P}(\bfr)\) $ has the following
three properties:
\nl$ {\bf Partition\ Property}:
\bfT(\bfr)=\cup_{j\in I} {\mf  p}_j, {\rm\  the\ interiors\ } \ca {\mf  p}_j {\rm\ are\ disjoint\ and}
$
each is equal to the closure of its interior: ${\mf  p}_j=\v{\ca {\mf  p}_j}.$
\nl {\bf Convexity Property}: For all $j\in I$ there exists a compact  convex set  $A_j$ in $\R^n$ with
$\varpi(A_j)={\mf  p}_j$ and $\varpi|{\ca A_j}$ is injective.  
\nl{\bf Diagonal Property }: Every diagonal $\varpi(\R \bfr)$ of $ \bfT(\bfr)$ intersects ${\mf  p}_j$
in an arc of length $2\pi r_j^2.$

 We want to show there is a uniform way of
partitioning any right toroid.  In other words, we want to find a {\it natural } 
procedure ${\textsf{P}}$
which, when applied to $\bfT(\bfr),$ yields a Pythagorean  partition ${\mf P}(\bfr)$ of  $\bfT(\bfr),$ 
i.e. 
\begin{equation}\label{PI functor}%еееееееееееееееееееееееееееееееее
{\mf T}\ni\bfT(\bfr){{\buildrel \textsf{P}\over\rightsquigarrow}} \(\bfT(\bfr), {\mf P}(\bfr)\) 
\in {\mf P\mf T}.
\end{equation}%еееееееееееееееееееееееееееееееее
We mean {\bf natural} in the   general categorical sense reviewed in 
Appendix~\ref{AppCats}, which in this specific  situation amounts to the  following
\nl {\bf Naturality Property}: If    $\bfT(\bfr){{\buildrel \iota\over\hra}}  \bfT(\bfr')$ then for all  
 $k\in I'$,  there exists a  $j\in I$ so that   $\bfT(\bfr)\cap{\mf  p}'_k\subset {\mf  p}_j(\bfr).$
  (Here we are denoting $ {\mf  p}_k(\bfr')$ by ${\mf  p}'_k.$)

All we really need are the cases of the Naturality Property  when $\bfT(\bfr)$ has dimension
$n<\infty$, and  $\bfT(\bfr)\cap{\mf  p}'_k$ also has dimension $n$.

  \nin{\bf Example $n=2$.} Refer to the rectangle OABC in Figure~\ref{figenvelope}, in which 
the sides have lengths $2\pi r_1,2\pi r_2$ in the golden ratio. 
We have  $\bfT(r_1,r_2) = \wt{\rm{OABC}}, C_1=\wt{\rm{OC}}, C_2=\wt{\rm{OA}},
\wt{D}=\wt{\rm{OB}}.$  Let $\rm{AE},\rm{CF}\perp \rm{OB}.$ 
We can take $\rm{OE'CF}$ for ${\bf A}_1$, and ${\bf A}_2=\rm{ OF'AE}.$ 
 Set ${\bf A}={\bf A}_1\cup {\bf A}_2 $
so that $\bfT(r_1,r_2) = \tilde{{\bf A}}.$  The Naturality Property above 
entails that $\bfT(r_1)={\mf  p}_1\subset{\mf  p}'_1(r_1,r_2)\cap \bfT(r_1,1)=\wt{\rm OC }.$ 
 It follows that 
$C_1\subset{\mf  p}'_1(r_1,r_2)\subset C_1,$ i.e. $C_1= {\mf  p}_1(r_1,r_2)\cap \bfT(r_1,1).$
 In particular, $C_1\subset {\mf  p}_1(r_1,r_2). $  This always
 happens: For all $k$, $C_k\subset {\mf  p}_k,$
whenever they exist (Lemma~\ref{diag9}). A similar argument applies to $C_2=\wt{\rm OA}.$

\begin{figure}
  \begin{center}
\includegraphics[scale=.75,trim=0in 0in 1in 2in]{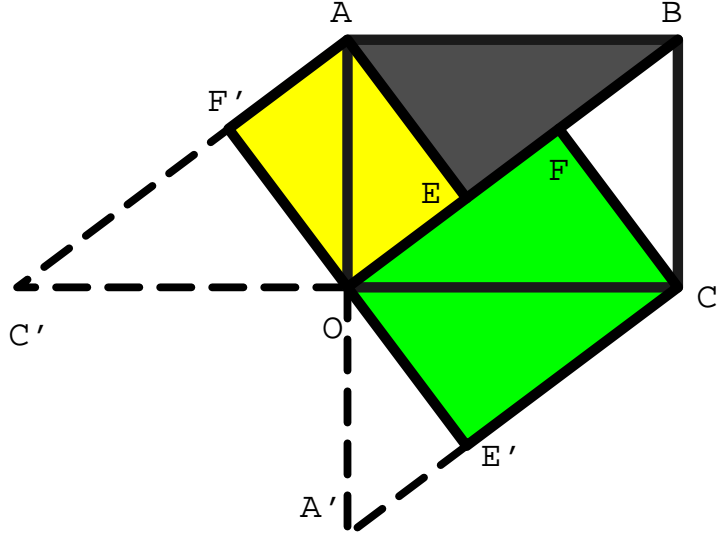}
  \end{center}
  \caption{
   Shown are 
$A_1={\rm OFCE'}$, $A_2={\rm OF'AE}$. Also shown are 
$ {\scriptstyle Sl}^\dagger(1,\b1)={\rm OFC}$, 
$ {\scriptstyle Sl}^\dagger(1,\sigma)={\rm OE'C}$, 
$ {\scriptstyle Sl}^\dagger(2,\b1)={\rm OF'A}$, 
$ {\scriptstyle Sl}^\dagger(2,\sigma)={\rm OEA}$, 
$B_1={\rm OA'CB}$, $B_2={\rm OC'AB},$ which are defined below.
}
  \label{figenvelope}
  \end{figure}

We want to prove the following result.
\begin{thm}\label{MainThm} %TTTTTTTTTTTTTTTTTTTTTTTT
There exists exactly one natural procedure  $\textsf{P}$ satisfying  the
 relation~(\ref{PI functor}), or in the language of Appendix~\ref{AppCats}:
there is a  manifest functor $\textsf{P}$ from ${\mf T} $ to ${\mf P\mf T}.$
\end{thm}

 The existence part of the proof begins with a description of the partition for dimension
$n.$
  We  then graphically illustrate it for $n=3$, as we have already done for $n=2.$ This 
gives the geometric  idea motivating the proof for general $n.$ Then we give an existence proof for any
$n$. By the naturality and  taking limits, this even holds for $n=\infty.$ 
We conclude with  a proof of the more difficult uniqueness
assertion.

\subsection{The construction of the Pythagorean partition.}\label{constr}

The idea behind the existence proof is the following: We start with an $n$-dimensional  box
 (rectangular parallelepiped) which covers the toroid almost isometrically. The box has a
natural partition into $n!$ simplices obtained by slicing it with $n$ 
hyperplanes through the main
diagonal  which are perpendicular 
to the faces. Then each simplex can be partitioned into $n$ convex subsets (slabs) by slicing it 
with hyperplanes perpendicular to the main diagonal, which acts as the hypotenuse.
 This hypotenuse  corresponds to a Hopf circle whose phase determines  
{which state  of  the subsystem $\ss_1$ obtains.}
The  slices perpendicular to the main diagonal are made at each vertex  in the given simplex. 
Now comes the {surprising} part (even for $n=3$): these slabs can be 
 translated to the generating
edges of the  box in only one way and when this is done, we are left with $n$ convex   
  neighborhoods of
these $n$  edges. The interiors of these parts map 
isometrically  to the desired members of the partition of the toroid.
The  desired partition is determined by the following definitions, where the $\bfe_j$ denotes
 the standard unit vectors.
\nl$
 s_j\deq2\pi r_j,%\bfe_j\deq(0,\cdots,\overbrace{1}^j,\cdots),
\bff_j\deq s_j\bfe_j,\,\bfs=\sum_j\bff_j,\,
\bfg_j\deq r_j^2\bfs-\bff_j.$
\nl ${\bf A}_k:={\bf A}^n_k\deq$ parallelotope generated by
 $\{\bfg_j,\bfg_j+r_k^2\bfs\;|\;1\le j\ne k\le n\}.$
\nl${\cal L}^n(s_1,\cdots,s_n):= {\cal L}^n(\bfs) \deq \bigoplus_j \Z\bff_j$, 
the lattice generated by the $\bff_j.$

\noindent $\varpi:\R^n\mapsto\bfT^n(r_1,\dots,r_n)\deq \R^n/{\cal L}^n(\bfs)$ is the
canonical local isometric surjection mentioned above.
\nl${\wt \bfx}\deq \varpi(\bfx)$ for any  $\bfx\in \R^n$ (so as to reduce the number of
$\varpi$'s and parentheses.)
\nl ${\mf p}_k\deq {\wt{\bf A}_k}.$
This defines the  partition $\mf P=\{\mf p_1, \mf p_2,\dots,\mf p_n\}.$ The reader
can check that Figure~\ref{figenvelope} comports with it, for $n=2.$
%\pagebreak
 Figure~\ref{GR} and Figure~\ref{YL} show the component parts, before 
translation,
 of $A_3$ for a box with edge-length ratios 5:4:3. 
\def\scal{.6}
 \begin{figure}[h]
  \begin{center}
\includegraphics[scale=.55]{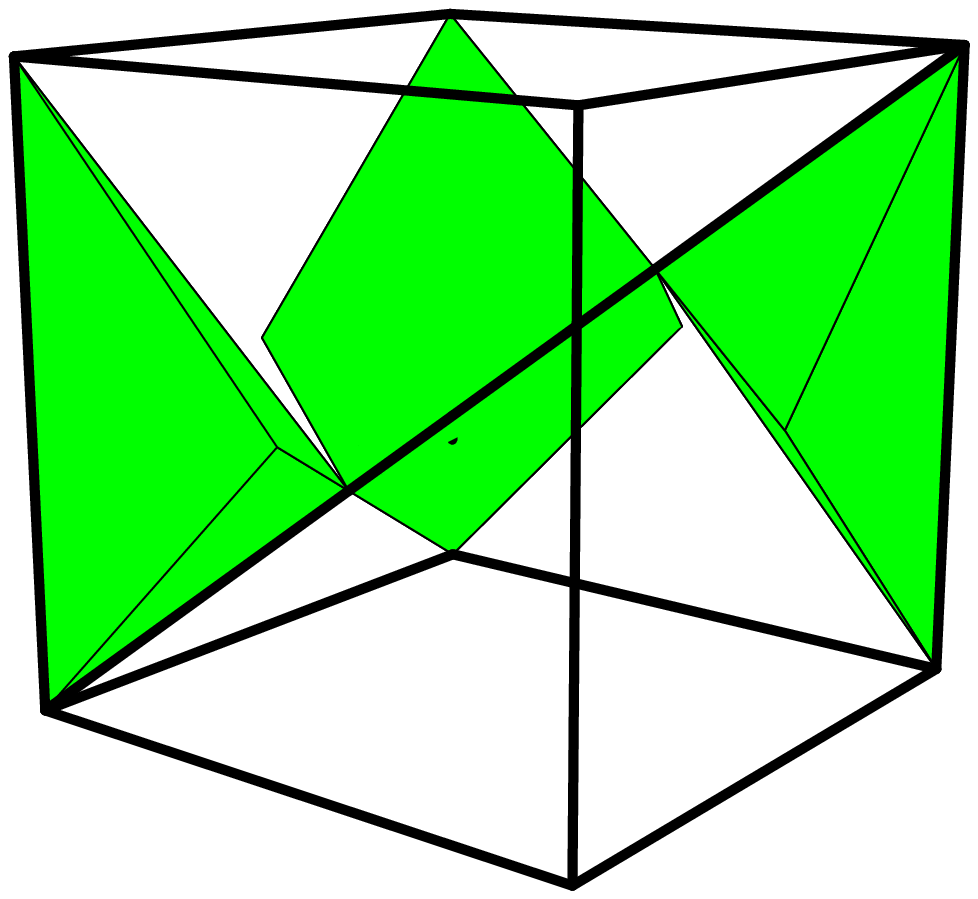}
 \end{center}
  \caption{
 ${\scriptstyle Sl}(3,312),{\scriptstyle Sl}(3,123),{\scriptstyle Sl}(3,231)$}
  \label{GR}
  \end{figure}

  \begin{figure}[h]
  \begin{center}
\includegraphics[scale=.55]{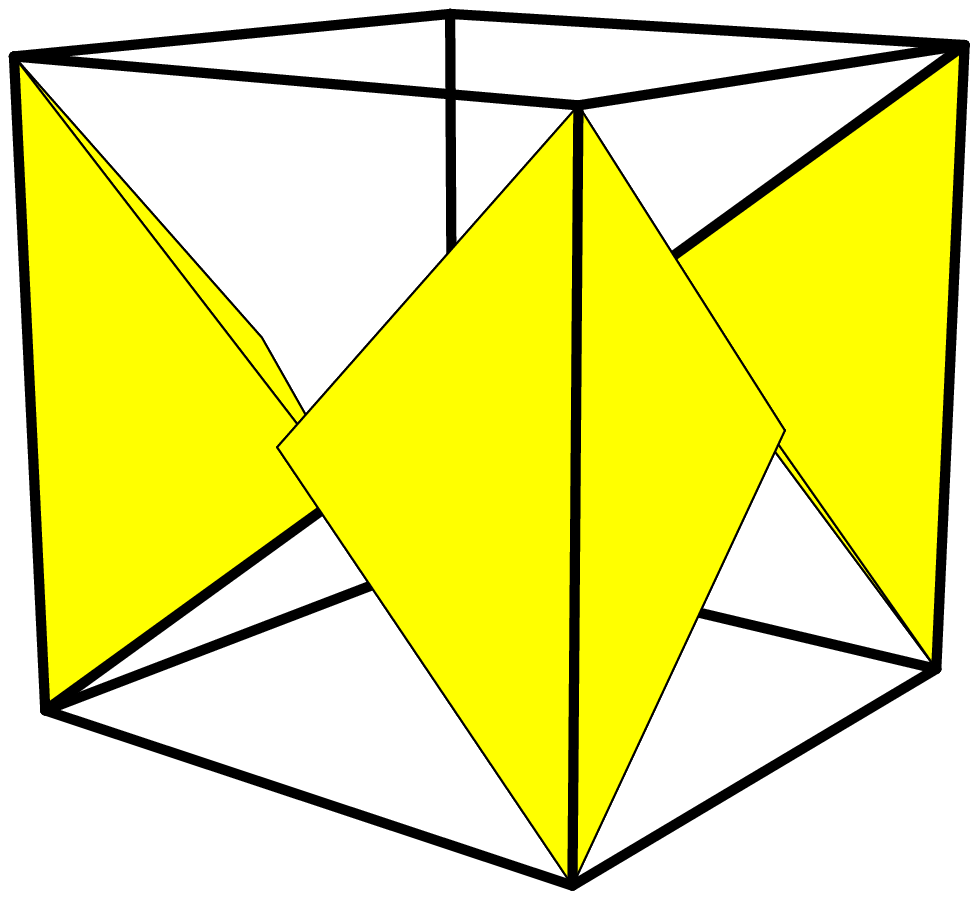} \end{center}
  \caption{
 ${\scriptstyle Sl}(3,132),{\scriptstyle Sl}(3,321),{\scriptstyle Sl}(3,213)$}
  \label{YL}
  \end{figure}

%   In Figure~\ref{YG}, the parts are 
%   lattice-translated to form the parallelepiped $A_3.$

      \begin{figure}[h]
      \begin{center}
 %\leavevmode\epsfysize=8.5cm\epsffile{YG.eps}  
 \includegraphics[scale=\scal,trim= 0 0 0 144]{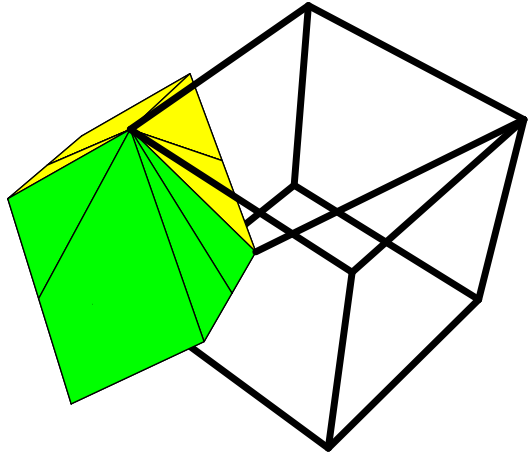}  
 \end{center}
      \caption{
     $A_3=\bigcup_{\sigma\in\Sigma_3}{\scriptstyle Sl}^\dagger(3,\sigma)$}
      \label{YG}
      \end{figure}
  \begin{figure}
  \begin{center}
 % \leavevmode\epsfysize=9.5cm\epsffile{AN111.eps}
%  \includegraphics[scale=\scal,trim= 0 0 0 144]{AN111.eps}  
  \includegraphics[scale=\scal,trim= 0 0 0 0]{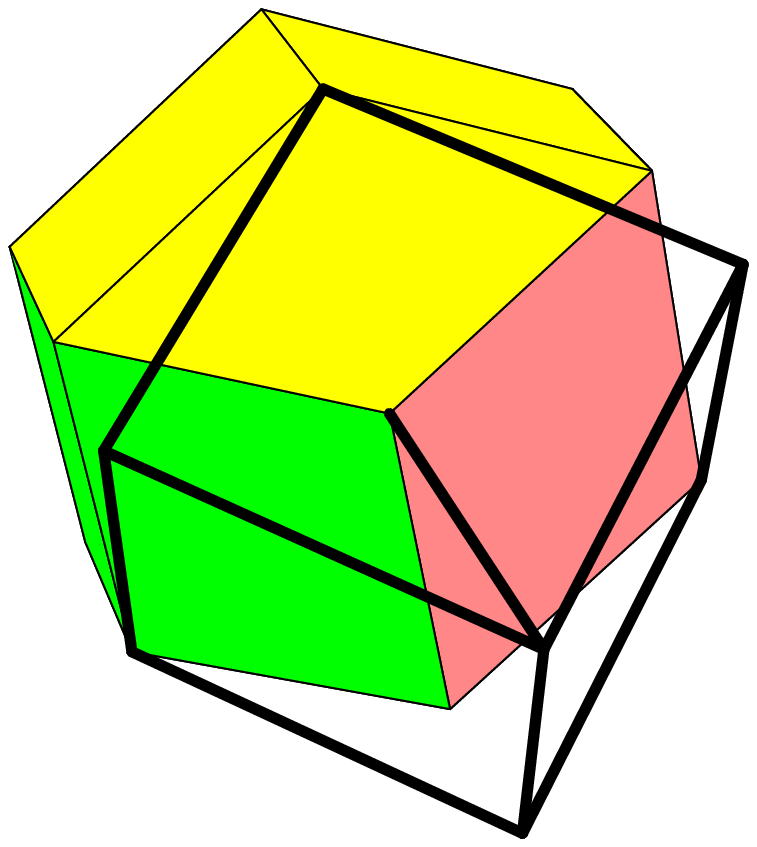}  
\end{center}
  \caption{
  The  partition for a cube.}
  \label{A111}
  \end{figure}

In Figure~\ref{YG}, the parts are 
 lattice-translated to form the parallelepiped $A_3.$

 In Figure~\ref{A111}, we illustrate  the  $A_k$  for the 3-cube. 
  \begin{figure}
  \begin{center}
%  \leavevmode\epsfysize=8.5cm\epsffile{AN543.eps}
% \includegraphics[scale=\scal,trim=0 0 0 144]{AN543.eps}
\includegraphics[scale=\scal]{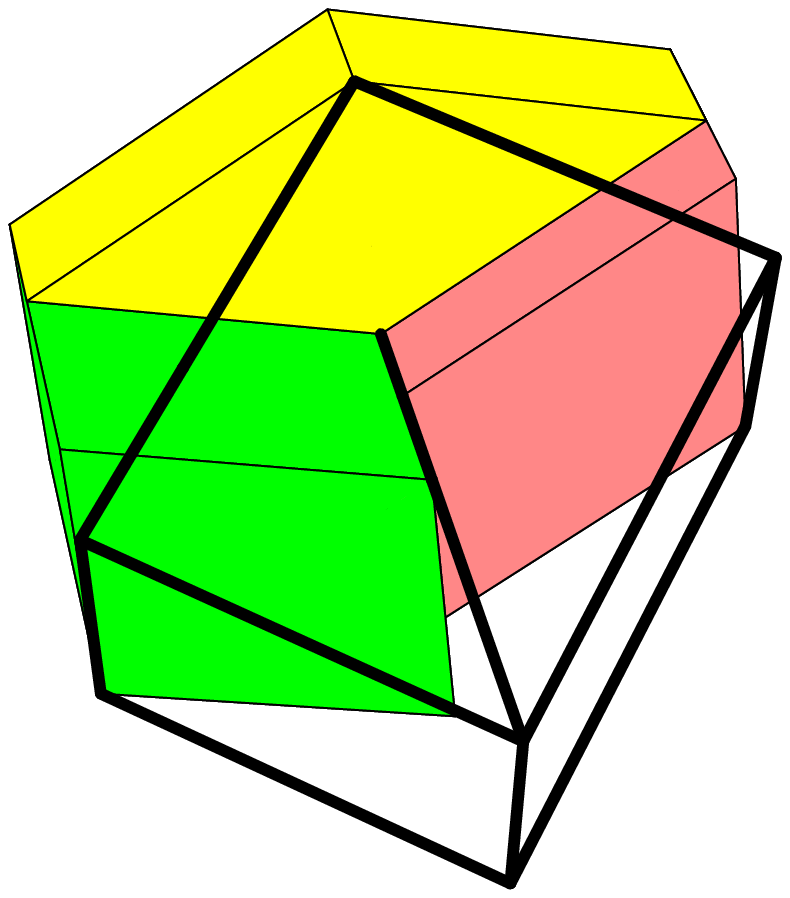}
  \end{center}
  \caption{
  The  partition for a box with edge ratios 5:4:3.}
  \label{A543}
  \end{figure}

In Figure~\ref{A543}, we illustrate  the $A_k$  for a $3$-dimensional box. 

  \begin{figure}
  \begin{center}
  %\leavevmode\epsfysize=8.5cm\epsffile{AN111s.eps}
%\includegraphics[scale=.5]{AN111s.eps}
\includegraphics[scale=.9,trim=0 0 0 72]{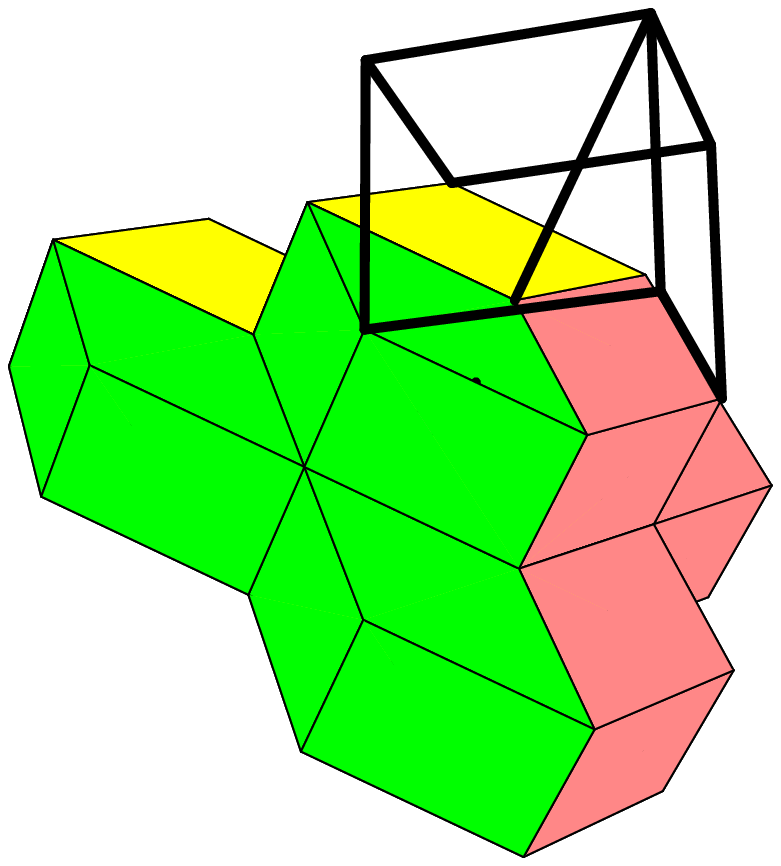}
  \end{center}
  \caption{
  The  tiling for a cube.}
  \label{A111s}
  \end{figure}
In Figure~\ref{A111s}, we illustrate  a portion of the  corresponding tiling.
  \begin{figure}
  \begin{center}
%  \leavevmode\epsfysize=8.5cm\epsffile{AN111sh.eps}
\includegraphics[scale=1]{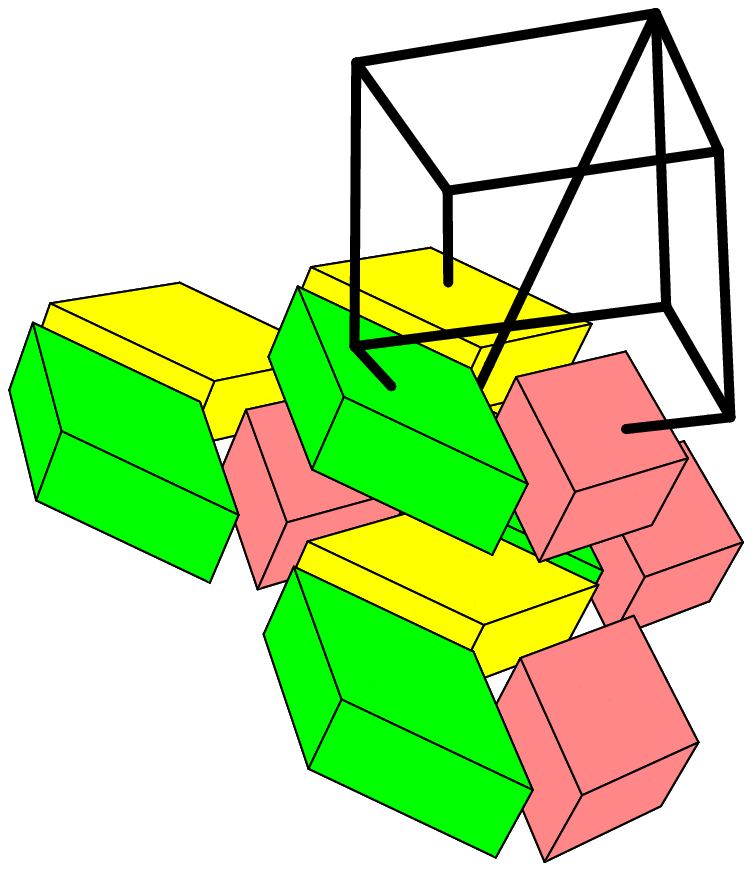}
  \end{center}
  \caption{
  The  shrunken tiling for a cube.}
  \label{A111sh}
  \end{figure}

In Figure~\ref{A111sh}, we
illustrate  the same  portion where each tile has been linearly shrunken towards its centroid by
a factor .8 so as to better reveal how successive layers appear twisted, although they 
 are obtained by lattice translations.

%%\newpage 

\subsection{The existence of a Pythagorean partition}\label{AppPart}

We will need some more definitions which we collect here. 
These are required for the  proofs of  existence and uniqueness.

 \begin{eqnarray*}%еееееееееееееее
 \Sigma_n  &\deq& {\rm symmetric\ group\ on\ }\{1,\cdots,n\},
 \b1:={\rm\ identity\ element\ of\ }\Sigma_n.\\
{\rm For\ all\ }  U\subset \R^n\;[U]&\deq& {\rm convex\ hull\ of\ } U.\\
U+v& \deq& \{u+v\;|\;u\in U\subset \R^n\}.\\
  D &\deq &[ \bo ,\bfs],\,\,\,\,D^\perp\deq\{\bfx\in\R^n\;|\;\bfs\cdot \bfx=0\}.\\
{\scriptstyle Slab}(v,w,U)&\deq & [D^\perp+v,D^\perp+w]\cap U\; \;\forall
U\subset\R^n.\\
\(\forall\, \sigma\in\Sigma_n\)\, S_\sigma& \deq&
[0,\bff_{\sigma(1)},\bff_{\sigma(1)}+\bff_{\sigma(2)},\cdots,
\sum_{\nu=1}^n\bff_{\sigma(\nu)}=\bfs]. \\%\;S_\b1=[\bo,\bff_1,\bff_1+\bff_2,\cdots,\bfs].\\
{\rm\ In\ particular\ } S_\b1&=&[\bo,\bff_1,\bff_1+\bff_2,\cdots,\bfs].\\
\bfv_{k\sigma}&\deq&\sum_{\nu<\sigma^{-1}(k)}\bff_{\sigma(\nu)}\\
{\scriptstyle Sl}(k,\sigma)&\deq&
  {\scriptstyle Slab}(\bfv_{k\sigma},\bfv_{k\sigma}+\bff_{k},S_\sigma)\\
{\scriptstyle Sl}^\dagger(k,\sigma)&\deq& {\scriptstyle Sl}(k,\sigma)-
\bfv_{k\sigma}
\end{eqnarray*}%еееееееееееееееееееееееееее
We say two sets in $\R^n$ of dimension $n$ are {\it quasi-disjoint} if their intersection 
 is of smaller dimension.
\pagebreak
\clearpage
\begin{lem}\label{simpartlem} %LLLLLLLLLLLLLLLLL
$\{S_\sigma\;|\;\sigma\in\Sigma_n\} $ is a partition of the box  $B$
generated by $\bff_1,\dots,\bff_n.$
 \end{lem}%ELELELELELELELELELELELELLE
{\bf Proof.} 
 The $S_\sigma$ are {quasi-disjoint}. Each
$S_\sigma\subset  B,$ so it suffices to
show 
$ B\subset   \cup_{\sigma\in\Sigma_n} S_\sigma.$ 
Let $\bfv\in B.$  Then $\bfv=\sum^n_{\nu=1} a_\nu \bff_\nu\,, a_\nu=\<\bfv,\bff_\nu\>.$
Also $a_\nu\in[0,1]$ and there exists $\sigma\in \Sigma_n$ so that
$$
a_{\sigma(1)}\ge a_{\sigma(2)}\ge\cdots \ge a_{\sigma(n)}\ge a_{\sigma(n+1)}:=0.
$$
$$
\tf \bfv=\sum^n_{\nu=1}  (a_{\sigma(\nu)}-a_{\sigma(\nu+1)})(
\bff_{\sigma(1)}+\cdots+\bff_{\sigma(\nu)})\in S_\sigma.
$$
$\qed$

\begin{prop}\label{simpart} %PPPPPPPPPPPP
$\forall k\le n\; \{ {\scriptstyle Sl}^\dagger(k,\sigma)\;|\;\sigma\in\Sigma_n\} $
 is a partition of ${\bf A}_k.$
 \end{prop}%ELELELELELELELELELELELELLE
{\bf Proof.}  
$${\scriptstyle Sl}(k,S\sigma)=[D^\perp+v_{k\sigma},
D^\perp+v_{k\sigma}+\bff_k]\cap S_\sigma
$$

$$\tf {\scriptstyle Sl}^\dagger(k,\sigma) = [D^\perp,
D^\perp+\bff_k]\cap( S_\sigma-v_{k\sigma})
$$
$$
 S_\sigma-\bfv_{k\sigma}= \left[-\sum_{\nu=1}^{\sigma^{-1}(k)}\bff_{\sigma(\nu)},
-\sum_{\nu=2}^{\sigma^{-1}(k)}\bff_{\sigma(\nu)},\dots, \bo ,\bff_k,
\sum_{\nu=\sigma^{-1}(k)}^{\sigma^{-1}(k)+1}\bff_{\sigma(\nu)},\dots,
\sum_{\nu=\sigma^{-1}(k)}^n\bff_{\sigma(\nu)}
\right]
$$

\nin $B_k:= \cup_\sigma   ( S_\sigma-v_{k\sigma}),\,E\{n,k\}:= \{1,\dots,n\}- \{k\}$
$$\tf B_k =\left [-\sum_{j\in M}\bff_j,\bff_k+\sum_{j\in M}\bff_j
\;{\Bigg |}\; M\subset E\{n,k\}\right]
$$ 
Set $\bfv_k=\sum_{j\ne k}\bff_j.$ Replacing $M$ by its complement in
$E\{n,k\}$ we get
$$ B_k =\left [\sum_{j\in M}\bff_j-\bfv_k,\bff_k+\sum_{j\in M}\bff_j
\;{\Bigg |}\; M\subset  E\{n,k\}\right].
$$
Set $ F_k =\left [\sum_{j\in M}\bff_j
\;{\Big|}\; M\subset E\{n,k\}\right],$ one of the $2$  facets of the box
$B$ which are perpendicular to $\bff_k.$ 
$$\tf B_k =\left [F_k-\bfv_k,F_k+\bff_k\right].
$$ 

Now $F_k+\bff_k$ is the other facet  of $B$ parallel to $F_k$ and $F_k-\bfv_k$ is a
parallel  facet of the
  box $B-\bfv_k$ (which also contains  $ \bo $ as a vertex.) 

\nin $ \bfv_M:=\sum_{j\in M}\bff_j-\bfv_k \in {\rm Vertices}(F_k-\bfv_k),\,
\bfw_M:=\sum_{j\in
M}\bff_j+\bff_k \in {\rm Vertices}(F_k+\bff_k)$

%\nin $E\{n,k\}\deq \{1,\dots,n\}- \{k\}$

Thus $B_k$ is a parallelotope
which is the convex hull of its edges 
$$ [\bfv_M, \bfw_M]
,\,\;  M\subset E\{n,k\}.$$
We have $ \bfw_M- \bfv_M= \bfv_k+ \bff_k=\bfs$ which implies  
 the $[\bfv_M, \bfw_M]$ are parallel translates of $D=[ \bo ,\bfs],$ \ie 
 $$
[\bfv_M, \bfw_M]=D+\bfv_M.
$$
$$
\tf \cup_{\sigma\in\Sigma_n} {\scriptstyle Sl}^\dagger(k,\sigma)
= [D^\perp,
D^\perp+\bff_k]\cap\cup_{\sigma\in\Sigma_n}( S_\sigma-v_{k\sigma})
= [D^\perp,
D^\perp+\bff_k]\cap B_k
$$
$$D^\perp\cap [\bfv_M, \bfw_M]\ne \emptyset\; \because
 \bfv_M\cdot \bfs\le 0\,,\bfw_M\cdot \bfs>0.
$$
$$(D^\perp+\bff_k)\cap [\bfv_M, \bfw_M]\ne \emptyset \;\because
 (\bfv_M-\bff_k)\cdot \bfs<0\,,(\bfw_M-\bff_k)\cdot \bfs\ge0.
$$
The last three relations imply
$$
 \cup_{\sigma\in\Sigma_n} {\scriptstyle Sl}^\dagger(k,\sigma)= [D^\perp\cap B_k,
(D^\perp+\bff_k)\cap B_k].
$$
$$
\tf \cup_{\sigma\in\Sigma_n} {\scriptstyle Sl}^\dagger(k,\sigma)=
\left [D^\perp\cap [\bfv_M, \bfw_M],
(D^\perp+\bff_k)\cap [\bfv_M, \bfw_M]\;\Big|\; M\subset E\{n,k\}\right]=
$$
$$
\left [D^\perp\cap(D+\bfv_M),
(D^\perp+\bff_k)\cap (D+\bfv_M)\;\Big|\; M\subset E\{n,k\}\right]
$$
$$
=
\left [D^\perp\cap(D-\sum_{j\in E\{n,k\}- M}\bff_j),
(D^\perp+\bff_k)\cap(D-\sum_{j\in E\{n,k\}- M}\bff_j)\;\Big|\; M\subset
E\{n,k\}\right]
$$
$$
=
\left [D^\perp\cap(D-\sum_{j\in  M}\bff_j),
(D^\perp+\bff_k)\cap(D-\sum_{j\in  M}\bff_j)\;\Big|\; M\subset
E\{n,k\}\right]
$$
$$
=
\left [(D^\perp+\sum_{j\in  M}\bff_j)\cap D-\sum_{j\in  M}\bff_j,
(D^\perp+\bff_k+\sum_{j\in  M}\bff_j)\cap D-\sum_{j\in  M}\bff_j\;\Big|\; M\subset
E\{n,k\}\right]
$$
$$
=
\left [\sum_{j\in  M}r_j^2\bfs-\sum_{j\in  M}\bff_j,
\sum_{j\in  M}r_j^2\bfs+r_k^2\bfs-\sum_{j\in  M}\bff_j\;\Big|\; M\subset
E\{n,k\}\right]
$$
$$
=
\left [\sum_{j\in  M}\bfg_j,
\sum_{j\in  M}\bfg_j+r_k^2\bfs\;\Big|\; M\subset
E\{n,k\}\right]={\bf A}_k, {\it i.e.} $$
$${\bf A}_k= \cup_{\sigma\in\Sigma_n} {\scriptstyle Sl}^\dagger(k,\sigma)$$
and the $ {\scriptstyle Sl}^\dagger(k,\sigma)$ are quasi-disjoint.

\nin $\qed$ 
\begin{thm}\label{existspart} %PPPPPPPPPPPP
$\varpi|{\bf\ca A}_k$ is injective (${\ca X}$ denotes the interior of $X$.) 
\newline${\mf P}\deq \{{\mf p}_k=\varpi({\bf A}_k)\;|\;k=1,\dots,n\}$ is a partition of 
$\bfT(\bfr).$
 \end{thm}%ELELELELELELELELELELELELLE
{\bf Proof.} Both assertions follow from the fact that the ${\bf A}_k$ are
unions of  translates, $ {\scriptstyle Sl}(j,S_\sigma)-\bfv_{k\sigma}$, by elements
of the lattice ${\cal L}(\bfs)$ of quasi-disjoint subsets 
${\scriptstyle Sl}(j,S_\sigma)$ of the box $B,$ which is a fundamental domain for
$\bfT(\bfr).$
\newline$\qed$

\nin $\Delta \deq  \varpi(D),$ the closed, diagonal subgroup of $\bfT(\bfr).$
\begin{lem}\label{diag} %LLLLLLLLLLLLLLLLL
$\forall k $ every coset $C$ of  $\Delta $ in $\bfT^n(\bfr)$ intersects ${\mf p}_k.$
 \end{lem}%ELELELELELELELELELELELELLE
{\bf Proof.}  By the theorem, there exists a line $ L \subset \R^n$, parallel to $D,$
such that $ \varpi(L)=C,$ which intersects some ${\bf A}_j.$ It follows that $L$ intersects the 
$(n-1)-$dimensional parallelotope generated by the 
 $\{\bfg_h\;|\;h\ne j\}.$ Thus $L +\bfg_j-\bfg_k$ intersects ${\bf A}_k$ and 
$$L +\bfg_j-\bfg_k=L+(r_j^2\bfs-\bff_j)-(r_k^2\bfs-\bff_k)
= L+\bff_k-\bff_j \equiv L\pmod{{\cal L}(\bfs)  }.
$$
$\qed$ 
\newline {\bf Remark.} $C$ is a geodesic circle contained in $\bfT(\bfr)$ which is ``parallel'' to $\Delta.$
\begin{lem}\label{diag6} %LLLLLLLLLLLLLLLLL
 For all $k$ every coset $C$ of  $\Delta $ in ${\mf p}_k\cap \bfT(\bfr)$ is a geodesic segment
of length at least $ 2\pi r_k^2.$
 \end{lem}%ELELELELELELELELELELELELLE
{\bf Proof.}  In the proof of the previous lemma, $\(L +\bfg_j-\bfg_k\)\cap {\ca {\bf A}_k}$ is an
interval of length $2\pi  r_k^2$ whose interior, by Theorem \ref{existspart}, is mapped injectively by
$\varpi$ into $\bfT(\bfr).$
\newline$\qed$
\newline $\breve{\mf p}_k\deq\varpi(\ca{\bf A}_k).$ Since $\varpi$ is a covering map, 
We have $\breve{\mf p}_k$ is open and ${\rm Dim}( {\mf p}_k-\breve{\mf p}_k)<n.$ 
Since $\sum_j 2\pi r_j^2 =2\pi  = {\rm Length}(D) = {\rm Length}(\Delta),$ and by the lemma,
$ {\rm Length}(\breve{\mf p}_k\cap C)\ge 2\pi  r_k^2,$
 we obtain
\begin{lem}\label{diag2} %LLLLLLLLLLLLLLLLL
 For all $k$ every coset $C$ of  $\Delta $ in $ \bfT(\bfr),  C\cap  \ca {\mf p}_k\,$ is either empty
or a geodesic segment of length  exactly $2\pi  r_k^2.$
 \end{lem}%ELELELELELELELELELELELELLE
 $\qed$

Each ${\mf p}_k$ is {\it toroidally convex} in the sense that it differs by a closed lower
dimensional set
 (namely $\varpi(\partial {\bf A}_k)$)
from a subset $\ca {\mf p}_k$, which is isometric via $\varpi^{-1}$ with an open convex subset $\ca{\bf 
A}_k$ in Euclidean space. We will just use the term {\bf convex} for this notion in the sequel.

We have an exact sequence
$$
 \bo \ra\Delta\ra\bfT(\bfr)\ra\bfT(\bfr)/\Delta\ra \bo .
$$
Now 
$$\bfT(\bfr)/\Delta\cong (\R^n/{\cal L}(\bfs))\Big/ \((\R\bfs+{\cal L}(\bfs))/{\cal L}(\bfs)\)
\cong  \R^n\Big/ (\R\bfs+{\cal L}(\bfs))$$
Let ${\mf p}_{D^\perp}$ denote the orthogonal projection on $D^\perp.$
\newline  Then $\bfg_j=-{\mf p}_{D^\perp}(\bff_j).$ Thus $\bfT(\bfr)/\Delta\cong
D^\perp\Big/{\cal  G},$ where ${\cal  G}$ is the lattice in  $ D^\perp$ generated by any $n-1$ of
the $\bfg_j.$ (Note $\sum_j \bfg_j = 0).$ Thus the cosets   of $ \Delta$ in $\bfT(\bfr)$ are given
by the points
 of the toroid ${G}^{n-1}\deq D^\perp\Big/{\cal  G}.$
We can now state
\begin{lem}\label{diag8} %LLLLLLLLLLLLLLLLL
For all $ k$ and for all but a lower dimensional set of 
cosets $C\in {G}^{n-1},$ 
$C\cap \ca {\mf p}_k\,$ is  a geodesic segment of length  exactly $2\pi  r_k^2.$
 \end{lem}%ELELELELELELELELELELELELLE
\nin $\qed$

\nin  In our present notation, we have $C_k= \varpi([ \bo ,\bff_k])$ 
and $\b1=\varpi(\bo).$

\begin{lem}\label{diag9} %LLLLLLLLLLLLLLLLL
$\forall k $  $C_k\subset {\mf p}_k.$ 
$\forall k $  $C_k-\varpi(\bo )\subset
\ca {\mf p}_k.$ 
 \end{lem}%ELELELELELELELELELELELELLE
This follows from the above construction of the ${\bf A}_k.$

\nin$\qed$

With notation as above, we can collect our results in the following statement.
\begin{thm}\label{existspart2} %TTTTTTTTTTTTTTTTTTTTTTTT
The above partition ${\mf P}\deq \{{\mf p}_k\;|\;k=1,\dots,n\}$ of 
$\bfT(\bfr)$ satisfies:
\newline {\rm (I)} $\forall k $  $C_k\subset {\mf p}_k.$ 
$\forall k $  $C_k-\varpi(\bo )\subset
\ca {\mf p}_k.$ 
\newline{\rm (II)} For all $ k$ and for all but a lower dimensional set of 
cosets $C\in {G}^{n-1},$ 
$C\cap \ca {\mf p}_k\,$ is  a geodesic segment of length  exactly $ 2\pi r_k^2.$
\newline{\rm (III)} ${\mf p}_k$ is convex.
\newline{\rm (IV)} There exist quasi-disjoint parallelotopes $ A_k={\bf A}_k\subset \R^n$ so
that
\newline\indent{\rm (o)} ${\mf p}_k=\varpi( { A}_k)\,;$ 
\newline\indent{\rm (i)} $[\bo,\bff_k]\subset { A}_k\,;$ 
\newline\indent{\rm (ii)} $\cup_k {A}_k$ is a fundamental domain for 
 $\R^n{\Big/}{\cal L}(\bfs)\,;$ 
\newline\indent{\rm (iii)} For every line L in $\R^n$ which is parallel to $D,$ 
 $L\cap \ca A_k=\emptyset$  or an interval of length $2\pi r_k^2.$
 \end{thm}%ELELELELELELELELELELELELLE
\nin$\qed$

We note that we have arrived at our partition of the toroid by means of a 
 new tiling of the covering space $\R^n$ by the translates of sets 
$\cup^{n}_{k=1} {\bf A}_k$, which is non-convex (unless the box is a cube). 
{In any case,} this  tiling is not face-to-face and
 projects onto $D^\perp$ to yield a tiling of $\R^{n-1}$ by {\it zonotopes}. For 
$n=3$, this is a tiling of $\R^2$ by hexagons as shown in Figures~\ref{A111} and
\ref{A543}. 
For $n=4$, it projects to a tiling of $\R^3$ by
 rhombic dodecahedra.

\subsection{Uniqueness of Natural Partitions of Right Toroids.}

\begin{thm}\label{WeakUniq} %PPPPPPPPPPPPPPPPPPP
The functor $\textsf{P}: {\mathfrak T}\rightsquigarrow\mf P \mf T$, defined by the 
procedure of Section~\ref{constr},
is inverse to the forgetful
functor and so is unique.
 \end{thm}%ELELELELELELELELELELELELLE
{\bf Proof.}
We have to show for each  toroid 
$T=\bfT(\bfr)$  with distinct $r_k$, in a Hilbert space ${\cal H}$, there
is only one  (functorial) way to endow it with a Pythagorean partition. We use the
functoriality  to argue inductively on the dimension $d$ of $T.$ But first, we note that the
partition of
$T$ induces a tiling of  $R^n$ by convex  compact sets, which must then be polytopes.

The rough idea of the proof is to start with the faces of the box, where the induction  
yields the desired induced partition. Then we must extend the unique  determination of 
the polytopic parts into the interior of the box, using some simple connectivity properties.

 Now if $d=1$, we must take $A_1=[0,\bff_1=2\pi\bfe_1]$ and the uniqueness is trivially true.
If $d>1$ then there exists a morphism $\bfS^1(r_j)\hra T .$ Then the partition of $T$ restricted to
$\bfS^1(r_j)$ must be the  partition associated to $\bfS^1(r_j)$, \ie the trivial partition. This means that 
$\bfS^1(r_j)\subset {\mf p}_i,$ for some $i.$ Since the morphism splits we must have $r_i=r_j$
and we can assume $i=j.$            
  Thus $\bfS^1(r_j)=C_j\subset {\mf p}_j.$ We have shown:

{\it 
\hskip1in A natural partition of $\bfT(\bfr)\in {\mathfrak T}$ must preserve its bouquet.
}
\vskip1mm

Let us now take $d=2.$ Then $\ca {\mf p}_j$ is isometric, via $\varpi$ to $\ca A_j\subset \R^2,$
where the  $\ca A_j$ are  convex in $\R^2.$ Moreover, we can assume 
$[\bo,\bff_j]\subset A_j, (\bo,\bff_j)\subset \ca A_j, \varpi(\bo)=\b1.$ We also know every
 diagonal line $L$ (a line parallel to D) in $\R^2$
must intersect $\ca A_j$ either trivially or in a line segment of length $2\pi r_j^2.$
It is this latter possibility which must hold for every diagonal through a $p\in (\bo,\bff_j). $
It follows that $\ca A_j$ is the union of open line segments of length $2\pi r_j^2,$ since the total area
(2-dimensional Lebesgue measure) these comprise is
 $2\pi r_j^2\times \|g_j\|=2\pi r_j^2\times2\pi  r_1 r_2$
and these sum to the  area $4\pi^2 r_1 r_2$ of $T.$  

We claim that $\bo $ is a vertex of the convex polygon $A_j.$ It is a point of  $\partial A_j,$
since it lies in $\bigcap_k A_k.$ So we must eliminate the possibility that $\bo$ 
is an interior point of  an edge $E$ of $A_j.$
To accomplish this, we resort again  to the functoriality of the tiling. Each $T$ possesses an 
isometric involution, namely its geodesic symmetry about $\b1,$ or,
more simply,
 the inverse operation. This isometry preserves the special circles $C_j$ and hence must
preserve the ${\mf p}_j.$ Thus $\wt{-A_j}=\wt{A_j}={\mf p}_j.$ If now
$\bo$ were interior to $ E$, then $\b1=\wt{\bo}$ would be interior
 to  $\varpi\({(-A_j)\cup A_j}\)={\mf p}_j,$
a contradiction.

We know that every diagonal $L$ which intersects $A_j$ must intersect 
 it in a closed interval of length $2\pi r_j^2.$ For the main diagonal $\R D,$ this segment is the edge $E$
which has $\bo$ as a vertex. There are thus two possibilities: $E=r_j^2 D$ or $E=-r_j^2 D.$ The
latter possibility can be eliminated using a variant of the ``inversion argument'' we used
before. Specifically, if $E=-r_j^2 D,$
 then the angle between $E$ and $ \bff_j$
would be obtuse. It follows that $-A_j$ would intersect $(\bo,\bff_{j'})$ non-trivially for
$j'\ne j,$ a contradiction.

We now know that $A_j$ contains the triangle  ${\mf t}_j$ with vertices 
$ \bo, r_j^2 \bfs, \bff_j$
Similar arguments show that $\bff_j$ must also be a vertex of $A_j$ and that
the triangle ${\mf t}'_j$  with vertices  $  \bff_j, \bff_j-r_j^2 \bfs,\bo$ is contained in $A_j.$
Since the total area of  ${\mf t}_j\cup  {\mf t}'_j$ is $4\pi^2 r_1 r_2 r_j^2,$ the area of $A_j$, 
therefore $ A_j={\mf t}_j\cup  {\mf t}'_j $ and we have shown
that $A_j$ is the one previously constructed, and so, unique.

We take $n>2$ and inductively assume we have proven the uniqueness of our tilings
for  all  toroids of dimension less than $n$. 
For any $n-$dimensional {right} toroid  $T=T^n$,
we set
 $T_{-k}:= T^{n-1}_{-k}:= $ the $(n-1)-$dimensional toroid 
generated by the $C_j\,,j\ne k.$
Suppose$\{{\mf p}_k\;|\;k=1,\dots,n\}$ is a Pythagorean
tiling of $T.$ Then, by functoriality,
${\mf p}_{-j}:={\mf p}^{n-1}_{-j}:= \{{\mf p}_k\cap T_{-j}\;|\;k=1,\cdots,n\,, k\ne
j\}$ is a tiling of
$T_{-j}.$
 We tentatively drop the assumption that $r_1>\dots> r_n$ while still requiring that the $r_k$ be
positive distinct reals. We do this so that we may consider that $T_{-n}$ is an arbitrary $T_{-j},$
for the sake of simplifying notation.

\nin{\bf The three remaining steps of the proof.}

We know from the existence proof, that
$$
\forall k\le n\; \;
{\bf A}_k=\ca{\bigcup}_{\sigma\in\Sigma_n} {\scriptstyle Sl}^\dagger(k,\sigma).
$$
 The rest of the proof consists of three steps, showing:
\nl 1) the ${\scriptstyle Sl}^\dagger(k,\sigma)$ are
the (convex) hulls of their intersections with $\partial B$ and $D;$
\nl 2)  $A_k\cap\partial B'=\bfA_k\cap\partial B',$  for any lattice translate $ B'$ of $ B;$
\nl 3)  {$A_k\cap D'=\bfA_k\cap D'$,  for any lattice translate $ D'$ of $ D.$}

Since the $A_k$ are convex the theorem will then follow.
 These three steps will be established in the following lemmas. 
 In accordance with the notation introduced in Section~\ref{AppPart}, $S_\b1$ is the
 simplex corresponding to the identity
permutation $\b1.$ 
 We sometimes abbreviate:
$$R_{k \sigma}:=\(\sum_{j\le \sigma^{-1}(k)}r^2_{\sigma(j)}\)\bfs,\,
\;R^-_{k \sigma}:=\(\sum_{j< \sigma^{-1}(k)}r^2_{\sigma(j)}\)\bfs,
\;\bfv_k:=\bfv_{k\b1},
\;D_{k{\sigma}}=[R^-_{k{\sigma}},R_{k{\sigma}}].$$
\begin{lem}\label{geom22} %LLLLLLLLLLLLLLLLLLLL
\begin{equation}%еееееееееееееееееееееееееееееееее
{\scriptstyle Sl}(k,{\bf 1})= \Big[\partial B\cap {\scriptstyle Sl}(k,{\bf 1})
,D_{k{\b1}}\Big]
\end{equation}%ееееееееееееееееее      еееееееееееееее
 \end{lem}%ELELELELELELELELELELELELLE
{\bf Proof.} ${\scriptstyle Sl}(k,{\bf 1})$ is convex since it is the intersection of a convex slab
and a simplex. Thus, it is the convex closure of its vertices which are the intersections of the edges
of $S_{\bf 1}$ with the hyperplanes $ D^\perp +\bfv_{k{\bf 1}}$ and $D^\perp +\bfv_{k{\bf
1}}+\bff_{k} .$ Now all the edges of $S_{\bf 1}$ are contained in $\partial B$ except for $D.$ To see
this, note that 
$j<k\Ra [\bfv_j,\bfv_k]\subset [\bfv_j,\bfv_{j+1}\d \bfv_k]$ which is a $(k-j)-$dimensional
face of $S_\b1$ and so contained in $\partial B$ unless $j=0$ and $k=n.$ The result follows since
$(D^\perp +{\bfv}_{k{\bf 1}}) \cap D=R^-_{k{\bf 1}}$
and  $(D^\perp +{\bfv}_{k{\bf 1}}+{\bff}_{k} )\cap D= R_{k{\bf 1}}.$
\newline $ \qed$

\begin{lem}\label{geom3} %LLLLLLLLLLLLLLLLLLLL
\begin{equation}%еееееееееееееееееееееееееееееееее
{\scriptstyle Sl}(k,\sigma)= \Big[\partial B\cap {\scriptstyle Sl}(k,{\sigma})
,D_{k{\sigma}}\Big]
\end{equation}%ееееееееееееееееее      еееееееееееееее
 \end{lem}%ELELELELELELELELELELELELLE
{\bf Proof.}  We have $R_{k \sigma}=\(\sum_{j< \sigma^{-1}(k)}r^2_{\sigma(j)}\)\bfs$
and  $R^-_{k \sigma}=
\(\sum_{j\le \sigma^{-1}(k)}r^2_{\sigma(j)}\)\bfs.$  So the result follows by permuting
the  $\bff_k$ with $\sigma.$
\newline $\qed$

This completes  the first step.

\begin{lem}\label{AllBoxIntersection} %LLLLLLLLLLLLLLLLLLLL
For any Pythagorean partition $\{{\mf p}_1\d  {\mf p}_n\}$ with corresponding $\{A_1\d  A_n\}$
and for any lattice translate $B'$ of $B$
\begin{equation}%еееееееееееееееееееееееееееееееее
\forall k\; { A}_k\cap \partial B'= {\bf  A}_k\cap \partial B'.
\end{equation}%ееееееееееееееееее      еееееееееееееее
 \end{lem}%ELELELELELELELELELELELELLE
{\bf Proof.} Let $E$ be a facet of $B'=B+\bfv,$ where $\bfv$ is a lattice vector.
$$
\wt{{ A}_k\cap E }\subset \wt{{ A}_k}\cap \wt{ E}= {\mf p}_k\cap T_{-j}
$$
for some $j$, namely that $j$ such that $\bff_j\perp E.$ 
By renumbering, we can take $j=n.$ 
The ${ A}_k\cap E$ are convex. $\cup_k ( {\mf p}_k\cap T_{-n}) = T_{-n}.$  
We can apply the functor F to  
$T_{-n}\hookrightarrow T $ yielding  $F(T_{-n})\hookrightarrow F(T). $
This implies $ {\mf p}^{n-1}_i={\mf p}^n_i\cap T_{-n}$ provided $i<n.$ Inductively,
$ {\mf p}^{n-1}_i=\wt{ {\bf A}^{n-1}_i}.$ Thus ${ A}_i\cap E$ and ${ \bfA}_i\cap E$
are both lifts of ${\mf p}^{n-1}_i.$ The interior of  ${ \bfA}_i\cap E$, which equals
${\ca \bfA}_i\cap E,$ is a lift of an open subset $U\subset {\mf p}^{n-1}_i$ and 
$U$ is isometric to the convex set ${\ca \bfA}_i\cap E.$
Thus, there is an open subset $\breve U\subset { A}_i\cap E$ which is also a lift of the 
simply connected set $U.$ Both $\breve U$ and ${\ca \bfA}_i\cap E$ contain  $(\bo,\bff_i].$
Since lifts, with a common starting point (say ${1\over2}\bff_i$), of simply 
 connected sets \wrt the
covering map
$\varpi$ are unique, we get  $\breve U={\ca \bfA}_i\cap E.$ Taking closures, we get
${A}_i\cap E={\bfA}_i\cap E.$ The lemma follows by applying this argument to an arbitrary facet
of $B'$
\newline$\qed$

This completes the second step. The third step will be more involved as we have
to investigate the properties of the $A_k$ in the interior of the $B '.$ 

\begin{prop}\label{geom5} %LLLLLLLLLLLLLLLLLLLL
$\partial B\cap {\scriptstyle Sl}(k,{\bf 1})$ 
has as vertices ${\cal V}_k \cup{\cal V}_{k+1},${ where the ${\cal V}_k$ are defined below.}
\end{prop}%ELELELELELELELELELELELELLE
{\bf Proof.}
$ S_{\bf 1}\cap \partial B$  is a union of some of the facets of $ S_{\bf 1}$,
its {\it external} facets. Every facet of $ S_{\bf 1}$ is  obtained by taking a subset
 $T\subset \{\bo=\bfv_1\d \bfv_{n+1}=\bfs\}$ of size $ n$ and forming
$[T];$ the external facets being those which do not contain $[\bo,\bfs],$
 \ie  $T= \{{\bo}={\bfv}_1\d {\bfv}_{n}\}$ or $T= \{\bfv_2\d \bfv_{n+1}=\bfs\}.$
\begin{equation}%еееееееееееееееееееееееееееееееее
\tf S_{\bf 1}=
\big[\bo=\bfv_1\d \bfv_{n+1}=\bfs\big].
\end{equation}%ееееееееееееееееее      еееееееееееееее
\begin{equation}\label{S1intbdyBox}%еееееееееееееееееееееееееееееееее
S_{\bf 1}\cap \partial B
=
\big[\bo=\bfv_1\d \bfv_{n}\big]\cup\big[\bfv_2\d \bfv_{n+1}=\bfs\big] .
\end{equation}%ееееееееееееееееее      еееееееееееееее
 As above, ${\scriptstyle Sl}(k,{\bf 1})=
{\scriptstyle Slab}({\bfv}_{k{\bf 1}},{\bfv}_{k{\bf 1}}+{\bff}_{k},S_{\bf 1}).$

\nin{\bf Note:} 
$$
(\forall U\subset\R^n) \,
(\forall \bfu_1,\bfu_2\in\R^n)\;
 [D^\perp+\bfu_1,D^\perp+\bfu_2]\cap U=
\{ \bfu\in U \;|\; \bfs\cdot\bfu_1\le \bfs\cdot\bfu\le \bfs\cdot\bfu_2 \}.
$$
$$
S_{\bf1}\cap \( D^\perp +\bfv_{k\bf1}\)= \Big\{\bfu\in S_{\bf1} \;{\Big| } \;\bfu.\bfs=2\pi\sum_{j<
k}r_j^2   \Big\}.
$$
Also
$$
\partial B\cap S_{\bf1}\cap \( D^\perp +\bfv_{k\bf1}\)
=\bigcup'_{i<h} [\bfv_{i\bf1},\bfv_{h\bf1}]\cap  \( D^\perp +\bfv_{k\bf1}\),
$$
where the prime over the union indicates that the inner edge $[\bo,\bfs]$ is not included.
$$
\tf{\cal V}_k \deq \partial B\cap S_{\bf1}\cap \( D^\perp +\bfv_{k\bf1}\)
=
\{ a_{ihk}\bfv_{i\bf1}+(1-a_{ihk})\bfv_{h\bf1}\;|\; i\le k\le h,  (i,h)\ne (1,n+1)\},
$$
where  $ a= a_{ihk}$ satisfies $a\sum_{j<i} r_j^2 +(1-a)\sum_{j<h} r_j^2=\sum_{j<k} r_j^2.  $ 
$$
\tf i<h\Ra a= {\sum_{k\le j<h} r_j^2\over \sum_{i\le j<h} r_j^2}.
$$

${\cal V}_k $ are vertices of $ \partial B\cap{\scriptstyle Sl}^\dagger(k,{\bf 1}) $
 which are on the face  of ${\scriptstyle Sl}^\dagger(k,{\bf 1}) $
  perpendicular  to $\bfs$ closest to
$\bo$, \ie the face contained in $D^\perp +\bfv_{k-1}$. The others are ${\cal V}_{k+1}. $
\begin{prop}\label{NoCrisscross} %PPPPPPPPPPPPPPPPPPP
Let $k<m\le n, 0<b\le c<\sum_{i\le n}r_i^2.$ Then
$[{\cal V}_m,b\bfs ]\cap [{\cal V}_{k},c\bfs ]\ne \emptyset.$
 \end{prop}%ELELELELELELELELELELELELLE
{\bf Proof.} We can assume $b<c$
 The hyperplane   $D^\perp + \bfv_{m\bf1}$ disconnects the simplex $S_\b1:$
$$
S_\b1 -(D^\perp + \bfv_{m\bf1})=
 \left\{\bfu\in S_\b1 \;|\; \bfu\cdot \bfs <2\pi\sum_{i\le m}r_i^2    \right\}
{{\buildrel \rm disjoint\over\bigcup}} 
\left \{\bfu\in S_\b1 \;|\; \bfu\cdot \bfs >2\pi\sum_{i\le m}r_i^2    \right\}.
$$
Now $S_\b1 \cap  (D^\perp + \bfv_{m\bf1})= [{\cal V}_m,a\bfs ] , a= \sum_{i\le m}r_i^2 .$
There exists  (\cf below) 
 a piecewise linear homeomorphism $\phi:S_\b1\ra S_\b1$  fixing $S_\b1\cap \partial B$  
such that $\phi(b\bfs)=a\bfs$ and such that   $\phi([{\cal V}_m,a\bfs ] ) = [{\cal V}_m,b\bfs ]$
 Thus $[{\cal V}_m,b\bfs ]$ also disconnects $S_\b1$ and ${\cal V}_{k}$ lies in one component
(the connected component of $\bo$) and $c\bfs$ lies in the other (the connected component of
$\bfs$). The result follows.
\newline{\bf The construction of the map $\phi$.} 
 By \eqref{S1intbdyBox}  ${\cal V}_k \subset \partial B \cap S_\b1=
\big[\bo=\bfv_1\d \bfv_{n}\big]\cup\big[\bfv_2\d \bfv_{n+1}=\bfs\big].$ 
\nl Set  ${\cal V}^-_k =  {\cal V}_k \cap \big[\bo=\bfv_1\d \bfv_{n}\big].$ Then $[{\cal V}^-_k] $
 is a $g-$simplex, $g\le n-2,$ with vertices ${\cal V}^-_k.$
 Set  ${\cal V}^+_k =  {\cal V}_k \cap \big[\bfv_2\d \bfv_{n+1}=\bfs\big].$ $[{\cal V}^+_k] $
 is a  $g'-$simplex, $g'\le n-2,$ with vertices ${\cal V}^+_k .$ If $ 1<k\le n,$ \eg when
$ k=m$,  then $\max\{ g,g'=n-2\}.$
 Then $(D^\perp + \bfv_k)\cap \partial B=[{\cal V}^-_k]\cup [{\cal V}^+_k],$ the union of
two simplices. Thus $[{\cal V}_m,a\bfs ]  =[{\cal V}^-_m,a\bfs ]\cup[{\cal V}^+_m,a\bfs ]$
is the union of two  simplices. Likewise,
$[{\cal V}_m,b\bfs ]  =[{\cal V}^-_m,b\bfs ]\cup[{\cal V}^+_m,b\bfs ].$ 
 Hence we can define $\phi$ piecewise by requiring $\phi|{\cal V}^+_m= I_{{\cal V}^+_m},$
$\phi(\bo)=\bo, \phi(a\bfs)=b\bfs.$ By linearity, $\phi$ extends uniquely to the  $n-$simplex
$ [\bo,a\bfs,{{\cal V}^+_m}],$  mapping it homeomorphically to $ [\bo,b\bfs,{{\cal V}^+_m}],$
 Similarly, $\phi$ extends uniquely to the  $n-$simplex
$ [\bo,a\bfs,{{\cal V}^-_m}],$ mapping it homeomorphically to $ [\bo,b\bfs,{{\cal V}^-_m}],$
  Moreover, these extensions agree on the
intersections of their domains, since they agree on the vertices ${{\cal V}^-_m}\cap{{\cal V}^+_m},$ 
namely the two extensions are the identity on this set. We complete the definition of 
$\phi:S_\b1\ra S_\b1$ 
by similarly defining  $\phi| [\bfs,a\bfs,{{\cal V}^-_m}]\ra  [\bfs,b\bfs,{{\cal V}^-_m}]$
and $\phi| [\bfs,a\bfs,{{\cal V}^+_m}]\ra  [\bfs,b\bfs,{{\cal V}^+_m}].$

\nin $\qed$

\begin{lem} %LLLLLLLLLLLLLLLLLLLL
$\forall k\; \;{ A}_k\cap D =[\bo,r_k^2\bfs] .$ 
 \end{lem}%ELELELELELELELELELELELELLE
{\bf Proof.}  $\bo\in { A}_k\cap D\Ra   { A}_k\cap D $ is an interval of length $2\pi r_k^2.$ 
It therefore suffices to prove that $t\bfs\in A_k\Ra t\ge 0. $ Assume $t<0$,
Then $A_k\supset [\bo, t\bfs, \bff_k].$ We can take $k=n.$ Then $[\bo,\bfg_1]\subset \bfA_n.$

\nin$\qed$

\begin{lem} %LLLLLLLLLLLLLLLLLLLL
$\(\forall k\)\;\epsilon\in\R\,,\;\epsilon\bfs \in{ A}_k\Ra\epsilon\ge 0. $ 
 \end{lem}%ELELELELELELELELELELELELLE
{\bf Proof.}  We can take $k=n.$ We know inductively that
 $\delta(\bfs-\bff_1)\in A_1$ where 
$$ \delta={r_n^2\over\sum_{\mu=2}^{n} r_\mu^2}>0.$$
So $\epsilon\,\bfs \in{ A}_k,t\in[0,1]\Ra \Xi\deq t\epsilon\,\bfs
+(1-t)\delta(\bfs-\bff_1)
\in{A}_k.$ If $\epsilon<0$ we can take $t={\delta\over \delta-\epsilon},$
then $A_k\ni \Xi ={\delta\over \delta-\epsilon}\epsilon \bff_1
\deq \xi \bff_1.$ 
Since $\xi\in(0,1)$ this
contradicts the interior disjointness of $A_1, A_n.$
  \newline$\qed$

\begin{lem} %LLLLLLLLLLLLLLLLLLLL
$\forall k\; { A}_k \cap\R\bfs= [\bo,r_k^2 \bfs]$
 \end{lem}%ELELELELELELELELELELELELLE
{\bf Proof.} We know $\bo\in A_k,$ so the result follows from
the preceding lemma.
\newline$\qed$

\begin{lem} %LLLLLLLLLLLLLLLLLLLL
$\forall k\; { A}_k \cap B= { \bfA}_k \cap B$
 \end{lem}%ELELELELELELELELELELELELLE
{\bf Proof.} We know from Lemma \ref{geom3},
${\scriptstyle Sl}(k,{\bf 1})= \Big[\partial B\cap {\scriptstyle Sl}(k,{\bf 1})
,[R^-_{k{\bf 1}},R_{k{\bf 1}}]\Big]$ In particular, 
${\scriptstyle Sl}(1,{\bf 1})= \Big[\partial B\cap {\scriptstyle Sl}(1,{\bf 1})
,[R^-_{1{\bf 1}},R_{1{\bf 1}}]\Big]= \Big[\partial B\cap {\scriptstyle Sl}(1,{\bf 1})
,[\bo,r_1^2\bfs]\Big].$ Thus $A_1\cap B \supset {\scriptstyle Sl}(1,{\bf 1}).$
Similarly, for any permutation $\sigma$  such that $\sigma(1)=1$ we get 
 $A_1\cap B \supset {\scriptstyle Sl}(1,{\sigma}).$ Thus 
 $A_1\cap B \supset \bfA_1\cap B .$ We can similarly show $A_k\cap B \supset \bfA_k\cap B .$
It now follows that none of these inclusions can be proper:
 $\forall k\; { A}_k \cap B= { \bfA}_k \cap B.$
\newline$\qed$

\begin{lem} %LLLLLLLLLLLLLLLLLLLL
$\forall j,k\;  { A}_k \cap {\scriptstyle Sl}^\dagger(j,{\b1}) = { \bfA}_k \cap {\scriptstyle
Sl}^\dagger(j,{\b1}).$
 \end{lem}%ELELELELELELELELELELELELLE
{\bf Proof.} The translates $A'_k$  of $A_k$ exhaust $S_\b1$ and hence the 
$ {\scriptstyle Sl}^\dagger(j,{\b1}) .$ By Lemma~\ref{AllBoxIntersection},
we know  $({ A}_k+\bfv) \cap {\scriptstyle Sl}(j,{\b1})\cap \partial B =
( { \bfA}_k+\bfv)\cap {\scriptstyle Sl}(j,{\b1})\cap \partial B$ for any lattice vector $\bfv.$
Therefore, by Lemma~\ref{geom3},
 it suffices to show $({ A}_k+\bfv) \cap D=({ \bfA}_k+\bfv) \cap D.$ For this, it suffices to take
$\bfv= \bfv_{j\b1}$ for some $j=1\d n,$ since $ {\scriptstyle Sl}(j,{\b1})= {\scriptstyle
Sl}^\dagger(j,{\b1})+\bfv_{j\b1}.$ ((For $j=1,$ this follows from the previous lemma.))
If $({ A}_k+\bfv_{j\b1}) \cap D\ne\emptyset,$ then we know it is a subinterval of length
$2\pi r_k^2.$ Since these intersections must exhaust $D,$ it must be that
  $({ A}_k+\bfv_{j\b1}) \cap D= D_{k \sigma}$ for some permutation $\sigma.$
By Proposition~\ref{NoCrisscross}, the disjointness of the (interiors of the) 
 ${ A}_k+\bfv_{j\b1}$ implies $\sigma=\b1.$ Thus 
 $({ A}_k+\bfv_{j\b1}) \cap D= D_{k \b1}= ({ \bfA}_k+\bfv_{j\b1}) \cap D.$
The result now follows.

\nin$\qed$  
\begin{lem} %LLLLLLLLLLLLLLLLLLLL
$\forall j,k,\sigma\;  { A}_k \cap {\scriptstyle Sl}^\dagger(j,{\sigma}) 
= { \bfA}_k \cap {\scriptstyle
Sl}^\dagger(j,{\sigma}).$
 \end{lem}%ELELELELELELELELELELELELLE
{\bf Proof.} We have only to reorder the $\bff_j$ in the previous lemma.

\nin$\qed$  
\begin{thm} %LLLLLLLLLLLLLLLLLLLL
For all $k$  we have  ${ A}_k = { \bfA}_k .$
 \end{thm}%ELELELELELELELELELELELELLE
{\bf Proof.} This follows from ${ \bfA}_k=
 \bigcup_{\sigma} {\scriptstyle Sl}^\dagger(k,{\sigma})  .$

\nin$\qed$ 

We have thus established the uniqueness of the functor \textsf{P};  there is only
one natural way to  partition right toroids satisfying the diagonal property. 

  \section{Perspective States}\label{relstates}
\begin{quotation}
\dots each quality or 
property of a thing is, in reality, nothing else but its capability of exercising 
certain effects upon other things\dots, it can never depend upon the nature of one
agent alone, but exists only in relation to, and dependent on, the nature of some
second object, which is acted upon. \par
     \rightline{\emph{Helmholtz} } 
\end{quotation}

The root change we are making in going from SQM to IQM
 is in the concept of {\it state}. Classical mechanics and SQM share the concept of state
as adhering to a system {\it simpliciter}, without reference to other systems.
 In our view, the difficulties in the application of QM to individuals mandate relativizing 
the notion of state. In fact we deny the existence of 
an absolute state of an individual system.
 In order to be precise in a confusing area,  we need to make some formal definitions.

\nin{\bf Definition.} The  {\bf polar states} of $\ss=\ss_1+\ss_2$ are the elements 
 of the polar bundle ${\po}$.

An element of ${\po}$ can be written 
$\bfp:= (\gg, {\bfq})\in \bfS\times \bfT(\bfr,\gg).$ Here, we take
$\bfq$ from a  polar decomposition
$$
\gg=\sum_l\lam_k\,\gg_k
=\sum_l\lam_k\,\phi_k\!\ot\!\psi_k.
$$
The notation $\bfp= (\gg, {\bfq})$ is slightly redundant since the absolute values 
$r_k=|q_k|$ are already determined by $\gg.$ We could replace $\bfq$ by 
 $\vec\theta$ or $\vec\zeta$ 
 where $q_k= e^{i \theta_k}r_k=\zeta_k
r_k.$ However, the  
 notation $ (\gg, {\bfq})$ is
 more direct and seems to cause no problem.

If   a  regular $\gg$  is given, then $\bfq$ and  $\(\gg_k\)_k$ determine one
another. Another name we sometimes use for the polar state is {\bf joint state} to emphasize 
the analogy of the wave function with  a  probability  density.

The $(\gg,\bfq)$    parameterize the new 
 phase space of the composite system. They  give
extra phase angle data compared with SQM,   the arguments $\theta_k$ of the
$\lam_k.$ These phases, combined with the Pythagorean partition of the toroidal fibers, 
give a classical 
way of specifying a particular SQM state $[\phi_k]$ of $\ss_1,$
 the conditional state of $\ss_1.$ 
Namely, the map $P_{\phi_k} \mapsto {\mf p_k}\in{\mf P}$
extends to a faithful representation of the Boolean algebra generated by the projections 
$P_{\phi_k}$ onto a field of subsets of the toroidal fiber.
This comprises 
a bridge between quantum  and classical logic, but it is contextually
 restricted to the Boolean
algebra of subsets of $\{[\phi_k]\;|\;k=1,2,\cdots\}.$
The same applies to $\ss_2$  and  $\{[\psi_k]\;|\;k=1,2,\cdots\}.$

To get a dynamical description, we need the further specification of
a (possibly time-dependent) \ham\  $H$ for \ss. 
Once $(\gg,\bfq)$ is given at time $t$, it is determined for all $t$
 by the Hamiltonian evolution on $\bfS$ induced by $H$  and the dynamical connection $A^H.$

This completes the  description of the polar state of an interacting pair of systems.

\nin{\bf Definition.} The {\bf {conditional} spectral state of $\ss_1$} \wrt the
polar state $(\gg, \bfq,H)$ of \ss\ is $[\phi_{k(t)}].$ Symbolically,  
we  write this as
$[\phi_{k(t)}]=[ (\gg,\bfq)|\ss_1]$, defining the value of the function $t\mapsto k(t)$ as the index $k$ for
which 
$\lam(t) \in  {\mf p}_k(t).$

 A crucial element of the extension of SQM we are describing here consists of 
 the existence and
determination of the function $t\ra k(t).$ 
 The ray $[\phi_{k(t)}]$ can be identified with a {\it spectral}
projection
$P_{\C\phi_{k(t)}}$ of
the mixed state  ${\rm Tred}_1(P_{\C\gg})$ assigned  by SQM to $\ss_1.$
Now  ${\rm Tred}_1$, which is an orthogonal projection of the 
(Jordan) algebra of observables of \ss\ to that of $\ss_1$, 
 is the
non-commutative analogue of  the
projection of algebras of random variables used in the theory of conditional probabilities. 
This non-commutative analogue can also be expressed in the terminology of {\it partial} 
Boolean algebras, as in \cite[(iv)]{Kochenpar}.
 The conditional spectral state $[(\gg,\bfq)|\ss_1]$ is 
 the IQM  {extension} of this notion. 

We now make some technical remarks concerning the  considerations required to 
handle the lower dimensional  situations where the polar decomposition 
$$
\gg=\sum_k q_k\,\phi_k\ot\psi_k
$$
has either the special property 1)  the $|q_k|$ are not distinct, or    
 2) $\bfq $ lies on the boundary of two ${\mf p}_k.$

In either case, the function $k(t)$ becomes undefined; in fact, in the first case  the $[\phi_k]$ 
are not well-defined.

For 1) we need to use the dynamical behavior and assume  that $\gg(t)$ is an analytic function 
of $t,$ \ie $\gg(0)$ is an analytic vector for the \ham. From perturbation theory 
\cite[Chap.II,Th.6.1]{Kato} it follows that the 
spectral states (\ie the spectral projections of the reduced density operator $\rho_1$) can be
analytically continued across any {\it isolated degeneracy} where the eigenvalues of $\rho_1$ are not 
all distinct. This restores the definition of $k(t).$ It is possible to encounter a {\it permanent degeneracy}
as in the case of  identical particles mentioned in Section~\ref{appident}. We do not elaborate here on 
 the  special considerations required for this case.

In  2) we have the  ambiguity of $k(t)$ when $\bfq$ lies on a lower dimensional 
boundary. We can 
invoke the analogy with classical  physical theories which sometimes regard the  functions and subsets
of phase spaces to be  just representative of the corresponding $\sigma$-Boolean algebra entities
formed by factoring  out sets of measure zero,  including the lower dimensional boundaries of the
${\mf p}_k$

\subsection{ Special case: $\ss_2$ is empty= SQM.}
The ordinary QM of \ss\ is the special case when either 
$\ss_1$ or $\ss_2$ is empty.
 Taking $\ss=\ss_1$,  we can take ${\cal
H}_1={\cal H}$ and  ${\cal H}_2=\C.$ We assume all the \ham s are time-independent.
Then the polar state of \ss\ is just $ (\gg, \bfq)$ with $\bfq\in\C$
and
the conditional spectral  state  of $\ss_1$ is  $[ (\gg, \bfq){|}\ss_1]=[\gg]$.
  This is ordinary quantum theory, \ie
the  quantum mechanics of an isolated system. In this case, we can speak about the 
state of $\ss_1=\ss,$ as is customary.

\subsection{ Compounding of perspectives.}\label{compound}

If we want to consider interactions of subsystems 
of $\ss_1$ or more general multiple interactions, then
we iteratively need more data at each new level, specifying the vector states
$\phi_k$, not just the  $[\phi_k].$ 
 A  grandiose example would be the situation, where 
$\ss,\ss_{1}$, and $\ss_{2}$ are respectively, the entire universe, the subsystem of bosons
and the subsystem of fermions.
Then one might want a representation of  some subsystem of $\ss_1$, \eg the
microwave background. 
The details are left to the reader.

An  iteratively complete  description of the state of $\ss=\ss_1+\ss_2$  
requires expanding the data above.
It is also necessary to specify \ham s $H_1$,  and $H_2 $ for $\ss_{1}$, and $\ss_{2}$ 
respectively.
These will usually be time-dependent. 
 Then  $H=H_0+H_1\ot \bfI_2+\bfI_1\ot H_2 $ 
is the total \ham\ of \ss, where the interaction \ham\ $H_0$ is thereby defined.

It is now also necessary to be given {\bf conditional vector states}, \ie 
particular $\phi_{k(t)}(t)\in [(\gg,\bfq){|}\ss_1]$ or, equivalently in the presence
 of $\gg_k(t)$, 
$\psi_{k(t)}(t)\in [(\gg,\bfq){|}\ss_2].$ The simplest way
 these can be determined is via
the
\ham s ${H}_i$ and the connections $A^{{H}_i},$ starting with initial values 
$\phi_{k}(0)$ for all $k.$  Now we have the necessary data to treat a 
decomposition $\ss_1=\ss_{11}+\ss_{12}$. This process can then be iterated, but always
 requiring additional information at each new stage. 

Simple behavior 
for conditional states is ruled out by the fact that a spectral projection of the reduced trace  
of a spectral projection of a density operator $\rho$ is not in general a 
spectral projection of the reduced trace of $\rho.$ We will elaborate on this
 cruel  fact  of life
 when we deal with  EPR in Section~\ref{epr}. {It seems that the closer one gets to the truth, 
the more relational is the required formulation.}

  \subsection{Evolution of conditional vectors states. }\label{condvecs}

The collective evolution of the $[\phi_k]$ is equivalent to the evolution of the density matrix
${\rm Tred}_1(\gg)$; it is a standard part of SQM. The particular one
 of these which obtains at a given time $t$, $[\phi_{k(t)}]$
 is given by the conditional spectral projective state $[(\gg,\vec q)|\ss_1].$ It is completely
 determined 
by the evolution of this joint state. However, if it is desired to iterate this procedure it is
then necessary to obtain a conditional spectral vector state, \ie a lift of 
$[(\gg,\vec q)|\ss_1]$ to $\bfS({\cal H}_1).$ 
 Another reason to study this situation is to facilitate
the proof that no interaction implies no jumping (\ie no change in the
 spectral states), without which IQM measurement
theory   would be of questionable meaning.
In this section we investigate the possibility of determining the
 evolution of conditional vector states.
 First
we show that essentially this requires being given separate or ``free'' \ham s $H_i$ for each
subsystem, together
 with an interaction \ham\ $H_0.$

We start with the usual composite system $\ss=\ss_1+\ss_2$ together with a \ham\ 
$H.$
% As we have shown in Appendix~\ref{dEigenVec}, using $\Im$ to denote the imaginary part, 
Let $t\mapsto\bfD_1(t)=:\bfD_1$ be the  curve of density operators on ${\cal H}_1$   given by
  ${\rm Tred}_1(P_{\gg(t)}).$  %{eq:deigen20}
Using $\Im$ to denote the imaginary part, we have by \eqref{eq:deigen20}
  \begin{equation}\label{deigen20}%еееееееееееееее
  \dot\phi_j= \sum_{k\ne j } {\<\phi_{k},\dot \bfD_1\phi_{j}\>\over x_j -x_{k}} \phi_{k}
  +\<\phi_j ,\dot\phi_j\>\phi_j= \sum_{k\ne j } {\<\phi_{k},\dot \bfD_1\phi_{j}\>\over x_j
  -x_{k}}
  \phi_{k} +i{\Im}\<\phi_j ,\dot\phi_j\>\phi_j.
  \end{equation}%еееееееееееееее
  We can write this as
  \begin{equation}\label{2deigen20}%еееееееееееееее
  \dot\phi_j= \sum_{k\ne j } {\beta_{kj}\over x_j -x_{k}} \phi_{k}
  +i{\Im}\<\phi_j ,\dot\phi_j\>\phi_j,
  \end{equation}%еееееееееееееее
  where the $ \beta_{kj}$ have been defined in \eqref{defbeta{ab}}.
  
  If we take the  $\phi_j$ and $\psi_j$ to be canonically horizontal over $\bfP({\cal H}_1)$ 
and  $\bfP({\cal H}_2)$, 
then we must have the 
  $\<\phi_j ,\dot\phi_j\>=\<\psi_j ,\dot\psi_j\>=0.$ This choice would lead to the
$\gg_j=\phi_j\ot\psi_j$
  being  canonically horizontal over $\bfP$. But we have required that the $\gg_j$ be
  $A^H$-horizontal. The most general way to satisfy this condition, while lifting the
  $\phi_k$ and the $\psi_k $ by connections on $\bfP({\cal H}_1)$ and
 $\bfP({\cal H}_2),$ is to require
   \begin{equation}\label{akeq2}%еееееееееееееееееееееееееее
{\dot\phi_a} = \sum_{k\ne a }
   {\beta_{ka}\over |\lam_a|^2-|\lam_k|^2}{\phi_k}+if_a\phi_a
\end{equation}%еееееееееееееееееееееееееее
$$%%еееееееееееееее
 {\dot\psi_a} = \sum_{k\ne a }
   {\beta'_{ka}\over  |\lam_a|^2-|\lam_k|^2}{\psi_k }+ig_a \psi_a
$$%еееееееееееееееееееееееееее
  where $f_j$ and $g_j$ are a real-valued smooth functions of $t,$ depending on the
  $\phi_j,$ and where
  $$
  \<\gg_j,H\gg_j\>= {\cal E}({\gg_j(t)}) = f_j(t)+g_j(t).
  $$ 
  To do this plausibly would require partitioning the expected energy 
   $ {\cal E}({\gg_j(t)})$ of $ {\gg_j(t)}$ between $\phi_j(t)$
  and  $\psi_j(t).$ The moral here is that $f_j$  and  $g_j$ cannot be
{\it separately} determined by any
    data we have so far considered, at least without some new principle.
That is why, when we need the separate evolutions of the 
${\phi_k}$ and ${\psi_k}$, we specify individual \ham s $H_i$ on ${\cal H}_i, $ 
 for $i=1,2.$

\subsection{The case of non-interacting subsystems.}\label{noint}

The important special case where $H_0=0$ is now treated. 
As we have said, we prove there is no jumping in this case. 
  
  Here we assume that the states of $\textsf{S}_1$ and $\textsf{S}_2$ evolve
  {\it separately}. 
  This means that the total Hamiltonian $H$  for ${\cal H}={\cal H}_1\ot{\cal H}_2$ is
  of the form
  $$
  H=H_1\ot\bfI_2+ \bfI_1\ot H_2.
  $$
  Then it is natural to lift the $[\phi_k]$ by the dynamical  connection $A^{H_1}$ defined in a 
  fashion analogous to \eqref{8CanonicalConnS}
   \begin{equation}\label{CanonicalConn1^S}%еееееееееееееееееееееееееееееееее
  A^{H_1}\deq\<\bfz,d\bfz\>+ i\<\bfz,H_1\bfz\> dt=: A^0+i{\cal E}^{H_1}(\bfz)dt,
     \end{equation}%ееееееееееееееееее      еееееееееееееее
and similarly for $[\psi_k].$
  
  \nin We now show that $\gg_k(t)\deq\phi^{H_1}_k(t)\ot\psi^{H_2}_k(t)$ is $A^H-$horizontal.
  \begin{equation}\label{42CanonicalConn1^S}%еееееееееееееееееееееееееееееееее
  \hskip-3.5in A^H(\dot\gg_k,\partial_t)=\<\gg_k,\dot\gg_k\>+ i\<\gg_k,H\gg_k\> 
  =
   \end{equation}%ееееееееееееееееее      еееееееееееееее
  $$
 \<\phi^{H_1}_k\ot\psi^{H_2}_k,\dot\gg_k\>+
  i\<\phi^{H_1}_k\ot\psi^{H_2}_k,H_1\phi^{H_1}_k\ot\psi^{H_2}_k\>+
  i\<\phi^{H_1}_k\ot\psi^{H_2}_k,\phi^{H_1}_k\ot H_2\psi^{H_2}_k\>
  =
  $$
  $$
  \<\phi^{H_1}_k\ot\psi^{H_2}_k,\dot\gg_k\>+  i\<\phi^{H_1}_k,H_1\phi^{H_1}_k\>+
  i\<\psi^{H_2}_k H_2\psi^{H_2}_k\>
  =
  \<\phi^{H_1}_k\ot\psi^{H_2}_k,\dot\gg_k\>+
  i{\cal E}^{H_1}(\phi^{H_1}_k)+
  i{\cal E}^{H_2}(\psi^{H_2}_k)
  =
  $$
  $$
  \<\phi^{H_1}_k\ot\psi^{H_2}_k,\partial_t(\phi^{H_1}_k)\ot\psi^{H_2}_k\>
  +\<\phi^{H_1}_k\ot\psi^{H_2}_k,\phi^{H_1}_k\ot\partial_t(\psi^{H_2}_k)\>+
  i{\cal E}^{H_1}(\phi^{H_1}_k)+
  i{\cal E}^{H_2}(\psi^{H_2}_k)=
  $$
  $$
  \<\phi^{H_1}_k,\partial_t(\phi^{H_1}_k)\>
  +\<\psi^{H_2}_k,\partial_t(\psi^{H_2}_k)\>+
  i{\cal E}^{H_1}(\phi^{H_1}_k)+i{\cal E}^{H_2}(\psi^{H_2}_k)=A^{H_1}
(\dot\phi^{H_1}_k,\partial_t)
  +A^{H_2}(\dot\psi^{H_2}_k,\partial_t).
  $$
  The last expression is zero because of the assumed horizontality of the  tensor factors
  \linebreak$\phi^{H_1}_k(t),\psi^{H_2}_k(t).$ Thus $ A^{H}(\dot\gg_k,\partial_t)$ is also zero,
\ie
  $ \gg_k$  is $A^{H}-$horizontal. 
  
  In this special situation, we arrived at the $ A^{H}-$horizontal $\gg_k$ by a different route
  than in Section~\ref{evolggk}. There we $ A^0-$horizontally lifted the $\phi_k,\psi_k$ and
then multiplied
  $\phi_k\ot\psi_k$ by a suitable phase factor function to get the
  $ A^{H}-$horizontal $\gg_k.$ Indeed, in general, the only situation in which it makes 
  sense to horizontally lift (by a dynamical  connection) the separate $\phi_k,\psi_k$ is when we have
``separate'' 
$H_1,H_2$
  as we naturally do in the present case.
  
  Thus, once the initial values of the $\gg(0),\lam_k(0),\phi_k(0),\psi_k(0)$
  satisfy
  \begin{equation}\label{PD(t)}%еееееееееееееееееееееееееееееееее
  \Gamma(t)=\sum_{k} \lam_{k}(t) \phi_{k}(t)\ot\psi_{k}(t)
  =\sum_{k} \lam_{k}(t) \Gamma_{k}(t)\,, 
  \end{equation}%ееееееееееееееееее      еееееееееееееее
  at $t=0$, and the $\phi_k$ are  $A^{H_1}-$horizontal, and the 
  $\psi_k$ are  $A^{H_2}-$horizontal, then 
  \begin{equation}\label{PD44}%еееееееееееееееееееееееееееееееее
  \sum_{k} \lam_{k}(0) \phi_{k}(t)\ot\psi_{k}(t)
  =\sum_{k} \lam_{k}(0) \Gamma_{k}(t)
  \end{equation}%ееееееееееееееееее      еееееееееееееее
  is automatically $A^H-$horizontal. It follows that the $\lam_k(t)$ in \eqref{PD(t)} are
  constants if, and only if, $\gg(t)$ is $A^H-$horizontal.
  
  We have just shown in the non-interacting  case, the map $t\ra\bfq(t)$ is a constant.
  In particular, the $r_k(t)=|\lam_k(t)|$ are constants, so that the right toroids
  $\bfT(\bfr)$ do not change shape and hence neither do the Pythagorean partitions
  ${\mf P}$. This implies (avoiding  the lower dimensional boundaries of ${\mf P}$)
  that the spectral states do not change. More precisely, the  index $k$ of the
realized  states $\phi^{H_1}_k(t),\psi^{H_2}_k(t)$ of $\textsf{ S}_1$, respectively 
    $\textsf{ S}_2$, relative to \ss\ never changes.  
  In other words,
 when the subsystems do not interact, there is no
jumping.

  \begin{thm}\label{nointeract} %%еееееееTHEOREMееееееееееTTTTTTTTTTTTT
For an interval of time, there is no jumping \ifff the interaction \ham\  $H_0=0.$ 
  \end{thm}
  {\bf Proof.} We have already seen that  $H_0=0$ implies no jumping. Suppose, there is no
 jumping, \ie for any $\gg$,  $\bfq$ stays within ${\mf p}_k$ for some $k.$ 
It must be then that the $r_k=|\lam_k|$ are constant or else  there would be jumping, 
even statistically.  Thus the ${\mf p}_k$ are time independent. 
 Even the $\lam_k$  must be
constant or else by moving
 $\bfq$ along a diagonal near a
boundary of two members of the partition $\mf P$, it can be arranged that  $\bfq$ 
crosses one of the fixed boundaries, at least for almost all $\gg.$ Thus almost all $\gg$  
have a polar decomposition $\sum_k\lam_k\,\phi_k\ot\psi_k$ with constant 
$\lam_k$.  It follows that $H$ has the non-interactive form 
$H=H_1\ot\bfI_2+ \bfI_1\ot H_2.$
\nl$\qed$
\nl{\bf Remark.} It can sometimes happen, that even with a non-trivial interaction,
the $r_k(t)$ are constant. Then there {\it is} jumping but no net or statistical  
jumping. This happens for certain values of the parameters in
 the hyperfine example of Appendix~\ref{AppHyperfine}, namely when $k=C l.$

\subsection{Interaction Hamiltonians}

We now consider the general case. We can  obtain evolutions of the conditional vector 
states when $H_0\ne 0$, 
 provided we have the proper sort of decomposition of the total \ham.
Using the bi-orthonormality\
property of polar decompositions we see
 \begin{lem}\label{L350} %%еееееееLEMMAееееееееееTTTTTTTTTTTTT
  Let
 $H=H_0+H_1\ot\bfI +\bfI\ot H_2.$
  \begin{equation}\label{C2S1L}%еееееееееееееее
j\ne k\Ra H_{{jj,kk}}= (H_{0})_{jj,kk}.
  \end{equation}%еееееееееееееее
  \end{lem}
$\qed$

We can apply this lemma and  Lemma~\ref{horlemgen} to obtain
\begin{thm}\label{intham}
 Let $\gg$ evolve according to the \SC\ equation in 
${\cal H}={\cal H}_1\ot{\cal H}_2$
\wrt\  the Hamiltonian \ $H=H_0+H_1\ot\bfI +\bfI\ot H_2.$

 Then $\gg(t)$ has a polar decomposition
$\gg(t)=\sum_k \lam_k(t)\gg_k(t),$ 
with $A^H-$\hor\  $\gg_k$ where $r_k=|\lam_k|$ and
    \begin{equation}\label{ShorizontalizingZeta41}%еееееееееееееееееееееееееееееееее
 \lam_k(t)  =r_k(t) e^{-i\int^t_0 \Upsilon_k(s)ds},\;
   \end{equation}%еееееееееееееееееееееееееееееееее
The $\Upsilon_k$ explicitly depend only upon $H_0$ and not $H_1$ and $H_2.$
Moreover, the $r_k$ satisfy
 \begin{equation}\label{againsmoothpdpos}%еееееееееееееее
\dot r_j=\sum_k \Im ((H_0)_{jj,kk})r_k.
 \end{equation}%еееееееееееееее

\end{thm}
{\bf Proof.}
Let the  $\phi_k$ be an $A^{H_1}-$horizontal lift of the
spectral projection    $[\phi_k]$  associated to $\gg(t) $ and
likewise for $\psi_k.$
Set $\gg_k^{12}:=\phi_k(t)\ot\psi_k(t) $  and $H_{12}:=H_1\ot\bfI +\bfI\ot H_2.$ 
By Section~\ref{noint}, the $\gg_k^{12} $ are $A^{H_{12}}{\rm-horizontal.}$
Using their bi-orthonormality  and ${ \dot \bfr}\cdot \bfr=\bo$, we get 
 \begin{equation}\label{intHam42}%еееееееееееееееееееееееееееееееее
\gg^{12}(t)=\sum_k r_k(t) \gg^{12}_k(t) {\rm\ is\ }A^{H_{12}}{\rm-horizontal.}
\end{equation}%еееееееееееееееееееееееееееееееее
 Using Lemma~\ref{horlemgen}
  \begin{equation}\label{intHamn42}%еееееееееееееееееееееееееееееееее
\tf \gg':= e^{-i\int^t_0 \<\gg^{12},H_0\gg^{12}\>ds}\gg^{12}(t)
{\rm\ is\  }A^{H}{\rm-horizontal.}
\end{equation}%еееееееееееееееееееееееееееееееее
Since $\gg$ and $\gg'$ have the same reduced traces and are
both   $A^H-$\hor, they differ by a constant phase factor which can be absorbed into
the $\gg_k.$ Thus we can take $\gg=\gg'.$ 
Again using Lemma~\ref{horlemgen}, we get
\begin{equation}\label{intHam44}%еееееееееееееееееееееееееееееееее
 \gg_k := e^{-i\int^t_0 \<\gg^{12}_k,H_0\gg^{12}_k\>ds}\gg_k^{12}(t)
{\rm\ is\ }A^{H}{\rm-horizontal.}
\end{equation}%еееееееееееееееееееееееееееееееее
Define the $\lam_k(t)$ implicitly by
\begin{equation}\label{intHamn45}%еееееееееееееееееееееееееееееееее
 \gg(t)=\sum_k \lam_k(t)\gg_k(t).
\end{equation}%еееееееееееееееееееееееееееееееее
Combining the last three equations, we see that \eqref{ShorizontalizingZeta41} holds with 
$$
\Upsilon_k(s)=\<\gg^{12},H_0\gg^{12}\>-\<\gg^{12}_k,H_0\gg^{12}_k\>=
\<\gg,H_0\gg\>-\<\gg_k,H_0\gg_k\>.
$$
Finally, from \eqref{smoothpdpos}
 \begin{equation}\label{Iagainsmoothpdpos}%еееееееееееееее
\dot r_j=\sum_k \Im ·(H_{jj,kk})r_k=\sum_k \Im ·((H_{0})_{jj,kk})r_k+
\sum_k \Im ·((H_{12})_{jj,kk})r_k.
 \end{equation}%еееееееееееееее
Now, in the non-interacting case, we know the  coefficients $\lam^{12}_j$ are constant
and so are their absolute values, $|\lam^{12}_j|$.  But
we also know from Lemma~\ref{lambdaEq}, that
$$\partial_t |\lam^{12}_j|=\sum_k \Im ·((H_{12})_{jj,kk})|\lam^{12}_k|$$
which suggests all the $\Im ·((H_{12})_{jj,kk})=0$ in \eqref{Iagainsmoothpdpos}.
This fact  follows directly from the definitions and  
establishes the last assertion of the theorem.
\nl$\qed$

\nin{\bf Conclusion:} All the  characteristics (such as the frequency) of the jumping between
states are determined by the interactive part $H_0$ of the \ham\ $H=H_0+H_1\ot\bfI +\bfI\ot H_2.$
This generalizes the  no-jumping result of Section~\ref{noint}.

\nin{\bf Remark:} Up to a scalar (multiple of the  identity operator) and
ignoring all analytic difficulties, $H$ alone determines natural choices for
$H_0, H_1$, and $H_2.$ Namely, let $H_1\ot\bfI_2$ be the orthogonal projection
 \wrt the \HS\ inner product of $H$ into the space of Hermitian operators on ${\cal H}$
of the form $h_1\ot\bfI_2$ where  $h_1$ is Hermitian.  Similarly, let
$\bfI_1\ot H_2$ be the orthogonal projection
 \wrt the \HS\ inner product of ${\cal H}$ into the space of Hermitian operators on ${\cal H}$
of the form $\bfI_1\ot h_2$, where   $h_2$ is Hermitian and traceless. Of course, these
definitions require that $H$ be \HS, and $h_2$ be trace-class, but we proceed formally.
Finally, we set $H_0=H-H_1\ot\bfI_2-\bfI_1\ot H_2.$ This gives  a ``minimal'' 
 interaction \ham.

The above mentioned orthogonal projections are strongly related to reduced traces.
For example, in finite dimensions it is not hard to see that
$H_1= {1\over n_2} {\rm Tred}_1(H),$ 
where $ n_2$ is the dimension of ${\cal H}_2$; this fact  follows from   Proposition~\ref{Tr1Proj}. 

\section{ The Interpretation of  IQM }\label{section5Discussion}
\setcounter{equation}{0}

In this section, we summarize the main features of the new model of
quantum mechanics.  We then compare 
IQM with
the standard treatments and discuss some of the ramifications. 
We conclude by applying it to the hyperfine splitting example in Section~\ref{ex2} .

With  reference to  Dirac's  dictum 
quoted in the introduction, we have
extended ``the mathematical formalism'' of SQM to include the right
  toroids $\bfT(\bfr)$ and  $\bfq(t)\in \bfT(\bfr)=\po_\gg$ for $\gg\in\bfS.$ The naturality of the
extension  leads to the existence of a natural evolution in the enlarged state space consisting
of the polar bundle $\po$. 
We can say
  this adds a  ``success'' in our ``attempts to  perfect and generalize the
   existing mathematical formalism''. As another step, we have  shown there is  
a natural way of partitioning the toroids. The corresponding tilings of Euclidean 
spaces 
appear to be
 mathematically  new.
 Our  approach to ``try to interpret the new mathematical
  formalism in terms of physical entities'' has  already been explicated,   starting with our
introductory remarks in Section~\ref{sqm+}.
 The main points will be discussed below.

\subsection{Summary of the extension of the mathematical formalism.}
The extension of the mathematical formalism of  SQM to IQM has
four main constituents, upon which we elaborate in
the ensuing discussion.
\nl  {\bf I}) Interacting systems have pure states, associated to
a particular  one of the eigenprojections of 
the density operator.
\nl \ \ {\bf II}) Unit vectors within  the rays 
assigned to the system by SQM
are incorporated into  the  representation of  pure states.
\nl  {\bf III}) These (eigen)vectors evolve by lifting the SQM evolution
 of the eigenprojections by the dynamical 
connection.
\nl  {\bf IV})  The choice of the  particular eigenprojection   
is  via the Pythagorean partition.

\subsubsection{I) The pure state of an interacting system.}

\nin The use of the polar decomposition to define a pure spectral projection for a
system in interaction with another was first proposed by Kochen in \cite{Kochen}.
The main objection raised to it was the lack of  dynamics, which
this paper is largely devoted to remedying.
The dynamics are summarized in  {\bf III}) and  {\bf IV}).
On the other  hand, its acceptance resolves the main paradox of SQM in
its attempted application to individual systems, the measurement problem.
The resolution is short enough to warrant repeating here.

A measurement of a system $\ss_1$ by an apparatus $\ss_2$ is viewed simply as
an interaction between two quantum systems, Neither the size
of $\ss_2,$ nor the observer plays any part in our analysis. We assume that throughout
the interaction, the system $\ss=\ss_1+\ss_2$ is isolated from
its environment.

 This means the SQM state  of  $\ss$ is of the form
$ [\gg(t)],$ which is undergoing \SC\ evolution. The more detailed IQM description 
requires the actual vector $\gg.$ 
It also requires a particular polar decomposition of $\gg,$ which we
write as  $\gg=\sum_k\lam_k\,\phi_k\ot\psi_k.$ 
 In \cite{Kochen}, Kochen proposed that 
$\ss_1$ has the pure state $[\phi_k],$ 
the measured state of $\ss_1$ 
 and synchronously $\ss_2$ has the pure state $[\psi_k],$ 
the corresponding state of the apparatus.  No mechanism for
determining which $k$ obtains was proposed except (implicitly) that the  choice should satisfy the
probabilistic requirements of SQM.
More recently in 
\cite[Baccigaluppi and the references therein]{Bacci} various stochastic evolutions (essentially
of the $k(t)$) have been considered. 

In the present theory,  $\bfq$ lies in a particular part ${\mf p}_k$ of the partition $\mf
P.$ This $k$ then determines the conditional spectral state $[\phi_k]=[(\gg,\bfq)|\ss_1].$ 
There is no collapse of $\gg$, \ie no mysterious transition from the pure state,
 $\gg$, to a mixed state, which is usually taken to
$\sum_k |q_k|^2 P_{\phi_k\ot\psi_k}$.
In fact, the only remnant of such a 
transition is in our {\it description} of the passage from SQM to IQM and this
transition
 goes in the opposite direction: we replace
the density operator
$\sum_k |\lam_k|^2 P_{\phi_k}$ 
(which is a  mixed state of $\ss_1$ and hence an inappropriate
 description of  an individual) by $[\phi_k]$.

\subsubsection{II) Using $\bfS$ instead of $\bfP.$}
The use of unit vectors in place of rays 
in representing  pure states of isolated systems is harmless and inessential, but
very useful. We employ it because it allows for a more uniform treatment of the
toroidal phases, which {\it are} essential to our treatment.

Actually, the use of unit vectors to represent pure states is a convenience  of which all
physicists avail themselves, while   occasionally paying lip service to $\bfP$ by noting that
$\<\phi,A\phi\>$ and $|\<\phi,\psi\>|^2$ are all that is measurable and these are independent
of the ``phases'' of $\phi$ and $\psi.$ On the other hand, the existence of  relative 
phases has long been noted  theoretically, \eg the geometric phase of Berry and its
generalizations, see \cite{Shapere}, and has  been measured in interference
experiments. It is true, however, that there is no generally accepted way of
associating a Hermitian operator to this measurable quantity 
(see \cite[Barnett-Pegg and its many references]{Barnett}.)
 This is a gap in the formalism of SQM applied to individuals (somewhat
patched up by the formalism of POVs as in \cite [Davies]{Davies},
and \cite[and the references therein]{Busch}),
since measurable quantities should be observables and observables should be
 represented by 
Hermitian operators, according to the formalism of SQM.
It is not coincidental that our  approach to unraveling  the mystery of 
the quantum theory of
individuals starts with   this loose thread of phases 
 in the cloak of the individual interpretation of SQM.

The transition from a relative to an absolute phase merely entails the choice of a basepoint,
\ie  a reference   phase.
The phase, relative to this basepoint, is an ``absolute'' phase. We can interpret
the experiments demonstrating interference between different 
laser sources, \eg   \cite{Louradour} and \cite{Rarity}, 
 as providing instances of such measurement.

Even if absolute phase is generally not directly measurable, that is not an absolute argument
against its existence or appropriateness. Many physical constructs have been introduced
without an expectation of being subject to direct measurement.
We  quote Feynman, Vol I, 38-8 in \cite{FeynmanL},
\begin{quote}
It is not true that we can pursue science by using only those concepts
which are directly subject to measurement.

In quantum mechanics itself there is a wave function amplitude, there
is a potential, and there are many constructs that we cannot measure
directly. 
\end{quote}

The use of the phase as a parameter to distinguish different individual systems belonging
 to an ensemble with the same pure state appears rather natural; for the ensemble of
systems has a statistical state represented by an ensemble of unit vectors in a ray.
However simply assigning a unit vector to an isolated system does not suffice to treat
interacting systems. In fact, the no-go theorems of \cite[Bell]{Bell}
and \cite[Kochen-Specker]{ KochenSpecker} show that such  non-contextual 
assignments of state run afoul of the predictions of SQM. To get a ``go'' theorem, 
we need to
combine the phases of the various polar components of the vector $\gg$
associated to a composite system $\ss= \ss_1+\ss_2.$

The ubiquity and utility of amplitudes suggest 
that possibly every individual physical system does indeed have a ``phase'',
 i.e. we can represent their states by unit vectors $\gg$ in ${\cal H}$, not just equivalence
classes   $[\gg]$ in $\bfP({\cal H})$; more precisely, that it is consistent to so model
individuals, which is what we end up doing. Indeed, the mathematical structures we employ
are a uniquely determined extension of the present mathematical structure of SQM.

\subsubsection{III) The evolution of the eigenvectors.}
The evolution of the reduced density  operator of an interacting system is
a standard part of SQM (see \cite{Blum}).
The evolution of its eigenprojections in $\bfP$ has been
increasingly used in, \eg  \cite{BarnettPhoenix,KnightPhoenix,EkertKnight,Furuichi}
 in connection with the Jaynes-Cummings model.  In IQM, 
 the individual state vectors  lie in the
polar bundle ${\po}$, and  the evolution of the eigenprojection in $\bfP$ lifts to ${\po}$ by means of the dynamical connection $A^H$. When the \ham\  $H$ is 0, the connection
reduces to the canonical connection used in obtaining the geometric phase of Berry. When the
\ham\  describes separate evolution for the subsystems, \ie when $H$
has the form $H_1\ot\bfI_2+\bfI_1\ot H_2, $ it turns out that there is no jumping
between the states vectors, as reviewed in the next section.

\subsubsection{IV) the Pythagorean partition $\mf P$  of the phases.}\label{}
 Assuming the extensions in ({\bf I-III})
 have been accepted,
we can argue for  the role of $\mf P$ as follows.
 We know from our assumption that 
$\bfq$ lies in the toroidal fiber ${\po}_\gg$ of ${\po}$ above $\gg$. This toroid
with base point is isometric to $\bfT(\bfr)=\prod_k \bfS^1(r_k).$ Hence the toroid must be
partitioned in subsets
${\mf p}_k^*$ {\it physically} by ({\bf I-III}). Indeed, $\bfq \in {\mf p}_k^* $ if, and
only if,
$\phi_k$ is, in fact, the state of $\ss_1.$
We should note, in this connection, that after a measurement, even a Copenhagen 
adherent would allow that 
 $\ss_1$ is in a  pure state. 
Returning to the physically defined partition $\mf P^*=\{\mf p^*_1,\cdots\}$, 
the diagonal property
of Section~\ref{Pyth} follows from
the probability requirements of SQM. 
 The naturality property of the partition says, in effect, that a partition of a  (right) toroid 
 is consistent with the partitions of the subtoroids it contains.
Thus, the partitions are
defined uniformly for all finite dimensions. 
 This implies the  results 
hold even for  $\infty$-dimensions {\it mutatis mutandis}  without dealing with it explicitly, 
Physically, and in $n$ dimensions,  the naturality property is a continuity property of
the partitions: the  partition of $\bfT(r_1,\cdots,r_n)$ approaches the partition of  
$\bfT(r_1,\cdots,r_{n-1})$ as $r_n\ra {0}.$

If we assume that $\mf P^*$ is a convex partition, then  Theorem~\ref{MainThm}
stated in Section~\ref{Pyth} and proved in
Section~\ref{AppPart}, shows  ${\mf P}^*={\mf P}.$

What is our rationale for assuming that we are dealing with convex partitions?
The convexity for the corresponding tiling of $\R^n$ is a simple and natural property
 that is usually assumed in mathematical discussions of tilings. We are however concerned
with  {\it physical} reasons for convexity. Consider the  special case in 
which  all the polar
vectors are energy eigenstates, \ie when for all  $k,$ there exist 
constants $r_k, E_k$ so that $\lam_k=r_k e^{-i E_k t}.$ Then the frequency of jumping 
of the  $k(t)$ specifying the state $\phi_{k(t)}(t)$ is minimized under the convexity 
hypothesis.  Conversely, this ``minimum jumping property''
 implies the convexity property of our partitions. This follows from the characterization of
convex subsets of a Euclidean space as those which intersect every line in an interval.  

It is possible that a deeper understanding of the physical processes that lead to the partition
would, in general,
allow the convexity to be derived from a physically natural 
variational problem.
 Mathematically, it is likely that these partitions minimize the co-dimension-one
volume of the boundary set, as is true when $n=2$ .
More speculatively, if the parts are analogous to  different thermodynamical phases 
there may be some physically plausible partition function whose critical values define the
boundary. Conceivably, such a function could lead to a non-convex partition
of the toroidal fiber above $\gg$, but one
which approaches $\mf P$ as $\<\Gamma, H\gg\>\ra 0.$

Here is one way we may intuitively think about the individual states of interacting systems.
First, an isolated system has a phase that changes uniformly with time at a rate
proportional to its energy. This phase can be regarded as an internal clock or 
pulse of the system (\cf Feynman \cite[QED]{QED}.) 
If the system interacts with another, the pulse quickens (if all energies are positive) and
becomes a complicated but still smooth function of time.
For instance, in the spin-spin interaction of the hydrogen atom, a discussion of which follows
in the next section, it is
shown that the phase of the electron is essentially an elliptic function of time.
 The partition
boundaries may be considered as thresholds between different quantum states. As such a
threshold is crossed, a different quantum state is assumed.
As an example, an excited state of an atom plus a weak exterior electromagnetic field
changes smoothly with time, but the condition of the 
field changes abruptly when the
atom decays and emits a photon.

\subsection{How IQM works in an example.}\label{ex2}
  \begin{quotation}The classification of the constituents of a chaos, nothing less is here
essayed.\par\rightline{\emph {from Moby Dick by Herman Melville }}
  \end{quotation}

We now apply the general IQM  theory to a specific case which is, essentially, 
the  simplest possible non-trivial example. We take the case of the spin of an 
electron in a hydrogen  atom, using the simplified model discussed in 
\cite[Feynman]{FeynmanL} and worked out in detail in  Appendix~\ref{AppHyperfine}.

It is  modeled by  two spin ${1\over2}$ systems, the  ``electron'' 
and the ``proton''. Thus ${\cal H}={\cal H}_1\ot{\cal H}_2\approx
\C^2\ot\C^2 \approx \C^4.$ The Hamiltonian $H$ is a pure interaction \ham\ 
with $H=\mu{\vec \sigma_1}\!\dot\otimes\! {\vec \sigma_2.}$ For notation and details not
given  here, see Appendix~\ref{AppHyperfine} and in particular, the glossary given in
\ref{gloss}.

We assume at  $t=0$, the data $\gg(0)$, and $\theta_\pm:=\arg \lam_\pm$ 
are given. These are enough to determine an initial  polar decomposition
$\gg=\lam_+\,\phi_+\ot\psi_++\lam_-\,\phi_-\ot\psi_-,$ 
up to reciprocal phase factors for the $\phi_\pm,\psi_\pm.$ These would be important
if we were interested in further details of the component systems, which are 
ruled-out by the simplicity of our model, \ie no subsystems of the component $\ss_i$ 
are possible here because $\dim {\cal H}_i=2.$ However, if we were to treat, say the
proton as a composite system, then the actual phase of the $\psi_\pm$ would come 
into play. For this simple example, we  are free to 
choose these original phases so as to make 
the $\phi_\pm,\psi_\pm$ real. This is accomplished in conjunction with assuming 
the $\lam_\pm$ are also real at $t=0,$ as in the appendix. We have also 
arranged the axes in $\R^3$ so that initially the $z-$axis bisects the  
spin vectors. We show in the appendix that the polarization vector 
representing the mixed state of the electron (and the proton) has an ellipse 
in a plane perpendicular to the $z-$axis as trajectory. We further  require 
that the $x,y-$axes are initially aligned along the major and minor 
axes of this ellipse, which implies that $q_\pm\in\R$.

Then we can determine the \pd for all $t$, using  the dynamical connection of IQM:
\begin{equation}\label{PDt}%еееееееееееееее
\gg(t)=\lam_+\,\phi_+\ot\psi_++\lam_-\,\phi_-\ot\psi_-\,,
\end{equation}%еееееееееееееее
where all the components are explicitly determined functions of $t.$
 In particular, we find
\begin{equation}\label{eq:H}%еееееееееееееее
\lam_\pm=\sqrt{{1\pm\sqrt{\Delta}\over 2}}
%\exp{i\(\nu+\sigma_\pm\mp {C^2l\over k}\Pi(e^2,S e,\omega t)-{\omega t\over2}\)}
\exp{i\(\nu+\sigma_\pm\mp {C^2l\over k}\Pi(e^2;\omega t|S^2 e^2)-{\omega t\over2}\)}
\end{equation}%еееееееееееееее
where we here briefly recall the  definitions of the functions involved in this
formula.
$$
\nu ={C^2\over \sqrt{1-S^2 e^2}}\arctan(\sqrt{1-S^2 e^2}\tan\omega t),
$$
and where the $\arctan$ is taken so that  the resulting function 
is a  continuous function of $t$, vanishing at 0, 
as is possible.
$$
\sigma_\pm =\arctan({C^2 l^2 \pm \sqrt{\Delta}\over k^2\sqrt{\Delta}}
\tan\omega t).
$$
Also
$$
\Pi(n;\varphi|m)=\int_0^\varphi {d\rho\over (1-n\sin^2\rho)\sqrt{1-m\sin^2\rho}}
$$
is Legendre's elliptic integral of the third kind.
$$
\phi_\pm=\pm e^{i\tau_\pm} \pmatrix {\alpha\cr \beta_\pm},\;
\psi_\pm= e^{i\tau_\pm} \pmatrix {\alpha\cr -\beta_\pm},
$$
where
$$
\tau_\pm= {l\over2}\arctan({C l \over k}\tan\omega t)
\pm{C l^2\over 2k}\Pi(e^2;\omega t|S^2 e^2).
$$

The spectral states $[\phi_\pm(t)]$ and $[\psi_\pm(t)]$ of the 
electron and proton give the axes of spin of the particles, and 
these two axes are antipodal on the ellipse, which is the 
trajectory of the SQM mixed states.  These spectral states do not tell when the 
spin is up or down for each axis (although they are synchronous). To find 
out when $[\phi_{k(t)}]$ is $[\phi_+(t)]$ or  $[\phi_-(t)]$, we must consider 
the complex vector $\bfq(t)= (\lam_+(t),\lam_-(t))=
 (r_+e^{i\theta_+} ,r_-e^{i\theta_-})$ and determine, for each $t$, whether
 $(S_+,S_-):=(r_+\theta_+ ,  r_-\theta_-)$ lies in $\varpi^{-1}({\mf p}_+)$ or $\varpi^{-1}({\mf p}_-).$
Here the  ${\mf p}_\pm={\mf p}_\pm(t)$ comprise the Pythagorean partition 
${\mf P}={\mf P}(t)$  constructed in Section~\ref{Pyth}.

In the terminology of that section, we have  ${\mf p}_\pm=\varpi(A_\pm).$
It will be convenient to replace the $A_\pm$ by the union 
$\underline{A}_\pm$  of
the appropriate basic building blocks ${\scriptstyle Sl}(k,\sigma)$ from which they
are constructed. This is justified since 
$\varpi(\underline{A}_\pm)=\varpi({A}_\pm).$ 
In this case, each $\underline{A}_\pm$ is just
the union of 2 triangles as in Figure~\ref{figenvelope}.
 We can take 
$\underline{A}_+=$OCF+ABE and $\underline{A}_-=$OAE+BCF. Thus, denoting the box OABC by $B$,
\begin{eqnarray}\label{pplus}
{
\underline{A}_+=\left\{(S_+,S_-)\in B\;{\Big |}\;
 S_- <  {r_- S_+\over r_+} ,\;
 {S_- r_-}+{S_+ r_+}< 2\pi  r_+^2\right\}\bigcup
}\nonumber
\\ \left\{(S_+,S_-)\in B\;{\Big |}\;
 S_- >  {r_- S_+\over r_+} ,\;
 {S_- r_-}+{S_+ r_+}> 2\pi  r_-^2\right\},
\end{eqnarray}
\begin{eqnarray}
{
\underline{A}_-=\left\{(S_+,S_-)\in B\;{\Big |}\;
 S_- >  {r_- S_+\over r_+} ,\;
 {S_- r_-}+{S_+ r_+}< 2\pi  r_-^2\right\}\bigcup
} \nonumber
\\ \left\{(S_+,S_-)\in B\;{\Big |}\;
 S_- <  {r_- S_+\over r_+} ,\;
 {S_- r_-}+{S_+ r_+}> 2\pi  r_+^2\right\}.
\end{eqnarray}
Of course, $S_\pm, r_\pm$, and even $B$ are all time-dependent.
 Let $\v{S_\pm}(t)$ denote the least non-negative residue of $S_\pm(t)$ modulo
$2\pi r_\pm(t)$. 
Then the condition that $k(t)=+$ is equivalent to 
 $\v{S_\pm}(t), r_\pm(t)з$ 
satisfying the inequalities coming from \eqref{pplus}.

We set 
$$c_1(t):=  {r_- \v{S_+}\over r_+}\; ,c_2(t):= 2\pi { r_+^2\over r_-}-{\v{S_+} r_+\over r_-}\;,
c_3(t):= 2\pi { r_-}-{\v{S_+} r_+\over r_-}.
$$
Then $\bfq$ lies in ${\mf p}_+$ \ifff either $\v{S_-}< c_1, c_2$ or 
 $\v{S_-}> c_1, c_3.$
For numerical values, say   $\theta={3\pi\over 7},\,  \lam_+(0)=.94,$ 
 we can compute these functions 
 as in Figure~\ref{conel}. 
 \begin{figure}
  \begin{center}
\includegraphics[scale=.8]{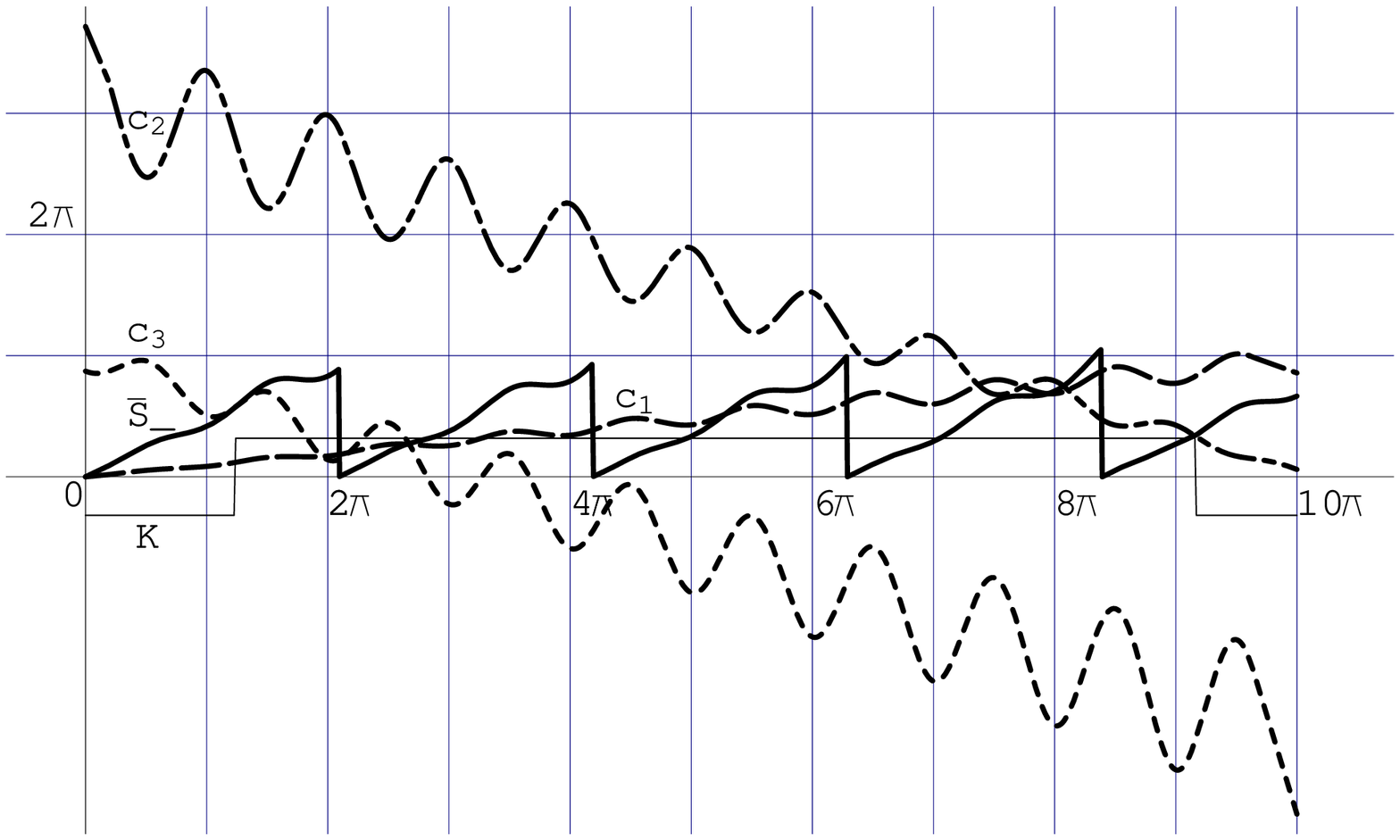}%{elspin.eps}%uses K(w)0606.nb 
 \end{center}
  \caption{
Here $\theta={3\pi\over 7},\,  \lam_+(0)=.94.$ 
${\rm Sign}(K(t))=k(t).$ 
%is $k(t)$   spectral state of electron spin.
}
  \label{conel}
  \end{figure}
 For example, restoring here $\hbar$, we can see that 
%$10\pi>4{\mu\over \hbar} t>3\pi\Ra \bfq\in{\mf p}_+$, \ie the conditional spectral state of
$4{\mu\over \hbar} t\in[3\pi,10\pi] \Ra \bfq\in{\mf p}_+$, \ie the conditional spectral state of
 the electron is $+$.

%  
%  We show in  Figure~\ref{lastFrame}, the polarization ball containing the ellipse which is the 
%  trajectory of the density operators represented by the blue (proton) and red (electron) balls. 
%   
%  \begin{figure}  %%%%  lastFrame
%    \begin{center}
%  %\includegraphics[scale=.7]{elspinS01.eps}%uses 060399A.nb  \!\(\(th = \(3  \[Pi]\)\/7;\)\)
%  \includegraphics[scale=.8]{lastFrame.eps}%uses K(w)0606.nb 
%   \end{center}
%    \caption{
%  Evolution of spectral states at $ t= 6.1{{\mf h}\over\mu}$ for 
%  $\theta={3\pi\over 7},\,  \lam_+(0)=.94.$
%  %$t=\pi  122 \hbar\over 4\times10 \mu=6.1 h\over\mu$
%  %$ (4{\mu\over \hbar} t   = \pi 122\over10 $  %+  .0001
%  %$t=\pi  122 \hbar\over 4\times10 \mu=6.1 h/\mu$
%  %${\rm Sign}(K(t))=k(t).$ 
%  %is $k(t)$   spectral state of electron spin.
%  }
%    \label{lastFrame}
%    \end{figure}

\section{Observables} \label{observables}

We have  avoided the use of observables; we now consider their place in IQM.
\subsection{States Versus Observables.}
The question of which to consider first, states or observables,
has been likened to the chicken and egg problem. 
 But there is really no doubt about priority, at least for IQM.
Our theory takes states as  primary. In fact, we have completed its basic structure
 without  even mentioning within it the observables which are
usually
 taken as basic. Lattice theory, favored by foundationalists, and 
 $C^*-$algebras predominating in QFT, are the main observable-first
 approaches. 
In fact, the {\it mathematical} notion of state is, essentially,
  a  positive linear functional on a
$C^*-$algebra.
 But, of course, this mathematical concept,  which
the algebra of observables must precede, derived from the standard
models of quantum mechanics. These approaches have their roots 
in Born and Jordan's version of  Heisenberg's matrix mechanics.

But physically, it is difficult to find a referent for an observable without
something to observe, \eg a weak beam of atoms. From this point of
view, state-first approaches appear to us to be more direct. Of course, the
original state-first approach was
 \SC's.

In such a formulation of SQM, the pure states are
identified with equivalence classes of irrefinable ensembles. 
These are the idealizations of the weak beams. The set 
  of these has a metric
 $m(\alpha,\beta)=1-p(\alpha,\beta), $  where $p(\alpha,\beta), $
is  the transition probability. This set turns out to be 
 metrically identifiable with  $\bfP=\bfP({\cal H})$  with
 $p(\alpha,\beta) = |\<a,b\>|^2,$
where
$a$ and $b$ are unit vectors representing $\alpha$ and $\beta$
Then $\theta(\alpha,\beta)= 2\arcsin(1-{p(a,b)})$ is 
also a metric, the Fubini-Study metric on $\bfP.$ 
Alternatively starting from the $\theta$-metric, we have that $\sin^2{\theta\over2}$ being 
differentiable, subadditive for $\theta\in [0,{\pi\over2}]$, and taking $0$ at $0$ transforms $\theta$ into
a smoothly equivalent  metric $1-p$  on $\bfP$ with the same invariance group.

Thus the basic projective Hilbert space model of SQM can be expressed 
in terms only involving the geometry of the
 physical states of  ensembles. An early  attempt at deriving
 the mathematical structure
of SQM through use of the geometry and symmetry was postulated by
Land\'e in \cite{Lande,Lande2}. 
His conjectures were settled, mostly affirmatively, by Ax in \cite{Ax}. More 
recent attempts at a geometric founding of SQM can be found in  
 \cite{Ashtekar} and  \cite{Hughston}.

\subsection{Measurability Versus Observability.}

Even though we take a state-first approach, we should have, by the end,
some correspondent to the  observables of SQM. In this subsection, we
show how to get the truly measurable observables. These are,
essentially, the { strongly repeatable instruments} of Davies in 
\cite[Chap.4]{Davies}.

The standard way, since von Neumann, of representing the measurement 
process is to consider a composite system $\ss=\ss_1 +\textsf S_2$,
where $\ss_1$  is the system to be measured and $\textsf S_2$ is the
apparatus.  
Given the observable A,  an experimental arrangement of \ss, including a Hamiltonian
$H$, is posited so as to model measurement. 

We reverse this procedure, by starting with essentially arbitrary
states of
$\ss=\ss_1 +\ss_2,$ and see what can be considered as being
measured. We assume, therefore, that  we have an evolving polar
decomposition for $t\ge 0.$ 
$$
\gg=\gg(t)=\sum_k \lam_k\,\phi_k\ot\psi_k.
$$
We require, for simplicity,  that it be regular at the time of measurement,
say $t=1.$
If the procedure is to have the usual repeatability requirement, then
we must require that the evolutions be free for $t\ge1$. Thus,
we  take the total Hamiltonian $H$ to be, for $t\ge1$,
 effectively of the
non-interacting form
$H=H_1\ot\bfI_2+\bfI_1\ot H_2.$  
By the result of Section~\ref{nointeract},
we have no jumping, \ie 
$r_k(t)$ is constant for $t\ge1.$ This guarantees the repeatability.
This procedure can be regarded as a method of measuring the
observable corresponding to the Hermitian Hilbert-Schmidt operator
$ A=\sum_k r_k P_{\phi_k},\; r_k:=|\lam_k|.$

Since we have assumed the $r_k$ distinct, this is equivalent to
giving a finite-dimensional-projection-valued measure on $\Z$. These operators
are dense in the strongly repeatable instruments, which by 
\cite[Th. 3.1]{Davies} can be identified with
 the projection-valued measures on $\Z.$ 

In this way, we recover the truly measurable observables. They have
the property that we can sensibly assign a state to the
result of an $A$-measurement  of a state. Indeed we have the 
formula 
%(the corrected Luder's rule) 
 that the result $\rho_1$ of measuring a state (represented by
the density operator) $\rho_0$  by the instrument
$A$ is the  state 
 \begin{equation}\label{luder}%еееееееееееееее
 \rho_1= \sum_k  P_{\phi_k}\rho_0 P_{\phi_k},
  \end{equation}%еееееееееееееее
as well as 
 \begin{equation}\label{luderax}%еееееееееееееее
 \rho_1 =\lim_{T\ra\infty}{1\over T}\int_0^T
 e^{i s A} \rho_0 e^{-i s A}ds =Q(\rho_0),
  \end{equation}%еееееееееееееее
Here $Q$
 denotes the projection (orthogonal \wrt the
Hilbert-Schmidt norm) on the set of density operators intersected with
 the commutant of  $A.$

As pointed out by Davies,  in general there is no comparable formula
to \eqref{luder} 
for arbitrary bounded Hermitian operators. 
The same applies to \eqref{luderax}. For example, suppose
the position operator (multiplication by $f(x)=x$) on $L^2[\R]$ 
commutes with some  density operator $\rho.$
Let $P_I$  denote the spectral
projection of the position operator corresponding to an interval $I.$
Then $P_I$ is multiplication by the characteristic function of $I.$
Let $v$ be an eigenfunction of $\rho$ with eigenvalue $r>0.$ Then all the 
$P_I(v)$ are eigenfunctions of $\rho$ with eigenvalue $r.$ There is an 
interval $I$ on which the essential infimum of $\pm v(x)$ is positive. 
Then using a subinterval $J$ of $I$, it follows that there must be an
infinite dimensional space of eigenvectors of $\rho$ with eigenvalue $r.$
This contradiction to $\rho$ being trace class, shows that the position
operator is not measurable within the framework of SQM.

In other words, only sufficiently ``small'' Hermitian operators correspond
to truly measurable quantities. Others have expectation values 
and can be approximated by instruments, {but they cannot sensibly
give an after-measurement state.}

\subsection{The  values of SQM   observables in IQM.}

Let $B$ be an Hermitian operator  on the \Hsp\ ${\cal H}_1.$
 For
simplicity, take $B$ to be 
positive
 with distinct eigenvalues $b_k$ and eigenvectors $\phi_k$.
 A basic assumption of SQM is that the  expected value of $B$ in any SQM state
$[\phi]$ is $\<B\phi,\phi\>.$ 

We want to compare
the SQM description of an observable with operator $B$
  with its IQM version. For this, we need
\nl1) another system $\ss_2$ with  \Hsp\ ${\cal H}_2$  and $\psi\in\bfS({\cal H}_2)$ 
\nl2) a \ham\  $H$ on ${\cal H}={\cal H}_1\ot{\cal H}_2$
and an \ON\ set $\{\psi_k\}$ in ${\cal H}_2$ such that
  \begin{equation}\label{measevol}%еееееееееееееее
U(\phi\ot\psi)= \phi_k\ot\psi_k,\;
{\rm where}\ U_t=e^{iHt} {\rm and}\ U=U_1.
  \end{equation}%еееееееееееееее
These are the same requirements as are invoked  in the theory of
 measurement in SQM. Discussions of this situation
with some proposals as to the formation of $H$ are contained in 
\cite{vonNeumann},
 \cite[Chap.4]{Davies}, and \cite{Busch}.

Now let $\phi\in \bfS({\cal H}_1.).$ Then we  have
$$
\phi=\sum_k\lam_k\,\phi_k\Ra U(\phi\ot\psi)=\sum_k\lam_k\,\phi_k\ot\psi_k.
$$
In this situation, SQM says that the probability that $B$ assumes the value $b_j$ in
state $[\phi_j]$ is
$|\<\phi,\phi_j\>|^2= |\lam_j|^2.$
In IQM more information is available because a particular $j$ is picked out by
virtue of the fact that $\ss_1$ is actually in one of the states $[\phi_j].$
Thus a definite value for $B$ is determined, namely $b_j.$ Moreover, 
for  $e^{i\theta}\in \bfS^1$,  the probability
 that $e^{i\theta} \phi$
would yield the value $b_j$ is $|\lam_j|^2,$  by the 
diagonal property of  Pythagorean partitions.
This probabilistic fact is independent of the choices 
made in 1) and 2) . If we make different choices 
of $H$ and $\psi$ or the 
 particular $\phi\in[\phi]$,  we would, in
general, get a different $j$ and so a different value of $B.$
 We note that in an important case,
the value of $B$ is independent of these choices. This happens when $[\phi]$ is one of the
$[\phi_k]$, say $[\phi_j].$ Then by \eqref{measevol}, $\phi\ot\psi$ evolves to
$\phi_j\ot\psi_j$ and so $B$ takes on the value $b_j,$ as it should.

Thus IQM can predict the actual value an observable attains in a state of $\ss_1$
provided that the complete experimental setup is specified, including the \ham\ $H.$
This added contextuality makes the set of dynamical variables on $\ss_1$
a much more complicated
object  than the algebra of
observables assigned to it by SQM.

Thus for IQM, the state-first approach appears natural. 
At any rate, the appropriate  notion of observable for individual systems would require 
some reworking  as did the notion of state.

\section{EPR }\label{epr}
\setcounter{equation}{0}

 The EPR paper \cite{EPR} raised  the  important issue 
of entanglement and the spectre of spooky  non-local
effects which  haunts physics to this day. 
EPR seems to present a truly {pu{\bf zz}ling} aspect 
 of the physical world, a ``{\bf Z}-mystery'' 
in the terminology of \cite[Penrose]{Penrose}. We agree that
it reveals a deep truth, but contend that this is
the inescapable relativity of states.  We 
follow Penrose in contrasting EPR with the Measurement Problem, a {parado{\bf x}}
 exemplified by \SC 's cat.  
We claim that the reduction of the state vector in the latter  is not a real physical effect.
It is, according to IQM, really a misapplication of a 
statistical theory to individuals. One should not be surprised at  inconsistencies 
arising from confusing states of ensembles with states used in  representing 
individuals.

We show how our interpretation leads to a description of the EPR experiment without
any paradoxical features. We shall study it in a general setting to emphasize that it is a 
pervasive effect of any multiple interactions, and not, as in \cite{EPR} and \cite{Bohmbook},
partially the result of degeneracies of state.

We first  describe the EPR  situation  using   SQM  absolute projective states, in order 
to examine its puzzling aspect. 

\subsection{The orthodox account of EPR.}

Two systems $\ss_1$ and $\ss_2$ interact and then are spatially
isolated from each other. 
We  use  parenthesized ligatures as variables for states
 of composite systems to keep track of which systems are involved at each stage.
 We denote the state after interaction, at $t=0$, of $\ss_1+\ss_2$ 
 by  $(\alpha\!\beta)_0$
and write this fact symbolically:
$$
(\ss_1+\ss_2)(0)=(\alpha\!\beta)_0.
$$
A measurement is going to be performed   on $\ss_2$ by an apparatus system $\ss_3$, 
which is initially in a state
$
\ss_3(0)=\gamma_0.
$
The state of $(\ss_1+\ss_2)+\ss_3$ is assumed to be, initially, 
$$
((\ss_1+\ss_2)+\ss_3)(0)=(\alpha\!\beta)_0\ot\gamma_0
$$
At the time $t=1$ of measurement and after, $\ss_2$ and $\ss_3$ are both spatially
separated from
$\ss_1,$ and so we can consider their compound state  as if they comprised an isolated
system. In polar form, we assume this state to be given by
  \begin{equation}\label{betameas}%еееееееееееееее
(\ss_2+\ss_3)(1) =\sum_i z_i\, \beta_i\ot\gamma_i
\end{equation}%еееееееееееееее
In other words, $\ss_3$ is  measuring  which $[\beta_i]$ is the state of $\ss_2.$ We also
have a polar decomposition
 \begin{equation}\label{eqre5new}%еееееееееееееее
(\alpha\!\beta\!\gamma):=
((\ss_1+\ss_2)+\ss_3))(1)=\sum_i y_i\,(\alpha\!\beta)_i\ot\gamma_i.
\end{equation}%еееееееееееееее
The parenthetical association of systems within $\ss=\ss_1+\ss_2+\ss_3$
 in the above equation serves only
as an additional mnemonic device to indicate which polar decomposition we are considering.
Indeed, we use below that $((\ss_1+\ss_2)+\ss_3))(1)=(\ss_1+(\ss_2+\ss_3))(1).$ 
While systems can be freely associated; the situation for their relative states is more
 complicated. \eqref{eqre5new} says that $\ss_3$ is (also) measuring which 
$[(\alpha\!\beta)_i]$ is the state of $\ss_1+\ss_2.$ Thus an outcome or 
reading  $[\gamma_i]$ for $\ss_3$ is synchronous with both $\ss_2$ being in state
$[\beta_i]$ and $\ss_1+\ss_2$ being in state
$[(\alpha\!\beta)_i)]$. It follows that $[(\alpha\!\beta)_i]$ has the form 
$[\alpha_i\ot\beta_i]$ for some $\alpha_i\in \bfS({\cal H}_1)$. For suppose 
$(\alpha\!\beta)_i$ has a general polar decomposition
$$
(\alpha\!\beta)_i=\sum_h \lam'_h\,\alpha'_h\ot\beta'_h.
$$ 
with non-zero $\lam'_h$. If there is more than one  non-zero $\lam'_h,$  say
for $h=1,2$, then at least one of $[\beta'_1],[\beta'_2]$, say $\beta'_1$ is unequal to $[\beta_i]$.
Thus there would be a positive probability that both $[\gamma_i]$ and $[\beta'_1]\ne [\beta_i]$
would  simultaneously occur, contradicting \eqref{betameas}.

{\bf According to the orthodox interpretation:} After a the measurement of $\ss_2$
 by $\ss_3,$  with $\ss_3$ ``reading'' $[\gamma_i]$  for some $i$,
the state of $\ss_2$ has collapsed to $[\beta_i]$, 
according to \eqref{betameas}. 
\nl {\it \hskip.5in Thus $\ss_1$ must be in state $[\alpha_i].$ } 
\nl The pu{zz}ling aspects are: 
\nl{\bf Z1}- that whereas
 $\ss_1$ was in no definite state after the interaction of $\ss_1$ and $\ss_2$ (since no 
observation
was made), $\ss_1$ suddenly enters the state $[\alpha_i]$ as a result of the
possibly distant 
interaction of $\ss_2$ and $\ss_3;$ 
\nl {\bf Z2}- that moreover, if $\ss_3$ were to be arranged differently,
 perhaps just rotated, then an entirely different $\alpha'_j$ would result, where 
$[\alpha'_j]$ need not even be among the $[\alpha_i]$. 

{This is indeed mysterious as long as one uses   absolute states.}

\subsection{The IQM resolution.}

{\bf Using the polar and conditional spectral states of IQM:} Now $(\ss_1+\ss_2+\ss_3,\ss_1)$
has a polar state
$((\alpha\!\beta\!\gamma)(t),\bfq(t)) $  at each time $t$  corresponding to a 
polar decomposition
        \begin{equation}\label{epreq6}%еееееееееееееее
       (\alpha\!\beta\!\gamma)=
       \sum_k \lam_k\,\alpha^1_k\ot(\beta\!\gamma)_k.
       \end{equation}%еееееееееееееее
from which the spectral projection $[\alpha^1]$ assigned to $\ss_1$ is derivable.
Indeed, $[\alpha^1]$ is the conditional state 
$[\alpha^1_{k(1)}]:=[((\alpha\!\beta\!\gamma)(1),\bfq(1))|\ss_1].$ 

Projection on  $[\alpha^1_{k(1)}]$ is an eigenprojection of 
 $A(t):={\rm Tred_1}\(P_{\C(\alpha\!\beta\!\gamma)(t)}(t)\),$
 at $t=1$. 
 Because of the assumed isolation of $\ss_1$ from $\ss_2+\ss_3$ for $t>0$, we can take
the total  \ham\ for $\(\ss_1+(\ss_2+\ss_3)\)$ to be of the form 
$H_1\ot\bfI_2+\bfI_1\ot H_{23}.$ 
 Thus we have
$$A(t) =e^{-i t H_1}A(0)e^{i t H_1}
=\sum_k |\lam_k|^2 e^{-i t H_1}P_{\alpha^1_k} e^{i t H_1}.
$$
It follows that for $t>0$ there exists $k(t)$ so that 
the conditional spectral state of $\ss_1$ relative to $\ss_1+\ss_2+\ss_3$ 
at  time $t$ is $[\alpha^1_{k(t)} (t)]=
 [e^{-i t H_1}\alpha^1_{k(t)}(0)].$

We now combine this  essentially standard analysis with the underlying hidden phases, which
according to IQM, determine the  particular value $k(1)$. Actually $k(t)$ is determined 
by IQM by  ${\bf\lam_k}\in{\mf p}_{k(t)}.$ 
It follows from 
Section~\ref{noint} that there is no jumping since there is no interaction
\ham. Thus $k(t)$ is a constant for $t>0.$ In particular it does not depend on
the (later, distant) interaction between $\ss_2$ and $\ss_3$ and certainly not on
 the particular outcome $[\beta_i]$ previously found for
 the spectral 
state of $\ss_2,$ or even whether any such measurement  is made.

Nevertheless, the  state $[\alpha_i]$, obtained as in $(*)$ above, 
 represents ``an element of reality'' since
we know empirically that if we test $\ss_1$ for the property $P_{\alpha_i}$ (after having 
obtained the ``reading'' $[\gamma_i]$ before), we will definitely obtain an affirmative answer. 

The IQM account is that the conditional spectral state of $\ss_1+\ss_2$ relative to 
$(\ss_1+\ss_2)+\ss_3$
at $t=1$ is given, for some $\bfq'$ by
$$
[((\alpha\!\beta\!\gamma),\bfq')|\ss_1+\ss_2]=[(\alpha\!\beta)_i]=[\alpha_i\ot\beta_i]
$$
from \eqref{eqre5new} and the ensuing discussion.
{ The index $i$ is determined by the toroidal part ${\mf p}_i$ which}
 {contains $\bfq'$ at $t=1.$}
 Then the conditional  spectral state of $\ss_1$ relative to
$\ss_1+\ss_2$ 
is derivable from  a polar decomposition of $\alpha_i\ot\beta_i,$ which is already in polar form.
 Hence it is just $P_{\alpha_i},$ \ie 
this iterated conditional spectral state is $[\alpha_i].$ So this ``element of reality'' is
faithfully represented in IQM as the iterated  conditional spectral state, 
 \begin{equation}\label{iteratecond}%еееееееееееееее
[\alpha_i]=[((\alpha\!\beta)_i,\bfq'')|\ss_1], {\rm\ where\ }
[(\alpha\!\beta)_i]=[((\alpha\!\beta\!\gamma),\bfq')|\ss_1+\ss_2]
 \end{equation}%еееееееееееееее
 for some $\bfq''$. 
{ This is an example of the compounding of perspectives discussed on Section~\ref{compound}.}
At the same time, the conditional spectral state of $\ss_1$
relative to
$\ss_1+\ss_2+\ss_3$ is
$[\alpha^1_{k(1)}(1)].$

Returning to the puzzling aspects mentioned above, we see that: 
\nl {\bf Z1} is obviated in IQM by $\ss_1$ always having a conditional state \wrt\  any
supersystem;
\nl {\bf Z2} is clarified in IQM by the  fact that these conditional states of
  $\ss_1$ \wrt \ss, actually do depend upon $\ss,$ which
may have distant parts.

This analysis of EPR enables us to answer a possible objection to our interpretation.
It has been argued in the literature that the spectral resolution of the density operator
 has no privileged role among the different decompositions of the operator into
convex combinations of one-dimensional projections. Fano \cite{Fano} in particular
argued that no such particular convex combination is ``intrinsically relevant apart 
from analytic convenience \dots''.

In Fano's striking example, a beam of atoms filtered to have total angular momentum
$J=1$ and $z$-component $J_z=0$ emits photons, after which the $z$-component of
angular momentum $J_z$ of the atom is measured by a Stern-Gerlach apparatus.
In the example, the eigenstates of the measured atoms are confined to a frame of 
eigenprojections of   $J_z$ in a
three-dimensional Hilbert space. The corresponding eigenvalues are $0,\pm1.$
According to which member $[\beta_i]$ of the frame occurs in the measurement,
 the polarization of the photon will be in a state $[\alpha_i]$, after (approximately) 
confining the photons to a given direction. Then the  $2\times2$ density matrix of the
 polarization  is, as Fano argues,  most naturally represented as a convex
combination of the three eigenprojections $P_{\alpha_i}.$  This is  not the
spectral decomposition (which has at most two summands), which would be 
relevant in an ordinary direct measurement of the polarization of the photon.

Now Fano's example is clearly  an experiment of the EPR type which we 
analyzed above, if we take $\ss_1$ to be the photon, $\ss_2$ the atom,
 and $\ss_3$ the Stern-Gerlach apparatus. The mixed state of $\ss_1$ is given in
our notation by the density matrix ${\rm Tred}_1(\alpha\!\beta\!\gamma)$, 
which has the non-spectral decomposition { $\sum^3_{i=1}|y_i|^2P_{\alpha_i}$. }
In our analysis, however, the rays $[\alpha_i]$ of the photon {\it do} arise 
as (iterated) spectral states: because the photon state is measured indirectly, 
via the spin of the atom with which it has previously interacted, the  ray 
$[\alpha_i]$  occurs as an iterated spectral state. It is actually,
in the terminology introduced in Section~\ref{relstates},
a conditional spectral state of $\ss_1$ \wrt a polar state of 
$\ss_1+\ss_2$ which itself represents a
 conditional spectral state of $\ss_1+\ss_2$ \wrt a polar state of 
$\ss_1+\ss_2+\ss_3,$ as is made explicit in \eqref{iteratecond}.

This is a case of a quite general situation. It is easy to prove that  
every convex decomposition of a density operator into one-dimensional 
projections  arises as a two-step  iterated spectral decomposition, as above.

It requires a shift in thinking to become reconciled to the necessary
{\bf perspectivity} of states. An analogy from 
special relativity may be helpful. Consider two  particles $\ss_1$ and
$\ss_2$ colliding and the subsequent distant collision of $\ss_2$ with a particle $\ss_3.$
 The  center-of-mass  inertial frame $F$ of $\ss_2+\ss_3$ does not
change as a result of the collision of $\ss_2$ and $\ss_3$; hence, any
characteristic property of $\ss_1$ 
{(corresponding to the state $[\alpha^1_{k(1)}]$)} such as mass is unchanged by this collision from
the point of view of an observer with frame $F.$  On the other hand, the inertial 
frame of $\ss_2$
is changed by this collision. Thus the mass of $\ss_1$ from the
perspective of $\ss_2$ suddenly changes {(corresponding to the state} $[\alpha_i]$).

 It is difficult to drop the idea
of the {absoluteness} 
of some quantities, such as mass, time and the state of a system.
The situation in QM is  actually more serious than in special relativity where
we have the possibility of passing from a description in
 one inertial frame to  that of another by means of 
the Poincar\'e group. The intransitivity of spectral projections prohibits so
 neat an extrication from  the 
%``Jewel Net of Indra'',  
net of compounded  perspectives.
This is also the reason that SQM is more tractable than IQM: reduced traces
{\it are} transitive. But if one is interested in modeling individual systems, then this extra
complication seems necessary. 
  \begin{quotation}Everything should be made as simple as possible, 
  but not simpler.\emph { Einstein}
  \end{quotation}
%  %  Iooss and Joseph

\newpage

{\centerline{\bf \huge APPENDICES}}

\appendix
\section{Polar Decompositions, Reduced Traces and Moment
Maps}\label{AppPolar}
\setcounter{equation}{0}

In this section we recall some basic facts about the polar decompositions
including the relation with the reduced trace. Then we show, that the reduced
trace is really a  moment map. The main fact we need is that the
natural  moment map associated with the action
of $\U(n_1)\times\U(n_2)$ on $\bfP(\C^{n_1\times n_2})$ is given by a pair of
reduced traces. 

\subsection{Polar decompositions and reduced traces}
\begin{lem}\label{polem1}
Let ${\cal H}_{1}$ and   ${\cal H}_2$ be Hilbert spaces. Let
$\gg \in  {\cal H}_{1}\ot {\cal H}_{2}.$ Then $\gg$ has a
 {\bf polar decomposition}, \ie there exist scalars $\lam_k$ and
orthonormal $\phi_k\in{\cal H}_1$ and orthonormal $\psi_k\in{\cal H}_2$
so that

\begin{equation}\label{polardecomp1}%еееееееееееееее
\gg=\sum_k \lam_k\,\phi_k\ot\psi_k
\end{equation}%еееееееееееееее
The $r_k:=|\lam_k|>0$ are unique. If they are distinct then
the
$\lam_k, \phi_k,\psi_k$ are all unique
up to phase factors. We sometimes also write $\gg_k=\phi_k\ot\psi_k.$ Then
\begin{equation}\label{polardecomp2}%еееееееееееееее
\gg=\sum_k \lam_k\,\gg_k.
\end{equation}%еееееееееееееее
Again, if the  $r_k$ are distinct, then the $\gg_k$ are unique up to phase
factors. If either the set of  $\lam_k $ or the set of $\gg_k$ is specified,
then the other set is uniquely determined.
We sometimes refer to the $\gg_k$ as {\bf bi-orthonormal}.
\end{lem}

The main content of this lemma is
\begin{lem}\label{polem2}
Let ${\cal H}_{1}$ and   ${\cal H}_2$ be Hilbert spaces. Let
$\gg \in  {\cal H}_{1}\ot {\cal H}_{2}.$ Then there exist  non-negative scalars
$\lam_k\in\R $ and orthonormal $\phi_k\in{\cal H}_1$ and orthonormal $\psi_k\in{\cal H}_2$
so that
\begin{equation}\label{polardecomp11}%еееееееееееееее
\gg=\sum_k \lam_k\, \phi_k\ot\psi_k.
\end{equation}%еееееееееееееее
\end{lem}
{\bf Proof.} This is well-known and can be found, \eg,  in \cite{Kochen},
but we include a proof which introduces  some maps used later.
 The map
 \begin{equation}\label{eq:tildeGamma}%еееееееееееееее
 {\cal H}_{1}\times {\cal H}_{2}\ra {\rm \overline{Hom}}({\cal H}_{2}, {\cal H}_{1}),
{\rm given\ by\ }(\phi,\psi)\mapsto[ \psi'\mapsto\<\psi',\psi\>\phi]
 \end{equation}%еееееееееееееее
 is $\C$-bilinear to the 
conjugate-linear maps. It therefore induces a map
 \begin{equation}\label{eq:2tildeGamma}%еееееееееееееее
 \widetilde{   } :{\cal H}_{1}\ot {\cal H}_{2}\ra {\rm
 \overline{Hom}}({\cal H}_{2}, {\cal H}_{1}),\;
 \phi\ot\psi{   \mapsto}[ \psi'\mapsto\<\psi',\psi\>\phi]
 \end{equation}%еееееееееееееее
 
 In particular, we have the conjugate-linear operator  $\widetilde
 \Gamma:{\cal H}_{2}\ra
 {\cal H}_{1}$  associated to any  $\Gamma\in {\cal H}_{1}\ot {\cal H}_{2}.$ 
 We also have a conjugate-linear bijective isometry $\dagger: {\cal H}^*_{2}\ra {\cal H}_{2},$
 whose inverse is explicitly induced by $\psi\mapsto [\psi'\mapsto\<\psi',\psi\> ].$
 Any bounded linear operator has a polar decomposition \cite[Theorem VI.10, p.197]{Reed1}.
 We can apply this fact to 
  $\tilde \Gamma\circ\dagger: {\cal H}^*_{2}\ra {\cal H}_{1}. $
 It follows that there exists a unique
 positive self-adjoint operator $V:{\cal H}_{2}\ra {\cal H}_{2}$ and 
 a partial conjugate-linear isometry
 $U:{\cal H}_{2}\ra {\cal H}_{1}$ with
 domain the image of $V$ so that 
 $\tilde \Gamma=UV. $
 \begin{equation}\label{eq:polar}%еееееееееееееее
 {\wt \Gamma}=UV\, , V=\sqrt{ ({\tilde \Gamma}^*\, \tilde\Gamma)}\, 
 , {\rm\ ker\ } U = {\rm\ ker\ } V = {\rm\ ker\ } {\wt \Gamma}\, , 
  {\rm\ ran\ }U = {\rm\ ran\ }\tilde \Gamma
 \end{equation}%еееееееееееееее
  $V$ is a positive self-adjoint Hilbert-Schmidt operator. It has an explicit
 spectral decomposition of the form 
 \begin{equation}\label{eq:spectaldecomp}%еееееееееееееее
  V=  \sum_j \lam_j  P_j\, ,\ \lam_j >0\, ,\  \sum_j \lam^2_j =1\, ,
 \end{equation}%еееееееееееееее
 where the $P_j$ are finite-dimensional mutually orthogonal projections.
 We are mostly interested in the case where  ${\rm\ dim\ }P_j=1$ for all $j.$ 
 Then there exist  $\psi_j\in {\cal H}_{2}\, , |\psi_j|=1$ so that
 \begin{equation}\label{eq:spectaldecomp2}%еееееееееееееее
  V(\psi)= \sum_j \lam_j \<\psi_j,\psi\>\psi_j ,\ \lam_j  >0\, ,\
   \sum_j \lam^2_j  =1\, 
 \end{equation}%еееееееееееееее.
 The 1-dimensional projections $P_j$ in \eqref{eq:spectaldecomp}  determine the 
 $\psi_j$  only  up to  phase factors which however do not change the value of V
 in the last formula. From it and \eqref{eq:polar}
 \begin{equation}\label{eq:uu}%еееееееееееееее
 {\wt \Gamma}(\psi_j )=UV(\psi_j )=U(\lam_j  \psi_j )= 
 \lam_j  \phi_j , {\rm\ where\ } \phi_j \,\defeq\, U(\psi_j) 
 \end{equation}%еееееееееееееее
 \begin{equation}\label{2tildeGamma}%еееееееееееееее
 \therefore\ \ {\wt \Gamma} (\psi)=  \sum_j \lam_j\<\psi,\psi_j\>\phi_j
 \end{equation}%еееееееееееееее
 since the operator defined by the RHS of this formula is conjugate linear and agrees with
 $ {\wt \Gamma} $ on the orthogonal complement of the kernel of  $ {\wt \Gamma}. $
 It follows now from the definition of $\ \wt{  }\ $ in
\eqref{eq:2tildeGamma}
 and its injectivity 
 \begin{equation}\label{eq:3tildeGamma}%еееееееееееееее
  \Gamma =  \sum_j \lam_j\,\phi_j\ot \psi_j
 \end{equation}%еееееееееееееее
which is \eqref{polardecomp11}. 
 {\bf This is the polar decomposition of $\Gamma.$}
 
 Moreover, if we change
  \begin{equation}\label{eq:4tildeGamma}%еееееееееееееее
 \psi_j\ra\tau_j\psi_j , {\rm\ then\ }
 \phi_j=U(\psi_j)\ra U(\tau_j\psi_j)=\overline{\tau_j}U(\psi_j)
 =\overline{\tau_j}\phi_j\,,
 \end{equation}\label{eq:5tildeGamma}%еееееееееееееее
 so that the only remaining indeterminacies are   reciprocal phase factors  in
 each pair $\phi_j,\psi_j .$
\nl$\qed$
\begin{lem}\label{lambdaEq}
 Let $ \Gamma(t) $
evolve according to the \SC\ equation (with $\hbar=1$) with \ham\ $H.$
 Let there be given a smooth curve of polar decompositions:
 \begin{equation}\label{smoothpd}%еееееееееееееее
  \Gamma(t) =  \sum_j \lam_j(t)\,\phi_j(t)\ot\psi_j(t)=
\sum_j \lam_j(t)\,\gg_j(t)
 \end{equation}%еееееееееееееее
Set $H_{jj,kk}=\<\gg_j,H\gg_k\>$ and $r_j=|\lam_j|.$ 
If each $\lam_j$ never vanishes, then
 \begin{equation}\label{smoothpdpos}%еееееееееееееее
\dot r_j=\sum_k \Im (H_{jj,kk})r_k.
 \end{equation}%еееееееееееееее
 If  the $\lam_j$ are all real
then
 \begin{equation}\label{smoothpdreal}%еееееееееееееее
\dot\lam_j=\sum_k \Im (H_{jj,kk})\lam_k.
 \end{equation}%еееееееееееееее
If the $\phi_k,\psi_k$ are \hor\ \wrt\ the canonical connection, then 
 \begin{equation}\label{smoothpdcan}%еееееееееееееее
\dot\lam_j=-i\sum_k H_{jj,kk}\lam_k.
 \end{equation}%еееееееееееееее
\end{lem}
{\bf Proof.}
The first assertion follows from the second, since if the 
$r_j=|\lam_j|$ never vanish, we can replace the $\lam_j$  
by the $r_j$ and smoothly compensate with a phase modification of the $\phi_j.$
 Differentiating \eqref{smoothpd},  we get 
\begin{equation}\label{diffPolar+}%еееееееееееееее
{-i} H \Gamma= \dot\Gamma=\sum_j \big(\dot\lam_j\,\phi_j\ot\psi_j
+
\lam_j\,\dot\phi_j\ot\psi_j + \lam_j\phi_j\ot\dot\psi_j\big),
\end{equation}%еееееееееееееее
Using inner products which are conjugate linear in the first variable, and making
use of the bi-orthonormality of the $\phi_j\ot\psi_j ,$ we
find by taking inner products of both sides with $\phi_j\ot\psi_j $
\begin{equation}\label{lambda}%еееееееееееееее
\< \phi_j\ot\psi_j,\dot\Gamma\> 
=\<\phi_j\ot\psi_j,{-i } H \Gamma\> 
            ={-i }\<\phi_j\ot\psi_j,H \Gamma\>
=-i\sum_k H_{jj,kk}\lam_k
\end{equation}%еееееееееееееее
\begin{equation}\label{lambda4}%еееееееееееееее
=\dot\lam_j+\lam_j\< \phi_j\ot\psi_j
,\dot\phi_j\ot\psi_j\>+\lam_j\< \phi_j\ot\psi_j
,\phi_j\ot\dot\psi_j\>
\end{equation}%еееееееееееееее
Since $||\phi_j||=||\psi_j||=1, \< \phi_j,\dot\phi_j\> , \< \psi_j,\dot\psi_j\> $
 are purely imaginary and thus so are the last two
terms of  \eqref{lambda4}.
If the $\lam_j$ are real, we deduce \eqref{smoothpdreal}.
 If the $\phi_k,\psi_k$ are \hor\ \wrt\ the canonical connection,
the last two
terms  of  \eqref{lambda4} vanish and  we deduce
\eqref{smoothpdcan}.
\nl$\qed$

\subsubsection{Quaternionic version}

\begin{quotation}
Quaternions came from Hamilton after his really good work had
         been done; and though beautifully ingenious, have been an unmixed
         evil to those who have touched them in any way. \par
     \rightline{\emph{Lord Kelvin} }  
\end{quotation}
 Suppose that the ${\cal H}_1$ and ${\cal H}_2$ are left Hilbert spaces
over the quaternions $\Q$. Then
 for $q\in \Q$ and $v\in {\cal H}_j,\; q\cdot v$ is defined  and satisfies
the standard rules, as in \cite[Vol.I]{Vara}. We can also regard them 
as right $\Q-$Hilbert spaces by using the definition $v\cdot q= \v{q}\cdot v.$
Of course, this works from right to left as well. 
This permits the formation of the tensor product  ${\cal H}_1\ot_\Q{\cal H}_2$
by using the rule $v q\ot_\Q w=v\ot_\Q q w.$

A $\Q$-Hilbert space ${\cal H}$  can be regarded as a $\C$-Hilbert space together with 
a  {\bf $\Q$-structure map} $j_{\cal H}$: this means $j_{\cal H}$ is $\R-$ isometric, 
conjugate-linear, and satisfies
$j_{\cal H}^2=-1.$ Here conjugation means for all $a,b,c,d\in\R,\;
\v{a+b\bfi+c\bfj +d\bfk}=a-b\bfi-c\bfj -d\bfk.$
Then $j_{\cal H}$ is just left multiplication by $\bfj.$ Also
$j:=j_{\cal H}$ is an isometry: $\<jv,jw\>=\<v,w\> $

\begin{lem}\label{qpolem1}
Lemma~\ref{polem2} holds for $\Q$-Hilbert spaces.
\end{lem}%еееееееееееееееееееее
{\bf Proofs.} Making the necessary changes of $\C-$(bi)linear to left($+$right bi)linear, 
{\it etc.},
including the quaternionic version of the polar decomposition, whose proof also
follows the complex version, the  proof of Lemma~\ref{polem2} works.

Another proof uses the following fact:
\nl We can identify ${\cal H}_1\ot_\Q{\cal H}_2$ with the $(-1)$-eigenspace of the operator
$j_{\cal H}=j_{{\cal H}_1}\ot_\C j_{{\cal H}_2}$ acting on ${\cal H}_1\ot_\C{\cal H}_2.$
This follows from \cite[p.30]{Adams}.

Now let $\gg\in {\cal H}_1\ot_\Q{\cal H}_2\subset{\cal H}_1\ot_\C{\cal H}_2$. Then
$\gg$ has a positive polar decomposition:
$\gg=\sum_k r_k \gg_k=\sum_k r_k \phi_k\ot_\C \psi_k.$
We first assume this polar decomposition has distinct positive $r_k.$ 
Applying $j_{\cal H}=j_{{\cal H}_1}\ot_\C j_{{\cal H}_2},$  we get 
$-\gg=\sum_k r_k(j_{{\cal H}_1}\ot_\C j_{{\cal H}_2})(\gg_k).$ Since
the $j_{{\cal H}_h}$ are isometric, the $j_{{\cal H}_1}( \phi_k)$ are orthonormal, 
as are the $ j_{{\cal H}_2}(\psi_k).$ Thus the $\gg_k':=(j_{{\cal H}_1}\ot_\C j_{{\cal H}_2})\gg_k$
 are bi-orthonormal. By the uniqueness of positive polar decompositions 
with distinct positive $r_k$, we must 
have $\gg_k'=-\gg_k.$ This means $\gg_k\in   {\cal H}_1\ot_\Q{\cal H}_2.$ This completes
the proof in the case of distinct positive $r_k$. The general case follows by taking limits.

\nin$\qed$
\begin{lem}\label{polem15}
Let ${\cal H}_{1}$ and   ${\cal H}_2$ be $\Q-$Hilbert spaces. Let
$\gg \in  {\cal H}_{1}\ot {\cal H}_{2}.$ Then $\gg$ has a
 {\bf polar decomposition}, \ie there exist scalars $\lam_k\in\Q$ and
orthonormal $\phi_k\in{\cal H}_1$ and orthonormal $\psi_k\in{\cal H}_2$
so that

\begin{equation}\label{polardecomp13}%еееееееееееееее
\gg=\sum_k \lam_k\,\phi_k\ot\psi_k
\end{equation}%еееееееееееееее
The $r_k:=|\lam_k|>0$ are unique. If they are distinct then
the
$\lam_k, \phi_k,\psi_k$ are all unique
up to $\Q-$phase factors, \ie unit quaternions. We sometimes also write
$\gg_k=\phi_k\ot\psi_k.$ Then
\begin{equation}\label{polardecomp14}%еееееееееееееее
\gg=\sum_k \lam_k\,\gg_k.
\end{equation}%еееееееееееееее
Again, if the  $r_k$ are distinct, then the $\gg_k$ are unique up to $\Q-$phase
factors. If either the set of  $\lam_k $ or the set of $\gg_k$ is specified,
then the other set is uniquely determined.
\end{lem}
\nin$\qed$

\nin{\bf Remark.} This allows the extension of our work to $\Q-$Hilbert spaces.
The $\Q-$version of the polar bundles
 will thereby have $\Q-$toroidal fibers
which are isometric to $\prod_k\S\U(2)(r_k),$ where
$\S\U(2)(r_k)$ is $\S\U(2)$ with
its invariant Riemannian metric normalized to give its
maximal tori (which look like $\bfS^1)$ total 
arclength $2\pi r_k.$ 

For now, we follow the advice implicit in Lord Kelvin's opinion, and avoid quaternions elsewhere 
in this paper.  
We remark for later work, however, that the crucial partitioning theorems extend to
$\prod_k\S\U(2)(r_k),$ because of  functoriality and the conjugacy of
 maximal tori.

\subsection{Reduced traces.}\label{tred}

A good reference for reduced traces is \cite{Davies}.
\nl{\bf Definition.} The reduced, or partial, trace,  
$\rho_1\deq  {\rm Tred}_{{\cal H}_{1}}(\rho)
:= {\rm Tred}_{1}(\rho)$
of a trace class operator $\rho$ in ${\rm Hom}({\cal H}_{1}\ot 
{\cal H}_{2},{\cal H}_{1}\ot {\cal H}_{2})$ is 
defined implicitly  via the sesquilinear form
\begin{equation}\label{eq:partTrace}%еееееееееееееее
\<{\rm Tred}_{1}(\rho)(\phi),\phi'\>\defeq
\sum_k\<\rho(\phi\ot\psi'_k),\phi'\ot{\psi}'_k\> ,
\end{equation}%еееееееееееееее
where the $\psi'_k$ comprise any \ON\ basis for ${\cal H}_{2}.$

If we apply the partial trace to $ \rho:=\rho_\Gamma:=P_\Gamma=$
orthogonal projection on $\C\gg$  and
use the  polar decomposition of $\Gamma,$ \eqref{polardecomp1}, we get 
\begin{equation}\label{eq:partTrace23}%еееееееееееееее
\sum_k\<P_\Gamma(\phi\ot\psi_k),\phi'\ot{\psi}_k\>
=\sum_k|\lam_k|^2\ \overline{\<\phi_k,\phi \>}\<\phi_k,\phi'\>.
\end{equation}%еееееееееееееее
Thus
\begin{equation}%еееееееееееееее
\<{\rm Tred}_{1}(\rho)(\phi),\phi\>
=\sum_k|\lam_k|^2\ |{\<\phi_k,\phi \>}|^2.
\end{equation}%еееееееееееееее
Set
\begin{equation}\label{eq:dnesityOp}%еееееееееееееее
  \rho_{1}\,\defeq\,\sum_j |\lam_j|^2 P_{\phi_j} ,
\end{equation}%еееееееееееееее
Then we see that
\begin{equation}\label{eq:2dnesityOp}%еееееееееееееее
  \rho_{1}={\rm Tred}_{1}(\rho).
\end{equation}%еееееееееееееее
So, explicitly, we have the following relation between polar decompositions
and reduced traces.
\begin{prop}\label{polar-Tr1}%% еееееPROPеееееееее
If the $\phi_j\ot \psi_j $ are bi-orthonormal then
 \begin{equation}\label{eq:polar-Tr1}%еееееееееееееее 
{\rm Tred}_{1}(P_{\sum_j \lam_j\,\phi_j\ot \psi_j})
=\sum_j |\lam_j|^2 P_{\phi_j}.
\end{equation}%еееееееееееееее
\end{prop}
{\bf Examples.} 1)
$ {\rm Tred}_{1}(P_{\phi_j\ot \psi_j})
= P_{\phi_j}.$
\nl 2) If ${\cal H}_1 = {\cal H}_2$ and $\phi_j$ is ON then
\begin{equation}\label{eq:adjTr1}%еееееееееееееее 
 {\rm Tred}_{1}(P_{{1\over\sqrt{2}}(\phi_1\ot \phi_2 \pm\phi_2\ot
\phi_1}) = {1\over 2}P_{\phi_1}+ {1\over 2}P_{\phi_2}.
\end{equation}%еееееееееееееее

 Let ${\rm Trace}^{j}$ denote the ordinary trace  for ${\cal H}_j,$
and  ${\rm Trace}$ denote the ordinary  trace for ${\cal H}_1\ot{\cal H}_2.$
\begin{prop}\label{adjointTr1}%% еееееPROPеееееееее
If the $A\in {\rm Hom}({\cal H}_1 , {\cal H}_1)$ is
bounded, then
 \begin{equation}%еееееееееееееее 
 \rho_{1}={\rm Tred}_{1}(\rho)\Ra{\rm Trace}^{1}(\rho_1 A)=
{\rm Trace}^{1}({\rm Tred}_{1}(\rho) A)=
{\rm Trace}(\rho\cdot( A\ot {\bf I}_2)) .
\end{equation}%еееееееееееееее
In other words, the reduced trace map is adjoint 
to $A\ra A\ot {\bf I}_2.$
\end{prop}
\begin{proof} Let $\phi_j,\psi_k$ be bases for ${\cal H}_1,{\cal H}_2.$
\begin{equation}\label{eq:1adjpf}%еееееееееееееее 
{\rm Trace}^{1}(\rho_1 A)=\sum_j \<\phi_j, \rho_1 A \phi_j\>=
\sum_j \sum_k\<\phi_j\ot\psi_k, \rho ((A \phi_j)\ot\psi_k)\>=
\end{equation}%еееееееееееееее
\begin{equation}\label{eq:2adjpf}%еееееееееееееее 
\sum_j \sum_k\<\phi_j\ot\psi_k, (\rho \cdot( A\ot {\bf I}_2 ))(\phi_j\ot\psi_k)\>=
{\rm Trace}(\rho( A\ot {\bf I}_2)) .
\end{equation}%еееееееееееееее
\end{proof}

For any Hilbert space ${\cal H}$, we let ${\cal L}({\cal H})$ denote the $C^*$-algebra
 of  bounded linear operators on ${\cal H}.$
If ${\cal H}={\cal H}_1\ot{\cal H}_2$, then we regard  ${\cal L}({\cal H}_1)$ as
a subalgebra of ${\cal L}({\cal H})$ via the injection 
$$
{\cal L}({\cal H}_1)\hra {\cal L}({\cal H}) {\rm\ given\ by\ } L_1\mapsto L_1\ot\bfI_2.
$$
More generally, if we are just given two Hilbert spaces and a unital $C^*$-algebra 
homomorphism ${\cal L}({\cal H}_1)\hra {\cal L}({\cal H})$, then there exists a Hilbert 
space ${\cal H}_2$ so that the situation is as above. This follows from
 \cite[Th.5.40]{Douglas}. This is a familiar fact for finite matrix algebras to 
 which we now confine
 our attention.

Thus we have ${\cal H}={\cal H}_1\ot{\cal H}_2$ with $\dim {\cal H}_i=n_i< \infty.$
We will also make use of the Hilbert-Schmidt inner product on ${\cal L}({\cal H})$ 
given by $\<L,L'\>\deq {\rm Trace}(\v{L} L').$

\begin{prop}\label{Tr1Proj}%% еееееPROPеееееееее
$P:={1\over n_2}{\rm Tred}_{1}:{\cal L}({\cal H})\ra {\cal L}({\cal H}_1)$ is orthogonal projection
on  $ {\cal L}({\cal H}_1). $
 \end{prop}
{\bf Proof.} From \eqref{eq:partTrace}, it follows that $P$ is idempotent. 
To see that it is self-adjoint \wrt the \HS\ inner product, we observe from
Proposition~\ref{adjointTr1}
$$
{\rm Trace}(\v{P(L)}\cdot L')={1\over n_2}{\rm Trace}(\v{{\rm Tred}_{1}(L)}L')
={1\over n_2}{{\rm Trace}^1(\v{{\rm Tred}_{1}(L)}{\rm Tred}_{1}(L'))}
$$
$$
={1\over n_2}\v{{\rm Trace}^1(\v{{\rm Tred}_{1}(L')} {{\rm Tred}_{1}(L)})}
=\v{{\rm Trace}(\v{P(L')}L)}={\rm Trace}(\v{L}\cdot{P(L')}).
$$
$\qed$

\subsection{Reduced traces and  moment maps}

Let ${\cal H}$ be a Hilbert space. It  has a natural symplectic
structure given by 
$$\omega(X,Y)=\Im\<X,Y\>,$$ where $X,Y\in {\cal H}.$ Here ${\cal H}$ is identified, as a
real vector space, with its own  tangent space. 
%Actually, Marsden and Ratiu, p.97, 
%use $\omega(X,Y)=-2 \hbar\Im\<X,Y\>.$

Let $G$ be a subgroup of ${\U}({\cal H})$, and $\mf g$ its Lie algebra which
consists of those skew-Hermitian operators $\xi$ for which $\exp \xi\in G$. We assume
$\mf g$ is a norm-closed Lie subalgebra of  the skew-Hermitian operators
$\mf u({\cal H})\subset
{\rm Hom}({\cal H},{\cal H}).$ 
A moment(um) map $\mu:{\cal H}\ra \mf g^*$ is characterized by
\begin{equation}\label{amom}%еееееееееееееееееееееееееееееееее
\forall \xi\in \mathfrak g,\;\mu(X)(\xi) =
-{i\over 2} \<\xi(X), X\> .
\end{equation}%ееееееееееееееееее      еееееееееееееее
Now we can identify $\mf g^*$ with $\mf g$ by means of the pairing on $(\mf g\times\mf g)$
\[
(\xi,\eta)\ra {\rm Trace}( \xi^*\eta)
\]
provided $\mf g$ is contained in the Hilbert-Schmidt operators. This suggests that
the ``right' group of unitaries are those which are logarithmically Hilbert-Schmidt.
We have from
\cite[p.163]{McDuff}, in our notation,  the moment map is given by the formula:
\begin{equation}%еееееееееееееееееееееееееееееееее
{\cal H}\ni X\ra\mu(X) = {i\over 2}\<X,\cdot\>X \in  \mathfrak g\cong \mathfrak g^*.
\end{equation}%ееееееееееееееееее      еееееееееееееее
The projective version of  \eqref{amom}  is
\begin{equation}%еееееееееееееееееееееееееееееееее
\forall \xi\in \mathfrak g,\;\mu([X])(\xi) =
-{i\over 2} {\<\xi(X), X\> \over ||X||^2}.
\end{equation}%ееееееееееееееееее      еееееееееееееее
which is in agreement with \cite[page 335]{Marsden}.

Suppose now that  ${\cal H}={\cal H}_1\ot{\cal H}_2$ and 
\begin{equation}%еееееееееееееееееееееееееееееееее
G={\U}({\cal H}_1)\times {\U}({\cal H}_2)\Ra 
{\mathfrak g}=  {\mathfrak u}({\cal H}_1)\times {\mathfrak u}({\cal H}_2).
\end{equation}%ееееееееееееееееее      еееееееееееееее
We can take the pairing on $\mf g$ to be 
\begin{equation}%еееееееееееееееееееееееееееееееее
\<(\chi_1,\chi_2),(\xi_1,\xi_2)\>_{\mathfrak g}=
 {\rm Trace}^1\(\chi_1^*\xi_1\)+ {\rm Trace}^2\(\chi_2^*\xi_2\) 
\end{equation}%ееееееееееееееееее      еееееееееееееее
where the last two traces are the ordinary  traces for trace  class operators on 
$ {\cal H}_1 ,{\cal H}_2.$

Applying  \eqref{amom} to this case, we get
\begin{equation}%еееееееееееееееееееееееееееееееее
 \mu(X)(\xi_1,\xi_2) =- {i\over 2}  \<(\xi_1,\xi_2)(X), X\> 
 =- {i\over 2}  \<(\xi_1\ot{\bf I}_2+{\bf I}_1\ot\xi_2)(X), X\> 
\end{equation}%ееееееееееееееееее      еееееееееееееее
$$
 =- {i\over 2} \left\{ \<(\xi_1\ot{\bf I}_2)(X), X\>+
  \<({\bf I}_1\ot\xi_2)(X), X\>\right \}
$$
Let ${\rm Trace}$ denote the usual  trace on the trace class operators 
 on ${\cal H}$.
\begin{equation}\label{trace10}%еееееееееееееееееееееееееееееееее
\tf  \mu(X)(\xi_1,\xi_2) 
=- {i\over 2} \left\{  {\rm Trace}\Big((\<X,\cdot\> X)(\xi_1\ot{\bf I}_2)\Big)+
 {\rm Trace}\Big((\<X,\cdot\> X)({\bf I}_1\ot \xi_2)\Big)\right\}
\end{equation}%ееееееееееееееееее      еееееееееееееее
Then, from Proposition~\ref{adjointTr1}
\begin{equation}%еееееееееееееееееееееееееееееееее
 \mu(X)(\xi_1,\xi_2) =- {i\over 2} \left\{  {\rm Trace}^1(  {\rm Tred}_{1}(\<X,\cdot\> X)\xi_1)+
 {\rm Trace}^2(  {\rm Tred}_{2}((\<X,\cdot\> X)\xi_2))\right\}
\end{equation}%ееееееееееееееееее      еееееееееееееее
\begin{equation}%еееееееееееееееееееееееееееееееее
\tf  \mu(X)\Big(\xi_1,\xi_2) =- {i\over 2}  \< ( {\rm Tred}_{1}(\<X,\cdot\> X),
 {\rm Tred}_{2}(\<X,\cdot\> X)(\xi_1,\xi_2\>\Big)
\end{equation}%ееееееееееееееееее      еееееееееееееее
This establishes
\begin{prop}\label{momprop} %PPPPPPPPPPPPPPPPPPP
$\mu(X)=-{ i\over 2}\({\rm Tred}_1(\<X,\cdot\> X), {\rm Tred}_2(\<X,\cdot\> X)\).$  
 \end{prop}%ELELELELELELELELELELELELLE

\begin{cor}\label{rangeMu} %PPPPPPPPPPPPPPPPPPP
The range of $\mu$ consists of those pairs $\(-{ i\over 2}B_1,-{ i\over 2}B_2 \)$ of
trace class skew-Hermitian matrices
 with  $B_1,B_2$ positive operators with  the same
non-zero spectral components.
 \end{cor}%ELELELELELELELELELELELELLE
$ \qed$ 
\begin{cor}\label{rangeMu2} %PPPPPPPPPPPPPPPPPPP
Using the notation of the last corollary, assume the non-zero eigenvalues
$r_1^2,r^2,\cdots$ of $B_1$ are distinct.
  These are also the non-zero eigenvalues of $B_2.$
Let $\phi_1,\cdots,\phi_n$ (respectively $\psi_1,\cdots,\psi_n$) be  corresponding  unit
eigenvectors for
$B_1$ (respectively $B_2$).
Let ${\cal L}_{2\pi\bfr}$ denote the lattice generated by the $r_k\bfe_k$. There
is an isometry 
$\iota$  to  the right toroid with radii $ r_k,$  given by 
$$
\iota( \sum_\nu \lam_\nu\,\phi_\nu\ot\psi_\nu)= 
{\(r_\nu\arg( \lam_\nu)\)_\nu}\in \R^n\Big/{\cal L}_{2\pi\bfr} = \bfT_{2\pi\bfr}.
$$
Then $\mu^{-1}\(\(-{ i\over 2}B_1,-{ i\over 2}B_2 \)\)= \Big\{ \sum_\nu
\lam_\nu\,\phi_\nu\ot\psi_\nu \;{\Big|}\;  |\lam_\nu|=r_\nu\Big\}\ {\buildrel {\rm
\iota}\over\cong}\ \bfT_{2\pi\bfr}.$
 \end{cor}%ELELELELELELELELELELELELLE
{\bf Proof.} Define $ \Lambda(r e^{i\theta})=
 \wt{r\theta }\in \R\Big/2\pi r\Z$ if $r>0;$ $\Lambda$ is a well-defined
 isometry which identifies  the circle of radius $r$  centered
at the origin (which is naturally a principal homogeneous space for the group
$S^1$ rather then a group itself) with the group
  $\R\Big/2\pi r\Z.$

Then $\iota( \sum_\nu \lam_\nu\,\phi_\nu\ot\psi_\nu)= \prod_\nu 
 \Lambda( \lam_\nu),  $ if we identify 
$\prod_\nu \(\R\Big/2\pi r_\nu\Z \)\approx \bfT_{2\pi \bfr}.$

The last statement follows from the
polar decomposition after  extending the 
$\phi_1,\phi_2,\cdots$ to an orthonormal basis for ${\cal H}_1.$

\nin $ \qed$ 

Recall that $\wt \gg$ was defined in the proof of Lemma~\ref{polem2}.
\begin{prop}\label{momFibres} %PPPPPPPPPPPPPPPPPPP
The fiber ${\cal C}(\Gamma)$  of the moment map above
$\mu(\Gamma)$  is isometric to a right toroid of dimension  ${\rm rank}(\wt{\Gamma}).$
 \end{prop}%ELELELELELELELELELELELELLE
\nin $ \qed$ 

If $\Gamma=\sum^n_{\nu=1}
\lam_\nu\,\phi_\nu\ot\psi_\nu$ with $r_\nu = |\lam_\nu|$ distinct for
$\nu=1\d n\le \infty$ then ${\cal C}(\Gamma) $ is an $n-$dimensional toroid with
$n$ uniquely-defined (1-dimensional) foliations by circles. Namely, the $k-$th
circle $C_k(\Gamma')\subset {\cal C}(\Gamma) $  through $\Gamma'=\sum^n_{\nu=1}
\lam'_\nu\,\phi_\nu\ot\psi_\nu$ consists of the elements of the form
$\Gamma''=\sum^n_{\nu=1}
\lam''_\nu\,\phi_\nu\ot\psi_\nu,$ where $\lam''_\nu=\lam'_\nu$
if $\nu\ne k$ and $\lam''_k=e^{i\tau}\lam'_k, \tau\in\R.$ These circular foliations
are well-defined in the case where the   $r_\nu$ are distinct because then the 
$\C \phi_\nu\,,\C \psi_\nu$ and hence the $\C (\phi_\nu\ot \psi_\nu)$ are unique.

\begin{prop}\label{distinctive1} %PPPPPPPPPPPPPPPPPPP
For all   $\Gamma$ with $r_\nu = |\lam_\nu| $ distinct and each  $X\in{\cal C}(\Gamma),$ 
 the  ${\rm rank}(\wt{\Gamma})-$toroid  ${\cal C}(\Gamma) $ contains a 
uniquely defined bouquet
$\{C_k(X)| k=1\d n\} $ of circles through $X.$
 \end{prop}%ELELELELELELELELELELELELLE
\nin $ \qed$ 
 
\section{Perturbation Theory and Connections on the Hopf Bundles
}\label{dEigenVec}
\setcounter{equation}{0}
Let $\rho=\rho(t)$ be a smoothly varying curve  of compact self-adjoint operators,
with a non-degenerate smoothly varying normalized eigensystem consisting of
the eigenvalues 
$x_j=x_j(t)$ and the eigenprojections $P_j=P_{\phi_j(t)}$ near $t=0.$  We 
use below  that
$\|\dot \rho(t)\|$  is bounded. The application we have in mind, is where 
$\rho(t)$ is a curve of reduced  density operators. % For historical reasons, 
 In this section we use 
 Dirac's suggestive  bra-ket notation.
\begin{thm}
For $t$ sufficiently close to $0,$
\begin{equation}\label{eq:deigen0}%еееееееееееееее
\dot P_j=\sum_{k\ne j }\bigg(
 {\<\phi_{k}|\dot \rho|\phi_{j}\>\over x_j -x_{k}} |\phi_{k}\>\<\phi_j|
+ {\<\phi_{j}|\dot \rho|\phi_{k}\>\over x_j -x_{k}} |\phi_{j}\>\<\phi_k|\bigg).
\end{equation}%еееееееееееееее
\end{thm}
{\bf Proof.} Near $0$, we can write, with $\rho'$ bounded,
\begin{equation}\label{eq:deigen2}%еееееееееееееее
 P_j(t)= {-1\over 2\pi i}\oint_{C_j} (\rho-\zeta)^{-1}d\zeta
= {-1\over 2\pi i}\oint_{C_j} (\rho(0) +t\rho'-\zeta)^{-1}d\zeta,
\end{equation}%еееееееееееееее
where $C_j$ is a sufficiently small circle in $\C$ centered at $x_j(0),$ and $\rho'\ra 
\dot \rho(0)$ as $t\ra0.$
\begin{equation}\label{eq:deigen4}%еееееееееееееее
\therefore  P_j(t)=  {-1\over 2\pi i}\oint_{C_j}(\rho(0) -\zeta)^{-1}
 \bigg(1 +t\rho'(\rho(0)-\zeta)^{-1}\bigg)^{-1}d\zeta=
\end{equation}%еееееееееееееее
\begin{equation}\label{eq:deigen108}%еееееееееееееее
  {-1\over 2\pi i}\oint_{C_j}(\rho(0) -\zeta)^{-1}
 \bigg(1 +\sum_{k>0}(-t)^k\big(\rho'(\rho(0)-\zeta)^{-1}\big)^k\bigg)d\zeta;
\end{equation}%еееееееееееееее
the series converges by the boundedness of $\rho'.$
\begin{equation}\label{eq:deigen6}%еееееееееееееее
\tf P_j(t)\equiv {t\over 2\pi i}\oint_{C_j}(\rho(0) -\zeta)^{-1}
\rho'(\rho(0)-\zeta)^{-1}d\zeta\ {\rm\ mod\ } t^2
\end{equation}%еееееееееееееее
Letting $t\ra0,$ we get
\begin{equation}\label{eq:deigen7}%еееееееееееееее
 \dot P_j(0)= {1\over 2\pi i}\oint_{C_j}(\rho(0) -\zeta)^{-1}
\dot \rho(0) (\rho(0)-\zeta)^{-1}d\zeta
\end{equation}%еееееееееееееее
In order to evaluate the integral, we now deface the pristine beauty of this operator equation, by
evaluating matrix elements \wrt unit eigenvectors. 
\begin{equation}\label{eq:deigen8}%еееееееееееееее
 \therefore\<\phi_k(0)|\dot P_j(0)|\phi_h(0)\>= {1\over 2\pi i}\oint_{C_j}\<\phi_k(0)|(\rho(0) -\zeta)^{-1}
\dot \rho(0) (\rho(0)-\zeta)^{-1}|\phi_h(0)\>d\zeta=
\end{equation}%еееееееееееееее
\begin{equation}\label{eq:deigen9}%еееееееееееееее
 {1\over 2\pi i}\oint_{C_j}\<(\rho(0) -{\bar\zeta})^{-1}\phi_k(0)|
\dot \rho(0) |(\rho(0)-\zeta)^{-1}\phi_j(0)\>d\zeta=
\end{equation}%еееееееееееееее
\begin{equation}\label{eq:deigen10}%еееееееееееееее
 {1\over 2\pi i}\oint_{C_j}\<(x_k(0) -{\bar\zeta})^{-1}\phi_k(0)|
\dot \rho(0) |(x_h(0)-\zeta)^{-1}\phi_h(0)\>d\zeta=
\end{equation}%еееееееееееееее
\begin{equation}\label{eq:deigen113}%еееееееееееееее
 {1\over 2\pi i}\oint_{C_j}{\<\phi_k(0)|
\dot \rho(0) |\phi_h(0)\>\over ( x_k(0)-\zeta)( x_h(0)-\zeta)}d\zeta
=\<\phi_k(0)|\dot \rho(0) |\phi_h(0)\>
{\rm\ res_{x_j}}{1\over ( x_k(0)-\zeta)( x_h(0)-\zeta)}
\end{equation}%еееееееееееееее
\begin{equation}\label{eq:deigen14}%еееееееееееееее
 \therefore\ \ \<\phi_k(0)|\dot P_j(0)|\phi_h(0)\>=
{\<\phi_k(0)|\dot \rho(0) |\phi_h(0)\>\over  x_h(0)- x_k(0)}\ 
\end{equation}%еееееееееееееее
if exactly one of $h,k$ equals $j.$ In the remaining equations of this proof, we omit the argument 0,
which can actually be any $t.$  We get 
\begin{equation}\label{eq:deigen15}%еееееееееееееее
\<\phi_k  |\dot P_j | \phi_h  \>=0
{\rm\ if\ } k,h\ne j{\rm\ or\ if\ } k=h= j.
\end{equation}%еееееееееееееее
\begin{equation}\label{eq:deigen16}%еееееееееееееее
P_j = |\phi_j\>\<\phi_j|\Ra \dot P_j= |\dot\phi_j\>\<\phi_j|+ |\phi_j\>\<\dot\phi_j|.
\end{equation}%еееееееееееееее
\begin{equation}\label{eq:deigen17}%еееееееееееееее
\tf  h\ne j\Ra\< \phi_h|\dot P_j|\phi_j\> =\<\phi_h|\dot\phi_j\>\ \&\ 
\< \phi_j|\dot P_j|\phi_h\> =\<\dot\phi_j|\phi_h\>.
\end{equation}%еееееееееееееее
\begin{equation}\label{eq:deigen18}%еееееееееееееее
\tf \ h\ne j\Ra \<\phi_h|\dot\phi_j\>=\< \phi_h|\dot P_j|\phi_j\> =
{\<\phi_h  |\dot  \rho|\phi_j  \>\over  x_j  - x_h  }.
\end{equation}%еееееееееееееее
{What about $\<\phi_j|\dot\phi_j\> ?$  From \eqref{eq:deigen16},\eqref{eq:deigen15}, 
we get only}
\begin{equation}\label{eq:deigen19}
0=\<\phi_j |\dot\phi_j\>\<\phi_j|\phi_j\>+ \<\phi_j|\phi_j\>\<\dot\phi_j|\phi_j\>
=\<\phi_j |\dot\phi_j\>+ \<\dot\phi_j|\phi_j\>=2{\rm\ Re}\<\phi_j |\dot\phi_j\>
\end{equation}
\begin{equation}\label{eq:deigen20}%еееееееееееееее
\tf\dot\phi_j= \sum_{k\ne j } {\<\phi_{k}|\dot \rho|\phi_{j}\>\over x_j -x_{k}}
\phi_{k} +\<\phi_j |\dot\phi_j\>\phi_j= \sum_{k\ne j } {\<\phi_{k}|\dot
\rho|\phi_{j}\>\over x_j -x_{k}}
\phi_{k} +i{\Im}\<\phi_j |\dot\phi_j\>\phi_j
\end{equation}%еееееееееееееее
Notice that changing the $\phi_{k}, (k\ne j),$ by a constant phase factor,  leaves the 
equation unchanged, 
while  if we so change $\phi_{j}$ all terms on both sides change by this same phase.

If  the  $\phi_{j}$ are horizontal \wrt the canonical connection $A^0$, then the last
term can be omitted and we get
\begin{equation}\label{eq:deigen21}%еееееееееееееее
\dot\phi_j= \sum_{k\ne j } {\<\phi_{k}|\dot \rho|\phi_{j}\>\over x_j -x_{k}} \phi_{k}
\end{equation}%еееееееееееееее
From \eqref{eq:deigen20},
\begin{equation}\label{eq:deigen223}%еееееееееееееее
\hskip-3in \dot P_j= |\dot\phi_j\>\<\phi_j|+ |\phi_j\>\<\dot\phi_j|=
\end{equation}%еееееееееееееее
\begin{equation}\label{eq:deigen233}%еееееееееееееее
 \sum_{k\ne j } {\<\phi_{k}|\dot \rho|\phi_{j}\>\over x_j -x_{k}} |\phi_{k}\>\<\phi_j|
+i{\Im}\<\phi_j |\dot\phi_j\>|\phi_j\>\<\phi_j| + {\rm\  HC},
\end{equation}%еееееееееееееее
where the last summand is the Hermitian Conjugate of the preceding summand.
The middle purely imaginary number times the  projection $P_j=|\phi_j\>\<\phi_j| $
disappears after being added to its HC and so we get:
\begin{equation}\label{eq:deigen243}%еееееееееееееее
\dot P_j=  \sum_{k\ne j } {\<\phi_{k}|\dot \rho|\phi_{j}\>\over x_j -x_{k}} |\phi_{k}\>\<\phi_j|
+{\rm\  HC}.
\end{equation}%еееееееееееееее
Each summand in this equation is independent of the phase of the $\phi_k.$
This equation does NOT depend on the horizontality of the $ \phi_{k}. $
\nl$\qed$ 

If we want a simple equation for the eigenvectors, rather than the  projections,
we have
\begin{cor} The  $ \phi_{j}(t) $ are $A^0-$horizontal \ifff
\begin{equation}\label{canconn}%еееееееееееееее
\dot\phi_j= \sum_{k\ne j } {\<\phi_{k}|\dot \rho|\phi_{j}\>\over x_j -x_{k}} \phi_{k}.
\end{equation}%еееееееееееееее
\end{cor}
$\qed$

This shows the relation between first order perturbation theory  
 and the canonical connection, which does
not seem to be explicitly mentioned in the literature 
 despite (or perhaps because of) its simplicity. We now extend this relation to the 
   dynamical  connection  defined in 
subsection~\ref{schrcon}.
\begin{cor} Let the $ \phi_{j}(t) $ br smoothly evolving eigenvectors of  $\rho(t)$ which
 is itself evolving  by
means of the 
\ham\ $H_1=H_1(t)$.  Then the   $ \phi_{j}(t) $  % in \eqref{eq:deigen20}
are $A^{H_1}-$horizontal  \ifff  they satisfy 
\begin{equation}\label{schrodconn26}%еееееееееееееее
\dot\phi_j=\sum_{k\ne j } {\<\phi_{k}|\dot \rho|\phi_{j}\>\over x_j -x_{k}} 
\phi_{k} -i\<\phi_j | H_1|\phi_j\>\phi_j .
\end{equation}%еееееееееееееее
\end{cor}
{\bf Proof.} By Lemma~\ref{lem:contact}, the   $ \phi_{j}(t) $   are
%\begin{cor}$\rho(t)$ is evolving  by means of the 
%\ham\ $H_1=H_1(t)$ \ifff the   $ \phi_{j}(t) $  in \eqref{eq:deigen20}
%are $A^{H_1}-$horizontal.
%\end{cor}
%{\bf Proof.} If the   $ \phi_{j}(t) $  in \eqref{eq:deigen20}
are $A^{H_1}-$horizontal \ifff
\begin{equation}%еееееееееееееее
\<\phi_j|\dot\phi_j\>=
-i\<\phi_j | H_1|\phi_j\>.
\end{equation}%еееееееееееееее
Thus the result follows from \eqref{eq:deigen20}.
%
%\begin{equation}%еееееееееееееее
%\tf\dot\phi_j
%=\sum_{k\ne j } 
%{\<\phi_{k}|\dot \rho|\phi_{j}\>\over x_j -x_{k}} \phi_{k}
%+i{\Im}\<\phi_j |-i H_1|\phi_j\>\phi_j .
%\end{equation}%еееееееееееееее
%\begin{equation}\label{schrodconn26}%еееееееееееееее
%\tf\dot\phi_j=
%\sum_{k\ne j } 
%{\<\phi_{k}|\dot \rho|\phi_{j}\>\over x_j -x_{k}} \phi_{k}
%-i\<\phi_j | H_1|\phi_j\>\phi_j .
%\end{equation}%еееееееееееееее
%It follows that the $\phi_j$ satisfy the condition  \eqref{9CanonicalConnS}
%to be  $A^{H_1}-$horizontal \wrt  the Hopf bundle
% $\bfS({\cal H}_1)\ra \bfP({\cal H}_1).$
%The converse is similar.
\nl$\qed$

\newpage
\section{Example: Two Spin 1/2 Systems}\label{AppHyperfine}
\setcounter{equation}{0}

We now give the details of  the  hyperfine splitting example.
 A nice treatment of the basics
appears in
\cite[Feynman,Vol.III, Chap.12]{FeynmanL}. The  polar decomposition in
this example was worked out in ~\cite{Kochen}. We shall use the
results  and notations of this last treatment, with some
minor modifications, to investigate the dynamical
behavior of the two subsystems. The definitions of most  of the many new variables 
introduced are collected in a glossary in Section~\ref{gloss}.

The Hamiltonian is given by
\begin{equation}%еееееееееееееее
\index{$H$}
H=\mu {\vec{\sigma}}\dot\otimes \vec{\sigma'}\ \defeq\
\mu\sigma_x\otimes \sigma'_x+\mu\sigma_y\otimes \sigma'_y+
\index{$\mu$}
\mu\sigma_z\otimes \sigma'_z
\end{equation}%еееееееееееееее
In terms of the standard notation of, \eg,  Feynman
\begin{equation}\label{eqH2}%еееееееееееееее
H=\mu\pmatrix{1&0&0&0\cr
 0&-1&2&0 \cr
 0&2&-1&0\cr
 0&0&0&1}
\end{equation}%еееееееееееееее
We find the eigenvalues $E_1= -3\mu, E_2=\mu$ with  respective eigenspaces with the
indicated eigenvectors
\begin{equation}\label{eq:Hnew1}%еееееееееееееее
< |+-\> - |-+\> > \leftrightarrow <\pmatrix{0\cr 1\cr -1\cr 0}>,
\end{equation}%еееееееееееееее
$$
 < |++\>,|+ -\>+ |-+\>, |--\> >\leftrightarrow   <\pmatrix{1\cr 0\cr 0\cr 0},
\pmatrix{0\cr 1\cr 1\cr 0},\pmatrix{0\cr 0\cr 0\cr 1}>
$$
The Schr\"{o}dinger equation (with $\hbar =1$)  is
\begin{equation}\label{eq:Hnew2}%еееееееееееееее
i \partial_t \Gamma= H \Gamma
\end{equation}%еееееееееееееее
and thus has for its general solution:
\begin{equation}\label{eq:Hnew3}%еееееееееееееее
\Gamma(t) =\pmatrix{d_1 e^{-i\mu t}\cr d_2 e^{-i\mu t}+d_3 e^{3 i\mu t}
\cr d_2 e^{-i\mu t}-d_3 e^{3 i\mu t}\cr d_4 e^{-i\mu t}}
\end{equation}%еееееееееееееее

We want to choose axes in physical space $\R^3$ so as to make 
$\gg(0)$ simple.  Since the
Hamiltonian is rotation invariant, we can keep its matrix \eqref{eqH2} while rotating
 the
$z$-axis to bisect the initial spin vectors (in $ \R^3  $) of the two spin systems. Let the    angle
between them be 
$2\theta.$
\n{$\theta$}
 Thus we have
%  \begin{equation}\label{H6}%еееееееееееееее
%  \Gamma(0)=\lam_+(0) \pmatrix{\cos{\theta\over2}\cr\cr \sin{\theta\over2}}\otimes
%\pmatrix{\cos{\theta\over2}\cr\cr -\sin{\theta\over2}}
%+\lam_-(0) \pmatrix{-\sin{\theta\over2}\cr\cr \cos{\theta\over2}}\otimes
%\pmatrix{\sin{\theta\over2}\cr\cr \cos{\theta\over2}}\deq
%\pmatrix{a_0\cr b_0\cr c_0\cr d_0}
%  \end{equation}%еееееееееееееее
  \begin{equation}\label{H6}%еееееееееееееее
  \Gamma(0)=\lam_+(0) \!\pmatrix{e^{i\varphi}\cos{\theta\over2}\cr\cr \sin{\theta\over2}}\!\ot\!
\pmatrix{e^{-i\varphi}\cos{\theta\over2}\cr\cr -\sin{\theta\over2}}
+\lam_-(0)\! \pmatrix{-e^{i\varphi}\sin{\theta\over2}\cr\cr \cos{\theta\over2}}\!\ot\!
\pmatrix{e^{-i\varphi}\sin{\theta\over2}\cr\cr \cos{\theta\over2}}=:
\pmatrix{a_0\cr b_0\cr c_0\cr d_0}
  \end{equation}%еееееееееееееее
for some real $\varphi$. To simplify  calculations, we restrict to the case where 
$\varphi=0.$ Then
  \begin{equation}\label{H7}%еееееееееееееее
\pmatrix{a_0\cr b_0\cr c_0\cr d_0}=
 \pmatrix{
\lam_+(0)\cos^2{\theta\over2}-\lam_-(0)\sin^2{\theta\over2}\cr
-(\lam_+(0)+\lam_-(0))\cos{\theta\over2}\sin{\theta\over2}\cr
(\lam_+(0)+\lam_-(0))\cos{\theta\over2}\sin{\theta\over2}\cr
-\lam_+(0)\sin^2{\theta\over2}+\lam_-(0)\cos^2{\theta\over2}
}
\n{$C$}
\n{$S$}
= \pmatrix{
{1\over2}(Cl+k)\cr
-{1\over2}Sl\cr
{1\over2}Sl\cr
{1\over2}(Cl-k)
},\;{\rm where}
  \end{equation}%еееееееееееееее
\begin{equation}\label{H8}%еееееееееееееее
 C:= \cos\theta, S:= \cos\theta, k:=  \lam_+(0)-\lam_-(0),
l:= \lam_+(0)+\lam_-(0).
\end{equation}%еееееееееееееее
Note that \eqref{H7} is a polar decomposition of $\gg(0).$ We are still free to rotate about
the $z-$axis. Later, we use this to make the $\lam_\pm(0)$ real.

From \eqref{H6},(\ref{H7}), we have
\begin{equation}\label{H9}%еееееееееееееее
\Gamma(t) =\pmatrix{a_0 e^{-i\mu t}\cr b_0 e^{3 i\mu t}
\cr -b_0 e^{3 i\mu t}\cr d_0 e^{-i\mu t}}
= \pmatrix{
{1\over2}(Cl+k)e^{-i\mu t}\cr
-{1\over2}Sle^{3i\mu t}\cr
{1\over2}Sle^{3i\mu t}\cr
{1\over2}(Cl-k)e^{-i\mu t}
}
\deq
\pmatrix{a\cr b\cr -b\cr d}.
\end{equation}%еееееееееееееее
We now calculate the polar decomposition of $\gg(t)$ via the prescription 
in Section~\ref{AppPolar}.
We use the  associated matrix operator $\wt {\gg(t)}.$

\begin{equation}\label{eq:Hnew8}%еееееееееееееее
\wt {\Gamma(t)} \deq 
\pmatrix{a & b
\cr -b & d}
\end{equation}%еееееееееееееее
$\wt {\Gamma(t)}$ is the matrix operator (\wrt the standard bases) of the
$\C$-linear  operator associated with $\Gamma(t)$  via the isomorphism
$ \H_1\otimes\H_2\cong {\rm Hom}(\H_2^*,\H_1)$  defined so that
\newline$\(\forall h^*\in \H^*_2\)\;\wt {h_1\otimes h_2}(h^*)= h^*(h_2)h_1. $ Here 
$\H^*_2$, the dual space of $\H_2$, can be
 $\C$-linearly identified with the set of $\v{h}$ for $h\in \H_2$ by means of 
$\v{h}(h') = \<h,h'\>$ for all $h'\in\H_2.$ Then
\begin{equation}\label{eq:Hnew58}%еееееееееееееее
\wt {\Gamma(t)}^*\wt {\Gamma(t)} 
=\pmatrix{\bar a & -\bar b\cr \bar b  &\bar d}\pmatrix{a & b\cr -b & d}=
\pmatrix{A & B\cr B^* & D},\ {\rm where}
\end{equation}%еееееееееееееее
\begin{equation}\label{eq:Hnew59}%еееееееееееееее
   A =\bar a a+\bar b b, D= \bar b b+\bar d d \in\R,\ A+D=1,\ B= \bar a b-\bar b d
\end{equation}%еееееееееееееее
We have Det$(\wt {\Gamma}^*\wt {\Gamma})  = AD-\bar B B=
|{\rm Det}(\wt {\Gamma})|^2= |ad+b^2|^2 $ and
Trace$(\wt {\Gamma}^*\wt {\Gamma})  = A+D=1.$ The eigenvalues $x_\pm$ satisfy
$  x_\pm^2 -x_\pm +(AD-\bar B B)=0. $ Thus they are 
\begin{equation}\label{eq:Hnew6}%еееееееееееееее
x_\pm = {1\pm \sqrt{\Delta}\over 2}  ,\ \Delta= 1-4(AD-B^*B)=1 - 4|ad+b^2|^2
\end{equation}%еееееееееееееее
As (unnormalized) eigenvectors for $\wt {\Gamma}^*\wt {\Gamma}$ we can
take
\begin{equation}\label{eq:Hnew11}%еееееееееееееее
\v{\psi_\pm} =-2\pmatrix{B\cr -A+x_\pm}= -\pmatrix{2B\cr D-A\pm\sqrt{\Delta}}=
 -\pmatrix{2(\bar a b-\bar b d)\cr |d|^2-|a|^2\pm\sqrt{\Delta}}\deq
\pmatrix{\bar \alpha\cr -\beta_\pm} .
\end{equation}%еееееееееееееее
The $\phi_j$ are  defined analogously to the 
$\psi_j$ using the matrix  operator 
$ \wt {\Gamma}\wt {\Gamma}^*.$ We can take
\begin{equation}\label{eq:Hnew15}%еееееееееееееее
\phi_\pm =\pm
 \pmatrix{2(b\bar d-a\bar b)\cr |d|^2-|a|^2\pm\sqrt{\Delta}}
=\pm\pmatrix{\alpha\cr \beta_\pm} .
\end{equation}%еееееееееееееее
Then a polar decomposition of $\gg(t)$ is given by
\begin{equation}\label{H16}%еееееееееееееее
{\Gamma}=\sum_{k=\pm}\lam_k\,{\phi_k\over
||\phi_k||}\otimes{\psi_k\over||\psi_k||}.
\end{equation}%еееееееееееееее
Note that this reduces to the polar decomposition
\eqref{H6} at $t=0.$ We have 
$$
||\psi_\pm||=||\v{\psi_\pm}||=||\phi_\pm||=||\v{\phi_\pm}||=
\sqrt{2\sqrt{\Delta}\(\sqrt{\Delta}\pm\(|d|^2-|a|^2\)\)}.
$$
 Using the equality of these norms and applying  equation~\eqref{H16} to $\psi_\pm,$ we
get 
\begin{equation}\label{eq:Hnew14}%еееееееееееееее
\lam_\pm\,{\phi_\pm}=\wt {\Gamma}(\v{\psi_\pm})=\pmatrix{a & b
\cr -b & d} \pmatrix{\v{\alpha}\cr -\beta_\pm} =\pmatrix{a\v{\alpha}-b\beta_\pm\cr
-b\v{\alpha}-d\beta_\pm}.
\end{equation}%еееееееееееееее
Thus
\begin{equation}\label{eq:Hnew18}%еееееееееееееее
\lam_\pm={\<e_j,\wt {\Gamma}(\v{\psi_\pm})\>\over\<e_j,{\phi_\pm}\>}, j=1,2.
\end{equation}%еееееееееееееее
Looking at the 2nd components in equation ~\eqref{eq:Hnew18}, we see
\begin{equation}\label{H19}%еееееееееееееее
\lam_\pm=
\mp\({b\v{\alpha}\over \beta_\pm } +d\).
\end{equation}%еееееееееееееее

\subsection{The Trajectories of the Spinors }
The spinors $\phi_\pm,\psi_\pm$ represent spins in various directions. In this
subsection we calculate these directions.

The operator representing (in SQM) the observable of $2\times$spin in the direction
$\bfn$ is 
$$
\sigma_\bfn= \bfn\cdot {\vec \sigma}=x\sigma_x+y\sigma_y+z\sigma_z,\;
x^2+y^2+z^2=1.
$$
It has unit column eigenvectors with eigenvalues $\pm1:$
$$
{1\over \sqrt{2(1\mp z)}}\pmatrix{x-iy\cr \pm1-z}
$$
The spin up or (+) state is represented by 
$$
{1\over \sqrt{2(1- z)}}\pmatrix{x-iy\cr 1-z}.
$$
The unit vectors $\phi_\pm  $  are of this form  if
we take
$$
x=\Re {\alpha\over \sqrt{\Delta}},\;y=-\Im {\alpha\over \sqrt{\Delta}},\;
z={|a|^2-|d|^2\over \sqrt{\Delta}}.
 $$
We continue to use the abbreviations:
$ C\deq \cos\theta $ and $ S\deq \cos\theta. $
From \eqref{H9},
$$
{|a|^2-|d|^2}={|a_0|^2-|d_0|^2}=C\(|\lam_+(0)|^2-|\lam_-(0)|^2\).
$$
Let $\mf Z$ denote the plane in $\R^3$ defined by $z={|a_0|^2-|d_0|^2}.$
The spin axis trajectory of $\phi_\pm(t)$ intersects $\mf Z$ in a curve $\mf C$ given by:
$$
{\mf C}:  x=\Re \alpha,\; y=-\Im \alpha
$$
We next find an equation in $x$ and $y$ for this curve in the plane $\mf Z.$ 
From \eqref{eq:Hnew15}, we have
$$
\alpha= 2(b\bar d-a\bar b)
=E\cos\omega t+F\sin\omega t-i G\sin\omega t, {\rm where\ }
$$
$$E=S\(|\lam_+(0)|^2-|\lam_-(0)|^2\),\;
F=2S\Im \(\lam_+(0)\v{\lam_-}(0)\),\; {\rm and}
$$
$$G=SC\(1+2\Re \(\lam_+(0)\lam_-(0)\)\)=
SC|\lam_+(0)+\lam_-(0)|^2.
$$
Hence the curve $\mf C$ is given by:
$$
x=E\cos\omega t+F\sin\omega t,\,
y=G\sin\omega t, z={|a_0|^2-|d_0|^2}.
$$

Eliminating the parameter $t$, gives
$$
G^2x^2-2FGxy+(E^2+F^2)y^2=E^2G^2.
$$
The discriminant of this conic is $-4G^2E ^2\le0,$ so if $GE\ne 0$ the curve $\mf C$ 
is an ellipse. The $x,y-$axes coincide with the elliptic axes if and only if
$ FG=0. $ Now $G=0$ means $\theta=0{\rm\ or\ } \theta={\pi\over2}$
when the curve is a single point, or $\lam_+(0)=-{\lam_-(0)} ,$
which is a case with a degenerate polar decomposition. Note that $GE=0$ also leads to similar
degenerate cases.
Thus, aside from degenerate cases, alignment of the $x,y-$axes with the elliptic axes 
is equivalent to $F=0$ or equivalently, $ \Im \(\lam_+(0)\v{\lam_-}(0)\)=0,$
i.e. $ \lam_+(0)$ and $\lam_-(0)$ have the same phase modulo $\pi.$ 
Hence, multiplying $\gg(0)$ by a phase factor, if necessary, we may assume
that  $ \lam_+(0)$ and $\lam_-(0)$ are real, $\lam_+(0)\ge0$, and that the $x-$axis is the
major axis and the $y-$axis is the
minor axis. This choice entails that $k^2\ge C^2 l^2.$

We can relate the present situation to the standard representation of mixed states
in $\C^2$ by means of polarization vectors in the  ball of radius ${1\over2}$ in $\R^3$, as in  [Blum, p.9]. 
To do this we will reconcile the differing conventions by contracting the unit sphere we have been
using to the boundary of the polarization ball, \ie multiply by ${1\over2}$.

The rays $[\phi_\pm]\in \bfP(\C^2)\approx \bf S^2$  give the pure projective states of
the electron which correspond to a pair of antipodal points on the  sphere of radius ${1\over2}$.
The line joining these antipodal points is called the spin
axis of the electron. The spin axis is determined by the
unique point ${\mf e}(t)$ in which it intersects the ellipse
${1\over2}{\mf C}$. This point, regarded as a vector from $\bo$, is
the polarization vector, \ie the mixed state of the electron,
  which is all that SQM accords to the subsystem\ $\ss_1.$
The same applies to the proton states, whose spin axis
passes through a point ${\mf p}(t)$ on
 the ellipse ${1\over2}{\mf C}$ antipodal (with respect to the center of the  ellipse) to  ${\mf e}(t).$

We show in  Figure~\ref{lastFrame}, the polarization ball containing the ellipse which is the 
trajectory of the density operators represented by the blue (proton) and red (electron) balls. 
 
\begin{figure}  %%%%  lastFrame
  \begin{center}
\includegraphics[scale=.8]{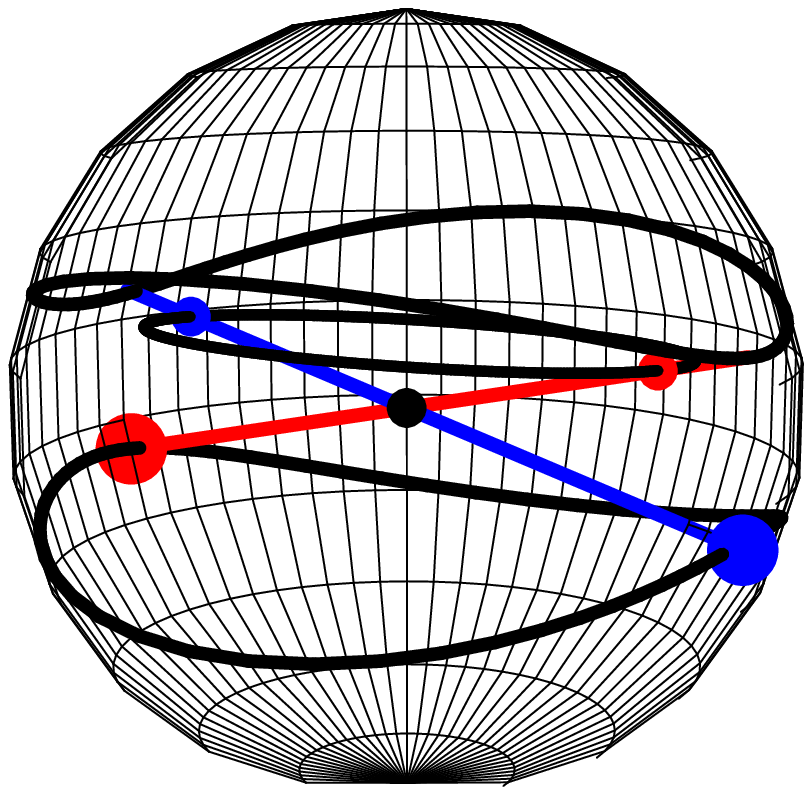}%{lastFrame.eps}%uses K(w)0606.nb 
 \end{center}
  \caption{
Evolution of spectral states at $ t= 6.1{{\mf h}\over\mu}$ for 
$\theta={3\pi\over 7},\,  \lam_+(0)=.94.$
An animation ending with this frame appears at 
 www.princeton.edu/$\wt{\hskip.1in }$jimax/iqm.html
}
  \label{lastFrame}
  \end{figure}

\subsection{Horizontalizing the Spin States}
We begin with two lemmas.

 \begin{lem}\label{lem:lhor} If $t\ra\chi=\chi(t)$ is any curve of non-zero vectors
 in a Hilbert space, then
\newline $e^{i \tau} \chi$ is horizontal with respect to the canonical connection if, and only if,
  \begin{equation}\label{eq:hor0}%еееееееееееееее
  \dot \tau = i{\<\chi,\dot\chi\>\over \|\chi\|^2}.
  \end{equation}%еееееееееееееееееееееееееее
  \end{lem}
{\bf Proof.} The condition for horizontality is:
 \begin{equation}\label{eq:hor1}%еееееееееееееее
0 = \<e^{i \tau}\chi,\partial_t (e^{i \tau}\chi)\>
=\<e^{i \tau}\chi, i \dot\tau e^{i \tau}\chi+e^{i \tau}\dot\chi\>
= i \dot\tau \|\chi\|^2+\<\chi,\dot\chi\>.
 \end{equation}%еееееееееееееееееееееееееее
$ \blacksquare$
\newline{\bf Remark.}  $\dot\tau $ is real \ifff  $\<\chi,\dot\chi\>$ is purely imaginary
 \ifff $\|\chi\|=$ constant.
 \begin{lem}\label{lhor2} If $t\ra\chi=\chi(t) =\pmatrix{\alpha\cr\beta}$ is a 
curve of non-zero vectors
 in the  Hilbert space $\C^2$, with $\beta$ real, then
 $e^{i \tau} {\chi\over \|\chi\|}$ is horizontal with respect to the canonical
connection if, and only if,
  \begin{equation}\label{eq:rhor0}%еееееееееееееее
  \dot \tau = - {\Im}\left({\bar\alpha \dot\alpha\over \|\chi\|^2}\right).
  \end{equation}%еееееееееееееееееееееееееее
  \end{lem}
{\bf Proof.}  Apply the preceding lemma to $\chi_1\deq {\chi\over \|\chi\|}.$
Let $L=\|\chi\|.$
 \begin{equation}%еееееееееееееее
\tf   i\dot \tau = -\<\chi_1,\dot\chi_1\> =- {\bar\alpha \dot\alpha\over
L^2}-{|\alpha|^2 \dot L\over L^3}+{\beta\over L }
{L\dot\beta-\beta\dot L\over L^2} =-
{\bar\alpha
\dot\alpha\over L^2}+ {\rm real}.
 \end{equation}%еееееееееееееееееееееееееее
Using the previous remark, we find that 
 \begin{equation}\label{eq:rhor1}%еееееееееееееее
  \dot \tau  = - {\Im}\left({\bar\alpha \dot\alpha\over
\|\chi\|^2}\right).
 \end{equation}%еееееееееееееееееееееееееее
$\qed$ 

We now want to horizontalize $\phi_\pm,\psi_\pm $ using 
Lemma \ref{lhor2} 
It follows from the lemma that the same factor
$e^{i\tau_+},$ can be used for $\phi_+,\psi_+ $ and similarly for $\phi_-,\psi_- ,$ where
 \begin{equation}\label{eq:H22}%еееееееееееееее
 \dot\tau_\pm = -{{\Im}(\dot\alpha\v{\alpha})\over L_\pm},\   L_\pm\deq || \phi_\pm||^2= || \psi_\pm||^2
  \end{equation}%еееееееееееееее
\begin{equation}\label{H23}%еееееееееееееее
-{\Im}(\dot\alpha\v{\alpha})=4\omega|b_0|^2(|a_0|^2-|d_0|^2),
  \end{equation}%еееееееееееееее
\begin{equation}\label{H28}%еееееееееееееее
 L_\pm=|\alpha|^2+|\beta_\pm|^2= 2\sqrt{\Delta}(\sqrt{\Delta}\pm(|d_0|^2-|a_0|^2))
= 2\sqrt{\Delta}(\sqrt{\Delta}\mp C k l).
  \end{equation}%еееееееееееееее
\begin{equation}\label{eq:H25}%еееееееееееееее
 \tf\dot\tau_\pm = {C S l^3 k \omega\over \sqrt{\Delta}(\sqrt{\Delta}\mp C k l)}.
  \end{equation}%еееееееееееееее
\begin{equation}\label{H30}%еееееееееееееее
\hskip-4in\tf\tau_\pm =
 \end{equation}%еееееееееееееее
$$ {C l\over2 k}\int_0^t{d(\omega t)\over 1+\(({C l\over k})^2-1\)\sin^2 \omega t}
\pm {C^2 l\over2 k}
\int_0^t{d(\omega t)\over \(1+\(({C l\over k})^2-1\)\sin^2 \omega t\)
\sqrt{1+S^2\(({C l\over k})^2-1\)\sin^2\omega t}}.
$$
Let $e={\sqrt{k^2-C^2l^2}\over  k}$ be the eccentricity and let  $\Pi$ be Legendre's elliptic integral 
of the third kind:  %{EllipticPi}(-n,\phi ,m^2)
$$
\Pi(n;\varphi|m)\deq\int_0^\varphi{d\theta\over (1-n \sin^2\theta)\sqrt{1-m\sin^2\theta}}
\;\; ( {\rm\ in\ one\ traditional\ notation.})
$$
\begin{equation}\label{H31}%еееееееееееееее
\tf \tau_\pm = { 1\over2 }\arctan({C l\over k}\tan\omega t)\pm {C^2 l\over2 k}\Pi(e^2;\omega t|S^2 e^2).
  \end{equation}%еееееееееееееее
We now calculate the corresponding canonically horizontalized $\lam_\pm^0.$ We have
$$
\lam_\pm^0=\lam_\pm e^{-2 i \tau_\pm}
$$
since
$$
\gg=\lam^0_+\phi_+^0\ot\psi_+^0+\lam^0_+\phi_-^0\ot\psi_-^0.
$$
Note that $\arctan({C l\over k}\tan\omega t)=-\arg\alpha.$ Thus, using \eqref{H31} and \eqref{H19},
\begin{equation}\label{H26}%еееееееееееееее
\hskip-2in\lam^0_\pm=\mp{\alpha\over |\alpha|}({b\v{\alpha}\over \beta_\pm}+d)
e^{\mp i{C^2l\over k}\Pi(e^2;\omega t|S^2 e^2)}=
  \end{equation}%еееееееееееееее
$$%еееееееееееееее
\pm {-l e^{-i\mu t}\over2k \sqrt{1-e^2\sin^2\omega  t}}
\((k^2\pm\sqrt{\Delta})\cos \omega t+i(C^2 l^2\pm\sqrt{\Delta})\sin\omega t\)
e^{\mp i{C^2l\over k}\Pi(e^2;\omega t|S^2 e^2)}.
$$%еееееееееееееее
$$%еееееееееееееее
\tf \arg \lam_\pm^0=\sigma_\pm\mp {C^2l\over k}\Pi(e^2;\omega t|S^2 e^2)-{\omega t\over4},
$$%еееееееееееееее
where
$$
\sigma_\pm =\arctan\({C^2l^2\pm\sqrt{\Delta}\over k^2\pm\sqrt{\Delta}}\tan\omega t\).
$$
Then for the dynamically horizontalized $ \lam_\pm^H$ we have
$$%еееееееееееееее
\arg \lam_\pm^H=
\nu+\sigma_\pm\mp {C^2l\over k}\Pi(e^2;\omega t|S^2 e^2)-{\omega t\over2},
$$%еееееееееееееее
where
$$
\nu ={C^2\over \sqrt{1-S^2 e^2}}\arctan(\sqrt{1-S^2 e^2}\tan\omega t).
$$

In terms of the angle $\eta$ between the electron and the spin axis (cf. below), we have
$$
\nu ={C^2\over\sqrt{1-S^2 e^2}}\eta.
$$

We have given the details in the case where $\varphi=0$. The calculations in the general case 
are more complicated, but lead to similar results. The polarizing vectors of the electron and the 
 proton again move antipodally on elliptic trajectories. The phases of the $q_\pm^H$ in the 
 general case are also given by elliptic integrals. The main difference is that the argument $\omega t$ 
of the elliptic function is replaced by $\omega (t-t_0)$ for some $t_0.$ 

\subsubsection{Hamiltonian in the polar basis.}

In Section~\ref{evolggk} we described the evolution of a composite system in the polar bundle. 
 This led us to an autonomous system of ODE's satisfied by the canonically horizontal spectral states
$\phi_k(t)$,   
 $\psi_k(t)$ together with  ${\bf\lam(t)}.$ These equations required us to give the Hamiltonian as 
 a matrix $\(H_{jk,mn}\)$ in the polar basis, as well as the related matrices 
$\(\beta_{ab}\).$ In this section and the next we calculate these two matrices for the two
spin-${1\over2}$ systems, allowing one to write the system
Eqs.~\ref{eqxakeq1},\ref{eqxakeq2},\ref{eqxakeq3} explicitly.

We  first  compute the change of basis matrix $Q$ from the standard basis to the basis
\begin{equation}\label{newbas}%еееееееееееееее
\({\phi_i^0\ot\psi_j^0\over\|\phi_i^0\ot\psi_j^0\|}\)_{i,j=\pm}.
\end{equation}%еееееееееееееее
We have
$$
\phi_\pm^0=\pm e^{i\tau_\pm}\pmatrix{\alpha\cr\beta_\pm}
,\;
\psi_\pm^0=e^{i\tau_\pm}\pmatrix{\alpha\cr-\beta_\pm}
$$
$$
\|\phi_\pm^0\|^2=\|\phi_\pm^0\|^2=|\alpha|^2+\beta_\pm^2=\beta_\pm(\beta_+-\beta_-).
$$
\begin{equation}\label{newbas++}%еееееееееееееее
{\phi_+^0\ot\psi_+^0\over\|\phi_+^0\ot\psi_+^0\|}
=
{e^{2i\tau_+}\over\beta_+(\beta_+-\beta_-)}
\pmatrix{\alpha\cr \beta_+}\ot\pmatrix{\alpha\cr-\beta_+}=
{e^{2i\tau_+}\over\beta_+-\beta_-}
\pmatrix{
\alpha^2\over\beta_+\cr-\alpha\cr\alpha\cr-\beta_+
}.
\end{equation}%еееееееееееееее

\begin{equation}\label{newbas+-}%еееееееееееееее
{\phi_+^0\ot\psi_-^0\over\|\phi_+^0\ot\psi_-^0\|}
=
{e^{i(\tau_++\tau_-)}\over|\alpha|(\beta_+-\beta_-)}
\pmatrix{\alpha\cr\beta_+}\ot\pmatrix{\alpha\cr-\beta_-}=
{e^{i(\tau_++\tau_-)}\over\beta_+-\beta_-}
\pmatrix{
\alpha^2\over|\alpha|\cr{-\alpha\beta_-\over|\alpha|}\cr
{\alpha\beta_+\over|\alpha|}\cr|\alpha|
}.
\end{equation}%еееееееееееееее
\begin{equation}\label{newbas2}%еееееееееееееее
{\phi_-^0\ot\psi_+^0\over\|\phi_-^0\ot\psi_+^0\|}
=
{e^{i(\tau_++\tau_-)}\over\beta_+-\beta_-}
\pmatrix{
-\alpha^2\over|\alpha|\cr{\alpha\beta_+\over|\alpha|}\cr
{-\alpha\beta_-\over|\alpha|}\cr-|\alpha|}\!,\;
{\phi_-^0\ot\psi_-^0\over\|\phi_-^0\ot\psi_-^0\|}
=
{e^{2i\tau_-}\over\beta_+-\beta_-}
\pmatrix{
\alpha^2\over\beta_-\cr-\alpha\cr\alpha\cr-\beta_-
}.
\end{equation}%еееееееееееееее

We find:
\begin{equation}\label{HQ}%еееееееееееееее
Q={1\over \beta_+-\beta_-}
\pmatrix{
{\alpha^2\over\beta_+}e^{2i\tau_+}&{\alpha^2\over|\alpha|}e^{i(\tau_++\tau_-)}&
-{\alpha^2\over|\alpha|}e^{i(\tau_++\tau_-)}&{\alpha^2\over\beta_-}e^{2i\tau_-}\cr
-{\alpha}e^{2i\tau_+}&-{\alpha\beta_-\over|\alpha|}e^{i(\tau_++\tau_-)}&
{\alpha\beta_+\over|\alpha|}e^{i(\tau_++\tau_-)}&-{\alpha}e^{2i\tau_-}\cr
{\alpha}e^{2i\tau_+}&{\alpha\beta_+\over|\alpha|}e^{i(\tau_++\tau_-)}&
-{\alpha\beta_-\over|\alpha|}e^{i(\tau_++\tau_-)}&{\alpha}e^{2i\tau_-}\cr
-{\beta_+}e^{2i\tau_+}&{|\alpha|}e^{i(\tau_++\tau_-)}&
-{|\alpha|}e^{i(\tau_++\tau_-)}&-{\beta_-}e^{2i\tau_-}
}.
  \end{equation}%еееееееееееееее
Set $B=-{\beta_++\beta_-\over2|\alpha|}={C k l\over\sqrt{\Delta-C^2 k^2 l^2}} $
and  $\tau=\tau_+-\tau_-={C^2 l\over k}\Pi(e^2;\omega t|S^2 e^2).$

We now transform
$$%еееееееееееееее
H=\mu\pmatrix{1&0&0&0\cr
 0&-1&2&0 \cr
 0&2&-1&0\cr
 0&0&0&1}
$$%еееееееееееееее
from the standard basis to the new basis \eqref{newbas} (see \eqref{defH_jkmn}):
\begin{equation}\label{newH}%еееееееееееееее
\(  H_{jk,mn}\)=Q^*HQ={\mu\over B^2+1}
\pmatrix{
B^2-1&2B e^{-2i\tau}&-2B e^{-2i\tau}&-2e^{-4i\tau}\cr
 2B e^{2i\tau}&1-B^2&-2B^2&2B e^{-2i\tau} \cr
 -2B e^{2i\tau}&-2B^2&1-B^2&-2B e^{-2i\tau}\cr
 -2e^{4i\tau}&2B e^{2i\tau}&-2B e^{2i\tau}&B^2-1}
\end{equation}%еееееееееееееее
We have that $B=\tan\eta$ where $\eta$ is the angle between the
spin axis of the electron and the $z-$axis, so that $2\eta$ is the
angle between the spin axes of the electron and the proton. We may then write:
\begin{equation}\label{2newH}%еееееееееееееее
Q^*HQ={\mu}
\pmatrix{
\cos2\eta&\sin2\eta\, e^{-2i\tau}&-\sin2\eta\, e^{-2i\tau}&(-1+\cos2\eta)e^{-4i\tau}\cr
 \sin2\eta\, e^{2i\tau}&-\cos2\eta&1+\cos2\eta&\sin2\eta \,e^{-2i\tau} \cr
 -\sin2\eta\, e^{2i\tau}&1+\cos2\eta&-\cos2\eta&-\sin2\eta\,  e^{-2i\tau}\cr
 (-1+\cos2\eta)e^{4i\tau}&\sin2\eta\, e^{2i\tau}&-\sin2\eta \, e^{2i\tau}&\cos2\eta}.
\end{equation}%еееееееееееееее

\subsubsection{Computation of the  matrix $\(\beta_{ab}\).$}
We  use  the equation $\dot\rho=\(\beta_{ab}\)$ in the basis
$${\phi_j^0\over\|\phi_j^0\|},\;j=\pm.$$
Let $R$ be the change of basis matrix from the standard basis to
$$
{\phi_\pm^0\over\|\phi_\pm^0\|}=
 \pm {e^{i\tau_\pm}\over \sqrt{\beta_\pm(\beta_\pm-\beta_\mp)}}
\pmatrix{\alpha\cr
\beta_\pm}.
$$
\begin{equation}\label{Rmat}%еееееееееееееее
\tf R={1\over \sqrt{\beta_+-\beta_-}}
\pmatrix{
{\alpha\over|\alpha|}\sqrt{-\beta_-} e^{i\tau_+}&-{\alpha\over|\alpha|}\sqrt{\beta_+} e^{i\tau_-}\cr
\sqrt{\beta_+} e^{i\tau_+}&\sqrt{-\beta_-} e^{i\tau_-}
}
  \end{equation}%еееееееееееееее
In the new basis
$$
\rho=\pmatrix{
|\lam_+|^2&0\cr
0&|\lam_-|^2
}.
$$
\begin{equation}\label{rhodot}%еееееееееееееее
\tf \dot \rho=R^*\(\partial_t{\{}R\pmatrix{
|\lam_+|^2&0\cr
0&|\lam_-|^2
}R^*{\}}\)R\;=\(\beta_{ab}\).
  \end{equation}%еееееееееееееее
\begin{equation}\label{rhbeta}%еееееееееееееее
\hskip-3in\tf \(\beta_{ab}\)={1\over2}R^*
\pmatrix
{
0&\dot\alpha\cr
{\v{\dot\alpha}}&0
}R.
  \end{equation}%еееееееееееееее
\begin{equation}\label{rhbeta2}%еееееееееееееее
\tf \(\beta_{ab}\)={\omega S^2 l^2\over\sqrt{\Delta}}
\pmatrix{
-k^2 e^2\cos\omega t\sin\omega t&
-{e^{2i\tau}}{C k l(k^2 e^2\cos\omega t\sin\omega t
-i\sqrt{\Delta})\over\sqrt{\Delta-C^2 k^2 l^2}}\cr
-{e^{-2i\tau}}{C k l(k^2 e^2\cos\omega t\sin\omega t+
i\sqrt{\Delta})\over\sqrt{\Delta-C^2 k^2 l^2}}&
k^2 e^2\cos\omega t\sin\omega t}.
  \end{equation}%еееееееееееееее

\subsection{Glossary}\label{gloss}
$C=\cos\theta$
\nl$ S=\sin\theta$
\nl$ k=\lam_+(0)-\lam_-(0)$
\nl$  l=\lam_+(0)+\lam_-(0)$
\nl$  a={1\over2}( C l + k)e^{-i\mu t}$
\nl$   d={1\over2}( C l - k)e^{-i\mu t}$
\nl$   b=-{1\over2}S l e^{3i\mu t}$
\nl$ \Delta=l^2\(k^2+S^2(C^2 l^2-k^2)\sin^2\omega t\)$
\nl$ r_\pm^2=|\lam_\pm|^2={1\pm\sqrt{\Delta}\over 2}$
\nl$ \|\phi_\pm\|^2= \|\psi_\pm\|^2=2\sqrt{\Delta}(\sqrt{\Delta}\mp C k l)$
\nl$\alpha=S l(k\cos\omega t-i C l \sin\omega t)$
\nl$\beta_\pm=-C k l \pm\sqrt{\Delta}$
\nl$e={\sqrt{k^2-C^2 l^2}\over k}=$ eccentricity of ellipse
\nl $\omega=4\mu$
\nl $|\alpha|^2=-\beta_+\beta_-$

\section{Categorical Naturality}\label{AppCats}

In this appendix, we briefly consider a few of the  basic concepts of category theory.
Our main use of them will be to make the notion of {\it naturality} precise.

\subsection{Some  concepts and examples from category theory.}

The basic concept of category theory can be axiomatically described.

\nin{\bf Definition.} A {\bf category}  C is a collection  of  primitive 
entities, called  ``arrows'',
 with
a partially defined associative multiplication. 

This means that when the arrows $ab$ and $bc$  are defined so are  $(ab)c$ and $a(bc)$ and they are
equal. It is also required that for every arrow $a$ there are right and left identities: $e_L,e_R$ so
that $e_L a=a e_R = a.$ We can then think of these identities as being or representing objects.

For example, we could take C to be the collection of continuous
maps between topological spaces; this category might be denoted
TOP. With the same ``objects'', but using arrows corresponding to 
homotopy classes of continuous maps we get, say  HTOP.  Here, the
arrows are not merely maps.
 Other examples are homomorphisms between groups, 
defining GROUP  and isometric maps between metric spaces.

\nin{\bf Definition.} A {\bf functor} $ F $ from ${\rm C}_1$ to  ${\rm C}_2$ is a homomorphic 
function $F:C_1\fra C_2$ in the sense that $F(ab)=F(a)F(b)$, when  $ab$ is defined. 

The maps which take  topological spaces to their (singular) homology groups define functors from 
TOP$\fra$GROUP and even HTOP$\fra$GROUP.  

 A {\bf forgetful} functor is one for which  some of the structure and attendant
restrictions on morphisms are omitted. An  example which makes the idea clear is the functor
$FDP: {\rm DIFF}\fra {\rm TOP}$ which assigns 
to differentiable manifolds the underlying topological space
 or in terms of the arrow representatives, regards each differentiable map as just a continuous one.
Usually, such functors have no inverse. For instance, $FDP$ does not:  there is
no natural way to put a differentiable structure on a topological space.

We are interested in some cases where the forgetful functor has an inverse, \ie the additional
structure is canonically  definable. 
\nl {\bf Definition.} A {\bf manifest} functor is a functor $E:C_1\fra C_2$ which  has
 a forgetful functor  $F:C_2\fra C_1$  as an inverse.        

In other words, $E$ adds structure in a natural way, so that every arrow in $C_1$ 
preserves the additional structure. $E$ makes manifest the hidden structure already
possessed by the objects of $C_1.$

A typical  example of a manifest functor can be obtained from the fact 
(see, \eg \cite[III.7]{Reid}) that every non-singular cubic surface  $S$ in $\bfP(\C^4)$ 
contains exactly 27 (complex) lines $L$. Let CS denote the category of such  $S$ and 
CSL the category of  pairs $(S,{\cal L})$ where ${\cal L}$ is the set of lines $L\subset S.$
The morphisms in each category are those induced by automorphisms of $\bfP(\C^4)$. 
Then the forgetful functor from CSL to CS has  an inverse, a manifest functor.

\subsection{The naturality of certain constructions.}

The {\it {raison d'\^etre}}  of our terminology is to make the
statement that the functor   \textsf{P} which assigns to a right toroid, a
Pythagorean partition is manifest.
 This is a precise way of saying that such partitions exist, are
natural, and unique.

As another usage of this terminology, we
 now justify our assertion, made at the end of Section~\ref{sqm+}
about naturally assigning
 projective states to density operators.
  
Let ${\rm MIX}$ denote the category of convex spaces arising as (mixed)
state spaces of algebras of the form ${\cal B}({\cal H})$, bounded operators
on Hilbert spaces of some specified range of dimensions. For example,
we could allow all Hilbert spaces or just those of dimension $n.$ 
 The morphisms are those affine maps induced by unital homomorphisms 
of algebras.  Let  ${\rm MIXR}$ denote the category of triples
$(S,e,P)$ where $S\in {\rm MIX}, P$ is the set of pure states in $S$, \ie\ 
the extreme points of $S$, and $e:S^{\rm reg}\ra P$ is any map of
 the regular mixed states (those with distinct positive eigenvalues) to the
pure states.
\nl{\bf Remark.} There exist manifest functors 
$E:{\rm MIX}\fra {\rm MIXR}$.

For instance, $E(S):= (S,e,P)$, where $e(\rho)$ is the 
 eigenprojection  corresponding to the  smallest eigenvalue of $M$ is
 such a functor.

\begin{thm}\label{mani} %%еееееееTHEOREMееееееееееTTTTTTTTTTTTT
For any manifest functor $E:{\rm MIX}\fra {\rm MIXR},$ and for any
 $\rho\in S\in {\rm MIX}$, 
 we have for the map $e$ of the triple $(S,e,P)= E(S),$  that $e(\rho)$
is an eigenprojection of $\rho.$
\end{thm}%%еееееееTHEOREMееееееееееTTTTTTTTTTTTT
{\bf Proof.} 
The  isomorphisms of MIX  correspond in the usual way to
conjugation by unitary and conjugate-unitary operators.  Suppose now that the functor $E$
assigns to $S$ the triple  $(S,e,P)$ and that $e(\rho)=\gamma$ for some 
regular state $\rho\in S$.
 Let $U$ be any unitary commuting with $\rho.$ Then the
isomorphism
$\iota:{\rm MIX}\ra{\rm MIX}$  defined by  $\iota(\sigma)=U^*\sigma U$ fixes $\rho$.
Thus $E(\iota)$ must fix $\gamma:=e(\rho).$ But $E(\iota)\((S,e,P)\)=(U^*SU,U^*eU,U^*PU).$
 It follows that $U$ commutes with $\gamma.$ Since this  holds for every
unitary commuting with $\rho$, it follows that $\gamma$ is in the double commutant of 
the  algebra $X$  generated by $\rho$. By the spectral theorem, $X$ is generated 
by the eigenprojections
$P_j$ of the self-adjoint operator $\rho.$ Since
$\gamma$ is a one-dimensional projection, it must be a sub-projection of  one of the $P_j.$ 
Thus $\gamma$ is an eigenprojection of $\rho.$ 

\nin$\qed$

\section{Two Identical Systems}\label{appident}
\setcounter{equation}{0}

There are situations where there is (something like) a subsystem $\ss_1$ of 
$\ss$, but no natural complementary subsystem. The most prominent of such 
situations is that of identical particles. In order to show that even such systems
 pose no insuperable obstacle to IQM, we adumbrate the simplest of such cases: 
two identical particles (systems) where the one particle state has 
space  ${\cal H}_1.$ The state space of the two particle system can  then 
be identified with either the  subspace of ${\cal H}={\cal H}_1\ot{\cal H}_1$ consisting 
of the
symmetric tensors ${\cal H}_+:={\cal H}_1\ot^+{\cal H}_1$ or  
the anti-symmetric tensors ${\cal H}_-:={\cal H}_1\ot^-{\cal H}_1.$ Here
 \begin{equation}\label{inde0}%еееееееееееееее
{\cal H}_\pm= \{\gg\in {\cal H}\;|\; \sigma(\gg)=\pm\gg\},
\end{equation}%еееееееееееееее
where $\sigma$ is the $\C-$linear map determined by
$\sigma(\phi\ot\psi)=\psi\ot \phi$ for all $\phi,\psi\in{\cal H}_1.$

Now we have, as usual,  some \pd 
 \begin{equation}\label{inde2}%еееееееееееееее
\gg=\sum_{k} \lam_k\, \phi_k\ot \psi_k.
\end{equation}%еееееееееееееее
Applying the linear map $\sigma$, we find 
that $\gg\in {\cal H}_\pm$ \ifff $\sigma(\gg)=\pm \gg$. This is equivalent, for
 regular $\gg$, to having for all $k$ that there exists a $j$ and a 
$\zeta_j\!\in \bfS^1$ so that $ \phi_k=\zeta_j\psi_j.$
In the regular boson case (distinct  positive  $|\lam_k|$), 
  using the uniqueness property of 
polar decompositions, it follows that there exists a polar
decomposition of the form
\begin{equation}\label{3inde2}%еееееееееееееее
\gg=\sum_{k} \lam_k \,\phi_k\ot \phi_k.
\end{equation}%еееееееееееееее
Thus we can proceed 
 as before in what amounts to a specialized polar decomposition.
 We can speak of an  $\ss_1$-spectral state $\phi_k$ of \ss\ 
but  not of the state of the first particle.

In the fermion case, using \eqref{inde0}, the pairing  $\phi_k,\phi_j$, yields a partition of the
 indices. By
re-indexing using positive integers we can write
 \begin{equation}\label{inde3}%еееееееееееееее
\gg=\sum_{k} \lam_k (\phi_{2k}\ot \phi_{2k-1}- \phi_{2k-1}\ot \phi_{2k})
=:\sum_{k} \lam_k  \gg_k.
\end{equation}%еееееееееееееее
We are in a situation of {\it permanent degeneracy}. No such
vector has a regular  polar decomposition. The notion of regularity must
 be redefined to  cover this case.

 We proceed no further here. We have shown enough to define 
the $\lam_k$ and hence the appropriate polar bundle ${\po}$.
 Moreover, the natural
connection
$A^H$  can be used to define the evolution within ${\po}$ when the
Hamiltonian has the  symmetry properties mandated by SQM, taking 
into account the assumed indistinguishability of the systems.
 Now we can no longer speak of a 
spectral vector state of $\ss_1$.  We {\it can} speak of an  $\ss_1$-spectral state 
$\{\phi_{2k},\phi_{2k-1} \}$ of \ss\ meaning that one of the two particles is in 
 spectral state
$\phi_{2k}$, and the other is in $\phi_{2k-1}. $ This assumes  the
 polar decompositions 
 are regular, which  means the    $|\lam_k|$ are distinct. 
In the anti-symmetric case, we must take extra care not to think we are 
asserting that the state of \ss\ is (collapsed to) $\gg_k.$ 

We have thus shown that although some circumlocution is required, especially in 
the anti-symmetric case, IQM can incorporate identical particles. 

\centerline{\bf\large Acknowledgements} 

The  authors  thank Frances Yu for  her helpful comments and  Steve Miller
 for computer help. The first author would also like 
to thank  Tony Phillips (SUNY Stony Brook, Mathematics) and 
 Peter Sarnak (Princeton Mathematics)
for  facilitating  computer access, including the web pages at 
{\bf www.princeton.edu/$\wt{\hskip.1in}$jimax/iqm.html},
which contain supplementary graphics, current information and related discussions.
% and coming soon: IQM--The Motion Picture.

\bibliographystyle{plain}
\bibliography{AXQMTIS}

\end{document}